\documentclass{aa}  

\usepackage{multirow}
\usepackage{enumitem}
\usepackage{graphicx}
\usepackage{natbib}
\usepackage{comment}
\usepackage[dvipsnames]{xcolor}
\usepackage{txfonts}
\usepackage[colorlinks = true,
            linkcolor = blue,
            urlcolor  = blue,
            citecolor = blue,
            anchorcolor = blue]{hyperref}
\begin{document}

  \title{Galactic-scale evolution of classical and complex radio galaxies.} \subtitle{Impact of ambient morphology and jet geometry}
  \titlerunning{Initial evolutionary phases of extragalactic jet morphologies}

   \author{Gourab Giri,\inst{1}
   Prajnadipt Ghosh,\inst{2}
   Ravi Joshi,\inst{2}
   Anderson Caproni,\inst{3}
   Paola Rossi,\inst{4}
   Gianluigi Bodo,\inst{4}
   Sayan Kundu,\inst{5}
   Kshitij Thorat,\inst{6}
   Swarna Chatterjee,\inst{7}
   Dario Borgogno,\inst{8}
   Valerio Vittorini,\inst{8}
   Marco Tavani\inst{8}
          }

   \institute{
   Istituto Nazionale di Astrofisica (INAF) – Istituto di Radioastronomia (IRA), via Gobetti 101, 40129 Bologna, Italy
   \and
   Indian Institute of Astrophysics, Sarjapur Rd., Koramangala, Bangalore-560034, India.
   \and
   N\'{u}cleo de Astrof\'{i}sica, Universidade Cidade de S\~{a}o Paulo R. Galv\~{a}o Bueno 868, Liberdade, S\~{a}o Paulo, SP, 01506-000, Brazil
   \and
   INAF/Osservatorio Astrofisico di Torino, via Osservatorio 20, 10025 Pino Torinese, Italy
   \and
   Department of Physics, University of Bath, Claverton Down, Bath BA2 7AY, United Kingdom
   \and
   Department of Physics, University of Pretoria, Private Bag X20, Hatfield 0028, South Africa
   \and
   Centre for Radio Astronomy Techniques and Technologies, Department of Physics and Electronics, Rhodes University, Makhanda 6140, South Africa
   \and
   Istituto Nazionale di Astrofisica (INAF) - Istituto di Astrofisica e Planetologia Spaziali (IAPS), via del Fosso del Cavaliere 100, 00133 Roma, Italy 
   \\
    \email{g.giri@ira.inaf.it}
             }

   \authorrunning{G. Giri et al.}
 
  \abstract
   {Extragalactic jets, following their launching, exhibit a wide range of propagation orientations relative to the host galaxy's principal axis. This initial (mis)alignment has been shown to exert a substantial influence on jet’s subsequent large-scale evolution.}
   {This study aims to investigate the spatiotemporal evolution of jets as a function of their propagation direction within their triaxial hosts—introducing varying degrees of environmental hindrance—and as a function of internal jet properties (while maintaining identical jet power: $3\times 10^{44}$ erg/s)—introducing varying collimation and thrust.}
   {Observational data on extended radio sources are re-analyzed to identify key morphological and dynamical traits arising from variations in jet orientation and intrinsic properties. These findings are then systematically tested using a suite of 3D relativistic magneto-hydrodynamic simulations, through the generation of dynamical and radiative maps.}
   {When a jet propagates along host's major axis (path of maximal environmental resistance), it produces an X-shaped morphology with secondary lobe aligns along the minor axis, co-evolving actively alongside the active jet. At intermediate angles to the major axis, the jet morphology transitions into a double-boomerang structure with notably curved lobes. Such lobes are interestingly regenerative through both backflow and jet precession mechanisms, making it difficult to disentangle their origin. Jets propagating along the minor axis (path of minimal resistance) exhibit faster propagation, forming classical double-lobed sources. With increased thrust and improved collimation (keeping jet power constant), these jets advance even more rapidly, potentially evolving into giant radio galaxy candidates. Counterexample sources that deviate from these traits were also modeled. The spatial variation of internal turbulence shows significant fluctuations below $\sim 1$ kpc, with stronger magnetic fields further suppressing these irregularities. Magnetic field plays a key role in the radiative appearance of these sources, modulating features like missing or one-sided (wing) lobe emission, filamentary structures, and warmspot versus hotspot formation.}
   {}


   \maketitle
%

\section{Introduction}
\label{introduction}

A small fraction of galactic nuclei exhibit active behavior, triggered by the accelerated inflow of gas to the central supermassive black hole \citep{Springel2005,Goulding2018}, forming what are known as active galactic nuclei \citep[AGN;][]{Elvis1994,Urry1995,Netzer2015}. Within this subset, a limited number display the presence of plasma outflows \citep{Padovani2011}, resulting from the interplay between magnetic fields and the rotation of either the black hole \citep{Blandford1977} or the accretion disk \citep{Blandford1982}. Evidence of such outflows has been reported as early as the 1950s \citep[e.g.,][]{Jennison1953}, prompting extensive studies by \citet{Fanaroff1974} that classified these radio galaxies (RGs) into two primary categories based on their radio luminosity at 178 MHz: FRI and FRII sources. These classes can also be distinguished by their morphological properties, such as the absence or the presence of active hot-spots, which are zones of enhanced emission at the jet termination points. The ejected outflows, or jets, have been observed to travel distances ranging from sub-kiloparsec to several megaparsec scales \citep[cf.][]{Baldi2023,Dabhade2023}, indicating that their evolution may be intertwined with the diverse ambient environments they traverse. For example, these jets contain plasma of lower density (contrast of $10^{-1} - 10^{-6}$) compared to the density of the surrounding environment \citep{Leahy1984,Rossi2017,Kharb2019}, which hinders the jets' free propagation through space. At the same time, the jets travel at relativistic/sub-relativistic speeds, allowing them to exert mechanical and thermal effects on the ambient medium, thereby exhibiting an interdependent behavior.

Such coevolution of AGN jets and their cosmological environments has been a subject of investigation for over three decades. These studies have demonstrated how powerful jets, as they propagate through the medium of galaxy clusters or rich groups, deposit their mechanical and thermal energy by forming cavity zones, shocks and turbulence \citep[cf.][]{McNamara2012}. This heating process helps to regulate the drastic radiative cooling of the environment \citep[e.g.,][]{Cielo2018}, prevents catastrophic cooling flows \citep[e.g.,][]{Vagshette2019}, and plays a crucial role in shaping the thermodynamic state and large-scale structure of the present-day universe \citep[cf.][]{Soker2016}. This connection is further evident from statistical studies confirming that the most powerful outflows are typically found in the strongest cooling systems \citep[e.g.,][]{Birzan2004}. 
 
Radio observations of these distinct jets, when combined with X-ray studies, reveal environmental imprints such as signatures of jet bending, re-orientation, restarting phases, and off-axis extensions of back-flowing plasma \citep{Smith2002,McNamara2009,David2011,Bogdan2014,Randall2015,Liu2019,Ubertosi2021,Bambic2023}. Studies examining the distribution of galaxies around jet structures also reveal correlations between jet morphologies and the varying degrees of jet-frustration imposed by the surrounding environment \citep{Molnar2017,Hota2022,Mahatma2023}, even on megaparsec scales \citep{Safourios2009,Malarecki2015,Chen2018}. Numerical modeling, as demonstrated by \citet{Falceta-Gonclaves2010,Mandygral2012,O'neill2019,Lalakos2022,Giri2023}, offers deeper physical insights into the various configurations where this intertwined nature has been or is expected to be observed. 

Survey results from contemporary radio telescopes, in this regard, are crucial for expanding the sample of observed sources \citep{Bruggen2021,Knowles2022,Yang2019,Shimwell2019}.  
To highlight a few cases of complexly bent radio sources, investigations such as: \citet{Bruggen2021,Riseley2022,Edler2022,Oei2023} have offered insights into their genesis within larger-scale environments, including the cosmic filaments and voids. 

The present work explores a set of different methodologies for investigating jet-environment interactions in extragalactic systems, particularly during the early evolutionary phases of radio galaxies. It examines the development of extended radio galaxies, highlighting how the geometry of the surrounding medium at galactic scales and variations in jet injection conditions influence their evolution. We delve into understanding whether giant radio galaxies \citep[GRGs; $\gtrsim 700$ kpc;][]{Dabhade2023} with their relatively straight, extended jet propagation, and X-shaped radio galaxies \citep[XRGs;][]{Giri2024} with their distinctive X-like morphology, are influenced by initial conditions within galactic extents. It puts effort to understand how these early conditions may facilitate the development of such large-scale structures beyond galactic scales. Another central objective is to investigate what distinguishes GRGs from classical double-lobed RGs during their incipient stages. While classical doubles are far more common, only about 5–7\% of them eventually grow into GRGs \citep{Dabhade2020_LotssI,Hardcastle2019,Iswara-chandra2020,Zheng2024}. This raises an important question: are there specific initial conditions—such as the nature of jet ejection or the structural dynamics of the host galaxy—that either promote or hinder such extreme growth?

To note, in this context, the triaxial nature of baryon distribution in galaxies, groups, and clusters has been a subject of investigation for a long time \citep{Heiligman1979}, with recent studies extending this focus to dark matter halos \citep{Emami2021,Baptista2023}. Considering the diverse triaxial characteristics of the ambient environment reported in recent years and their potential influence on the formation of various radio galaxies \citep{Hodges-Kluck2010,Bruggen2021,Pandge2022}, it is crucial to explore and expand upon the connections established in the extensive literature on this topic. 

In the following section (Section~\ref{Sec:Radio - Optical Axes Correlation of Extended Radio Galaxies}), we undertake an investigation into the likely effects imposed on jet propagation (defined as radio axis) by the triaxial nature of the ambient medium (measured through the stellar axes in optical bands). This analysis lays the groundwork for the discussions in Sections~\ref{Sec:Optical - Radio axes correlation of GRGs} and \ref{Sec:Optical - Radio axes correlation of XRGs}, where we explore the subset of radio morphologies, as observed in relation to the jet flow direction within their environmental dynamics. To rigorously test these aspects, we have also conducted numerical simulations, which are examined in Section~\ref{Sec:Input from Numerical Simulations}, with a focus on the insights gleaned from magnetohydrodynamical analysis. Finally, in Section~\ref{Sec:Summary}, we summarize the key findings of this study. Furthermore, to convey the potential dichotomy in the evolutionary pathways of classical and giant RGs, we have included a discussion in Appendix~\ref{Sec:Potential dichotomy between classical RGs and giant RGs} outlining the possible physical factors that may contribute to this divergence. Following this, Appendix~\ref{Sec:Role of magnetization in modulating radio galaxy properties} presents a concise discussion on the role of magnetization in shaping radio galaxy evolution and appearance, serving as a complementary extension to the main analysis in Section~\ref{Sec:Input from Numerical Simulations}.

\section{Optical - Radio axes correlation of classical double-lobed RGs}
\label{Sec:Radio - Optical Axes Correlation of Extended Radio Galaxies}

Considerable volume of works have been conducted to investigate whether there is a correlation between the jet flow direction in radio galaxies and the axes of their host galaxies. These investigations aimed to determine whether all AGNs follow a uniform accretion model or if multiple, diverse accretion models are at play, e.g., cold versus hot mode, chaotic versus smooth feeding \citep{Gaspari2020}.

Studies dating back to the 1970s have reported contrasting findings on the preferred orientation of extragalactic radio outflows. Some suggest alignment with the host galaxy’s major axis \citep{Mackay1971}, as traced by the stellar continuum, while others indicate a tendency for jets to propagate along the minor axis \citep{Palimaka1979}. Subsequent studies have rather indicated that no significant correlation exists between the propagation direction of the jet flow and the stellar axes of the host galaxy, suggesting a nearly random distribution of jet orientations relative to galactic disks \citep{Birkinshaw1985, Sansom1987, Kinney2000}. Nonetheless, evidence supporting earlier claims of jet alignment with galactic (minor) axes continued to emerge concurrently \citep{Guthrie1979, McCarthy1987, Condon1991}.

Building on these findings, various explanations have been put forth to support the individual conclusions. Several studies have suggested that jet propagation along the minor axis of the host galaxy’s stellar geometry is expected. This is based on the notion that gas from wet mergers settling in the galactic disk plane \citep[e.g.,][]{Wang2020}, subsequently channels gas towards the central SMBH along the same plane \citep{Liu2004}. Consequently, the spin axis of the SMBH (thus the jet ejection direction) is anticipated to align with the rotation axis of the galaxy. Furthermore, it is plausible that the host galaxy's own gas reservoir is funneling matter into the accretion disk of the SMBH, facilitating such alignment. Yet, detailed mappings of the inner (megamaser) disk, the outer (nuclear or external) disk, and the ejected jet direction in many studies, including those of Seyfert radio galaxies, have revealed significant misalignments \citep{Greene2013, Pjanka2017, Kamali2019}.

In contrast to the preferred orientation for the jet flow direction, insights from numerical simulations suggest that a lack of alignment between the galactic disk and the jet propagation path is rather a naturally occurring phenomenon. Studies by \citet{Fragile2007,Hopkins2012} indicate that off-axis mass infall to the SMBH, triggered by a galaxy merger or gravitational perturbations near the nuclear region, naturally leads to misalignment between the inner accretion disk and the galactic disk. Similarly, \citet{Lalakos2022} demonstrate that collapsing, rotating gas from the host galaxy induces a wobbling jet with a variable propagation direction, while \citet{Liska2018} show that a tilted accretion disk can drive jet precession, further reinforcing the non-alignment hypothesis.

Cases where the radio axis is observed to align with the galactic major axis may sometimes be biased due to contamination of the sample by star-forming galaxies \citep[e.g.,][]{Battye2009,Webster2021,Zheng2024}. This bias can be mitigated by applying a lower limit to the radio structure's extent taken into consideration. For instance, in Seyfert radio galaxies, setting a lower limit of 1 kpc reveals a random orientation of the radio axis relative to the optical major axis \citep{Kinney2000,Gillimore2006}. Similar findings have been reported for extended FR type II radio galaxies in 3CRR sample \citep{Saripalli2009}. However, this later study also establishes that as the overall length of the jet structure increases ($\gtrsim 250$ kpc), the alignment of its propagation path tends to coincide with the optical minor axis of the host galaxy ($\lesssim 30^{\circ}$), aligning with the conclusion, for example, of \citet{Baum1989} for radio extends exceeding 200 kpc.

Such preferential propagation path for extended jets was previously reported by \citet{Palimaka1979}, with subsequent studies by \citet{Condon1991} showing this alignment primarily in elliptical galaxies. \citet{Battye2009} further emphasized the preferred alignment with red, high-concentration galaxies (typically early-type), while \citet{Najar2019} noted a stronger correlation between jet alignment and the optical minor axis in giant-sized radio galaxies ($\ge$ 2 Mpc). Most recently, \citet{Zheng2024} extended these findings to a larger sample (totaling $\sim 3682$ RGs). 

The primary limitation of these studies is the challenge in determining the true three-dimensional axes of the host galaxy, as the observed two-dimensional projection may not accurately represent the intrinsic orientation \citep{Binney1985}. Studies such as \citet{Valtonen1983}, which utilized velocity measurements, and \citet{Sansom1987}, who accounted for the effects of projection in determining the host galaxy's axes, have provided support for the hypothesis of random alignment between jet propagation and the host galaxy's (intrinsic) rotation axis. An alternative approach to studying this correlation has thus focused on dust lane orientation, which is less affected by viewing effect. In this regard, \citet{Schmitt2002} reported randomness in the relative orientation of the jet axis with respect to the (dust) disk plane \citep[however, see,][]{vanDokkum1995,deRuiter2001}. A compilation of individual source studies demonstrating the randomness of jet propagation directions relative to galactic axes can be assembled, including examples such as Centaurus A \citep{Burns1983}, 3C 270 \citep{Mahabal1996}, NGC 1068 \citep{Raban2009}, and PKS 2014-55 \citep{Cotton2020}.

In summary, this discussion emphasizes the absence of definitive evidence supporting a preferred alignment between jet propagation and the optical axes of the host galaxy, challenging the notion of a unified accretion model for AGNs. At the same time, the observed tendency for extended jets ($\gtrsim 200$ kpc) to align with the vicinity of the optical minor axis complicates this issue, warranting further investigation through contemporary telescope data and detailed theoretical studies via numerical simulations.

\section{Optical - Radio axes correlation of `Giant' RGs}
\label{Sec:Optical - Radio axes correlation of GRGs}
The hosts of the majority of GRGs have been observed to be massive ellipticals, with a median mass of approximately $10^{11.5}\, M_{\odot}$ \citep{Oei2023}. This finding, when correlated with the study by \citet{Saripalli2009}, which identified a strong association between the jet propagation direction of GRGs and the minor axis of their host galaxies, suggests that GRG jets likely experience an initial boost in propagation speed along the minor axis due to the higher pressure gradient in that direction \citep[also see,][]{Lan2021}. This gradient facilitates faster jet propagation compared to the flow along the major axis. Recent numerical work by \citet{Giri2025_GRGsim} also demonstrated that GRG jets reach larger scales more quickly when aided by a favorable ambient pressure gradient. In some uncommon cases of GRG formation, spiral or lenticular galaxies host such structures. In these cases, the jets also propagate perpendicular to the galactic plane or perpendicular to the plane of the galaxy group in which the host resides, facilitating the above discussion \citep[e.g.,][]{Hota2011,Bagchi2014,Clarke2017}.

\begin{figure}
\centering
\includegraphics[width=\columnwidth]{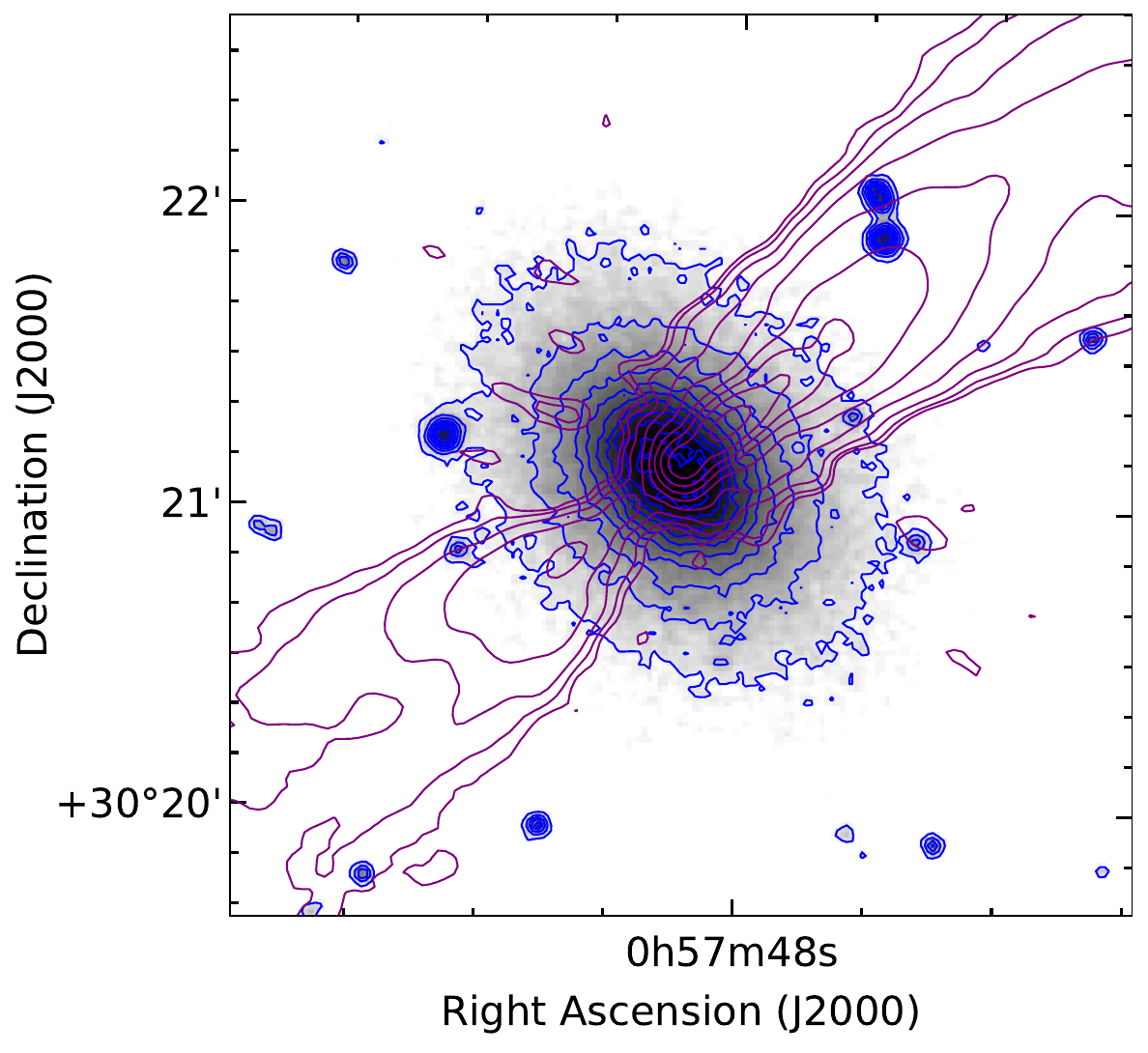}
\caption{NGC 315: a giant radio galaxy with the jet's propagation axis aligned along the minor axis of the host galaxy \citep[$\sim 10^{\circ}$ from the minor axis;][]{Saripalli2009}. The gray-scale DSS2 near infrared image, illustrating the galactic extent and geometry, is overlaid with blue contours from the same DSS2 image for better visualization. The purple contours from LoFAR 144 MHz radio observations are superimposed, indicating the jet's bulk flow direction.}
\label{Fig:GRG} 
\end{figure}

Early studies by \citet{Willis1974,Saripalli1994}, and subsequently by \citet{Subrahmanyan1996}, revealed that in several of their GRGs, the jets have been estimated to propagate along the path of least resistance within the host galaxy, aligning closely with its minor axis. Building on these initial findings, the statistical study by \citet{Saripalli2009} further established a strong preference for GRG jet alignment along the host galaxy’s minor axis. Notably, this trend is exemplified by NGC 315, whose jet orientation closely follows the host's minor axis—an alignment reproduced in this work and illustrated in Fig.~\ref{Fig:GRG}. This tendency was later reaffirmed in an extended GRG sample by \citet{Malarecki2015}. More recently, subtle trends in the studies by \citet{Andernach2021,Oei2022_5Mpc, Oei2023_spiralhost, Koribalski2025} appear consistent with this idea, showing indications that the initial jet propagation of GRGs may be influenced by regions of higher pressure gradients within the host galaxy, potentially aiding their early, relatively unhindered expansion.

GRGs, characterized by their enormous sizes ($\gtrsim 700$ kpc), are generally not found in rich galaxy clusters or dynamically evolved groups. Instead, they preferentially originate and evolve in lower-density environments—such as the outskirts of clusters \citep{Subrahmanyan2008,Bruggen2021}, cosmic filaments \citep{Malarecki2013,Oei2024_CosmicWeb}, or even voids \citep{Oei2022_5Mpc,Oei2024_7Mpc}. This tendency has been confirmed through multiple approaches, including studies of galaxy distributions surrounding GRG hosts \citep[e.g.,][]{Delhaize2021}, group membership assessments cross-matching with cluster and filament catalogs \citep[e.g.,][]{Simonte2024}, and polarization-based analyses \citep[e.g.,][]{Stuardi2020}. Notably, nearly 90\% of GRGs are found in sparse environments, and only about 0.34\% of Brightest Cluster Galaxies host GRGs \citep{Dabhade2020_LotssI,Dabhade2020_SaganI}. These statistics strongly suggest that the evolution of GRGs is intimately connected to their ambient environment, with their growth favored in regions offering minimal resistance and structural hindrance. It therefore aligns with the idea that during their early stages of evolution—while still confined to galactic scales—GRGs may receive additional support from internal pressure gradients within their host galaxies, aiding their rapid early-growth.

Besides environmental effects, GRG jets are also suspected to have well-collimated and sustainable high thrust over large distances, requiring numerical verification. Observationally, this is predicted from their predominantly FR-II appearance \citep{Dabhade2020_LotssI} and the observed collimation in most GRGs, including the record 7 Mpc-long jet \citep{Oei2024_7Mpc}.

\section{Optical - Radio axes correlation of `X-shaped' RGs}
\label{Sec:Optical - Radio axes correlation of XRGs}
\begin{figure}
\centering
\includegraphics[width=\columnwidth]{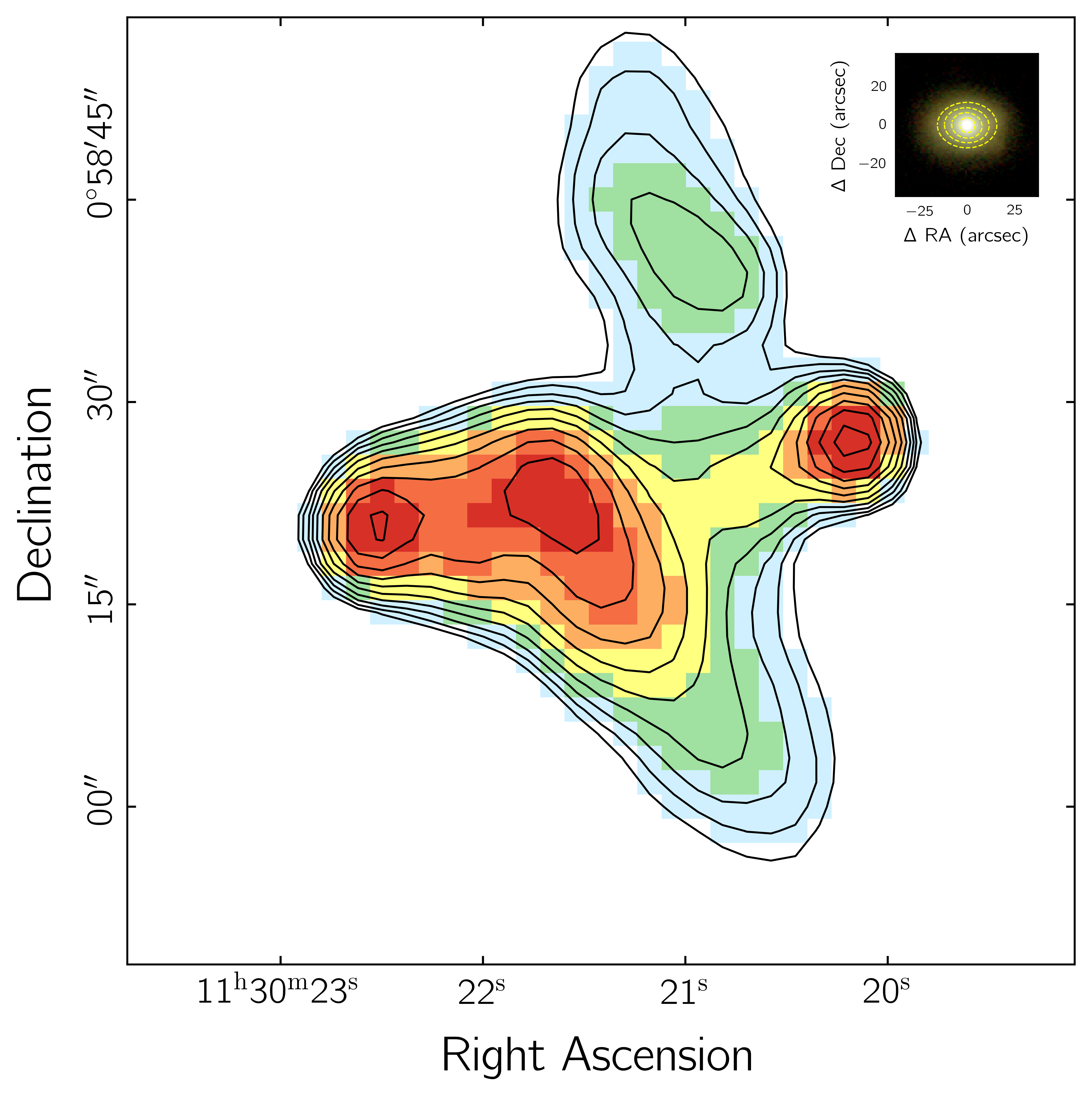}
\caption{4C +01.30: X-shaped jetted structure with the active jet traced by the yellow–red regions, and a nearly orthogonal secondary feature (wings) traced by the green–cyan regions \citep[1.4 GHz VLA FIRST survey map;][]{Wang2003}. The inset displays the host galaxy in optical bands (DECaLS DR10 $grz$ composite), with ellipsoidal isophotal contours overlaid to illustrate its geometric axes: the active jet follows the major axis, wings follow the minor axis.}
\label{Fig:X} 
\end{figure}

X-shaped radio galaxies consist of an active jet lobe, where ongoing jet activity is observed, along with a passively evolving lobe, known as the wing—a fainter secondary structure that co-evolves with the active lobe \citep[cf.][]{Giri2024}. The two lobes are typically oriented nearly orthogonal to each other \citep{Bera2020}. Notably, when examining the alignment of the radio wing relative to the host galaxy's optical axes, a strong correlation emerges, with the wing predominantly aligning along the minor axis of the host \citep[$< 10^{\circ}$ from the minor axis;][]{Saripalli2009}. Fig.~\ref{Fig:X} depicts such a characteristic configuration featuring both active and passive lobes. 
X-ray observations of such radio structure's immediate environment have further supported this conclusion \citep{Hodges-Kluck2010}, reinforcing that optical light distribution is a reliable proxy for tracing interstellar hot gas in galaxies. Individual cases demonstrating these characteristic alignments include, for instance, \citet{Bruno2019,Ignesti2020,Bruni2021}. 

This has motivated the development of models such as those proposed by \citet{Capetti2002}. These models suggest that when an active jet propagates along the major axis of its host galaxy, it encounters significant resistance. This hindrance to free propagation causes part of the jet material to flow backward toward the galaxy's center. The minor axis of the galaxy then becomes instrumental in redirecting this backflow. As a result, the deflected plasma moves along the minor axis, eventually forming the wings observed in some radio galaxies.

Since then, numerous numerical studies have explored various aspects of these radio galaxies, further strengthening this model and its ability to explain their characteristics across multiple scales \citep{Hodges-kluck2011,Rossi2017,Giri2022_XRG,Giri2025_GRGsim}. Observational studies on statistical samples of XRGs have further reinforced this trend, suggesting that both the active lobes and the wings appear to be aware of the geometry of the host galaxy \citep{Gillone2016}. While a handful exceptions exist \citep[e.g.,][]{Hodges-kluck2010_MinorAxisJet,Joshi2019,Yang2022}, they may result from projection effects (since the projected galactic axis may not reflect the true intrinsic rotation axis, as discussed in Section~\ref{Sec:Radio - Optical Axes Correlation of Extended Radio Galaxies}), warranting further investigation. The possibility of alternative models, such as jet reorientation \citep{Giri2024}, can also not be ignored when explaining such counter-examples.

\begin{figure}
\centering
\includegraphics[width=\columnwidth]{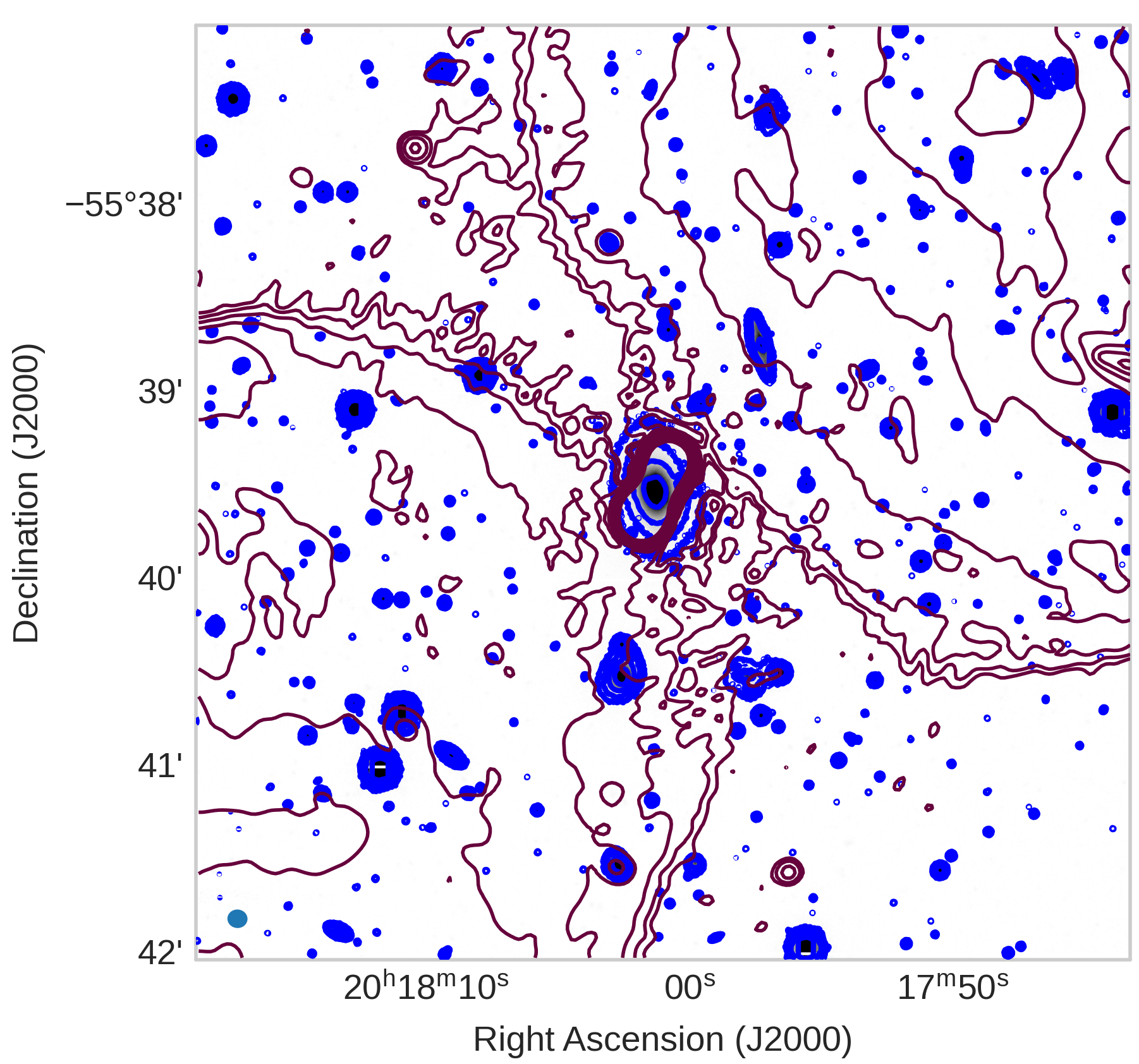}
\caption{PKS 2014-55: an X-shaped (double-boomerang) radio galaxy with multiple epochs of jet emergence \citep[jets from all epochs align at $\sim 40^{\circ}$ to the major axis of the host galaxy;][]{Cotton2020}. This configuration is shown with blue contours from R-band DES observation (concentric ellipses at the center) representing the galaxy isophotes, and purple contours for radio emission from L-band MeerKAT observations, illustrating the phases of jet activity with a `Double-boomerang' shape extending outward \citep[see,][]{Giri2024}. The renewed jet ejection near the center is indicated by the concentration of purple contours.}
\label{Fig:XRG} 
\end{figure}

\begin{figure}
\centering
\includegraphics[width=\columnwidth]{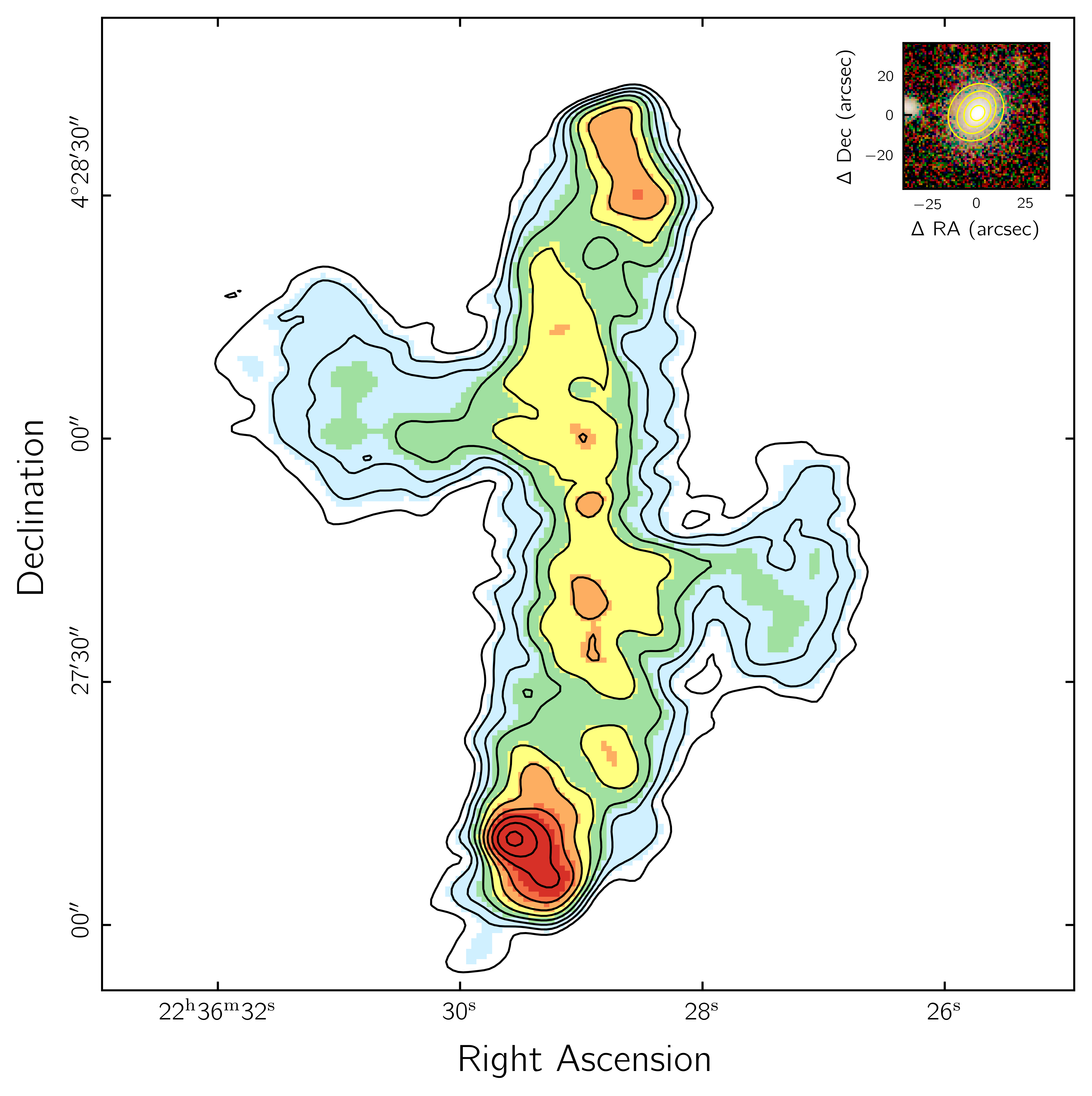}
\caption{J1340+5035: a representative example of an X-shaped radio structure (double-boomerang type), as revealed by its 1.4 GHz uGMRT map. The less intense wings are oriented close ($\lesssim 10^{\circ}$) to the optical minor axis of the host galaxy, shown in the inset figure (a color composite DECaLS DR10 $grz$ image). In the inset, yellow contours trace the elliptical shape of the galaxy. The more active jet propagation (active lobes) is aligned along the north–south direction making an angle to galaxy's major axis. The radio galaxy is part of the XRG sample reported in \citet{Yang2019}.}
\label{Fig:X-wing} 
\end{figure}

It is noteworthy to consider cases where the active jets propagate along an intermediate direction between the host galaxy's major and minor axes. Such configurations often produce classic examples of XRGs, although their morphology more closely resembles what is commonly referred to as a `Double-boomerang' structure \citep{Leahy1984}. A prominent example was reported by \citet{Cotton2020} in a GRG-XRG, where the jet was observed at an angle of approximately $40^{\circ}$ to the major axis. This radio galaxy PKS 2014-55 is reproduced in this study, with a zoomed-in view of the central region shown in Fig.~\ref{Fig:XRG} to highlight the jet flow direction relative to the host galaxy axes. Another example is shown in Fig.~\ref{Fig:X-wing}, where we present both the host galaxy’s geometric structure and the corresponding large-scale radio morphology. This morphology aligns well with predictions from the backflow model, as discussed in dynamical modeling studies by \citet{Rossi2017,Giri2022_XRG}. Several similar XRG cases have been highlighted by \citet{Kraft2005,Hodges-kluck2011,Gopal-krishna2022}, who also supported a backflow origin for these sources. However, alternative interpretations exist—most notably, \citet{Gower1982,Gong2011,Sebastian2024}, proposed that such structures could arise from precessing jets, offering a different perspective on their formation. 

This ambiguity surrounding the origin of double-boomerang structures calls for additional numerical studies that jointly trace their dynamical evolution and radiative fingerprints, helping to disentangle competing models across a broad parameter space. Here, we numerically investigate one such model — the backflow scenario — from both dynamical and radiative perspectives, as discussed in Section~\ref{Sec:Input from Numerical Simulations}. To explore alternative interpretations, we also incorporated analytical predictions from the jet precession model, evaluating the extent to which the simulated structures align with competing formation scenarios.

\section{Input from numerical simulations}
\label{Sec:Input from Numerical Simulations}
Motivated by discussions in Section~\ref{Sec:Radio - Optical Axes Correlation of Extended Radio Galaxies}, \ref{Sec:Optical - Radio axes correlation of GRGs}, and \ref{Sec:Optical - Radio axes correlation of XRGs}—where observational trends suggest that varying levels of environmental hindrance affect a jet's free propagation and subsequent evolution—we aim to explore this phenomenon through high-resolution numerical simulations. By probing a broader parameter space, we seek to examine how these environmental influences and jet parameters shape jet morphology and impact their overall extent. 

In this section, we first outline the numerical setup that forms the backbone of our study (\S~\ref{Sec:Numerical setup}). We then present the resulting morphological and geometrical outcomes, including both dynamical and radiative appearances (\S~\ref{Sec:Morphological distinctions}). This is followed by quantitative analyses of the emergent morphologies (\S~\ref{Sec:Morphological quantification}) and, finally, an exploration of the parameter space relevant to counterexample sources that warrant special attention (\S~\ref{Sec:Numerical insights into the Counterexamples}).

\subsection{Numerical setup}\label{Sec:Numerical setup}
We performed an extensive (relativistic) magneto-hydrodynamical parameter study of different jet-ambient medium configurations using the PLUTO astrophysical code \citep{Mignone2007}. Given our focus on studying jet evolution at galactic scales using high-resolution numerical grids, our analysis concentrates on jets evolving over tens of kiloparsecs, representing their emergent evolutionary phases. However, we note that these simulations are performed in scale-free units, allowing extension to larger scale inferences through appropriate scale-conversion. Here, we assume the following units: length ($L_0$) as 4 kpc \citep[core radius of a typical early-type galaxy;][]{Forman1985}, gas density ($\rho_0$) as 1 amu/cc \citep[typical core value for ellipticals;][]{Baum1989}, and velocity as the speed of light ($c$). These units are also consistent with those adopted by \citet{Rossi2017} for simulating radio galaxies on comparable scales. These three units further facilitate determining units for other variables used to convert the scale-free simulation data cubes into physical forms, such as pressure ($\rho_0 c^2$), time ($L_0/c$), and magnetic field ($\sqrt{4\pi\rho_0 c^2}$).

\begin{figure*}
\centering
\includegraphics[width=2\columnwidth]{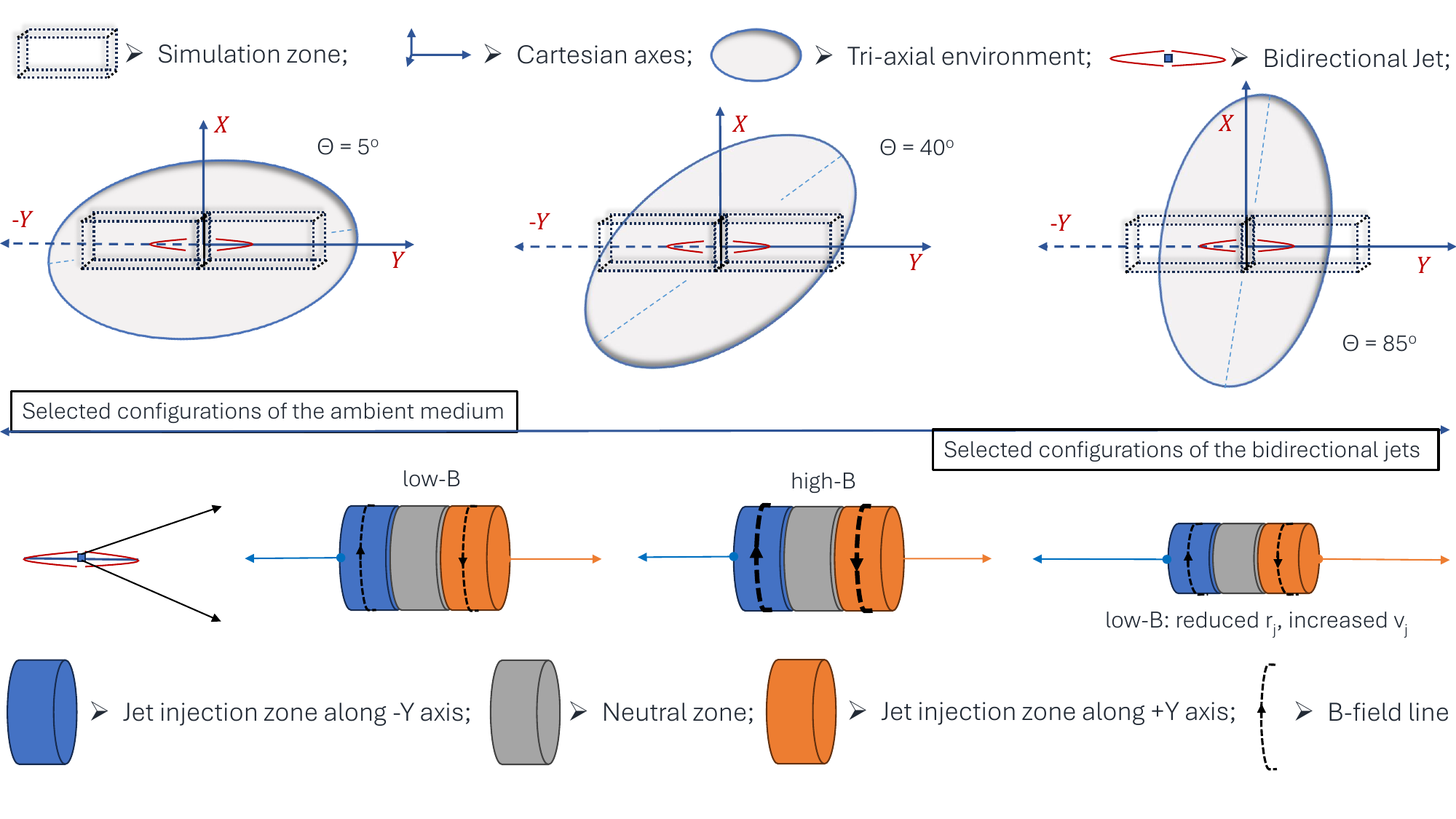}
\caption{Schematic diagram illustrating the various ambient medium and jet configurations used in the simulations. The top panel depicts the jet's propagation at varying angles ($\Theta$) to the major axis (dashed blue lines) within a triaxial medium, viewed along the $z-$axis. The bottom panel highlights different jet injection conditions, where parameters such as jet radius, velocity, and magnetic field vary while maintaining constant jet power across all cases. This comprehensive set of extrinsic and intrinsic parameters seeks to understand the jet-environment interaction during the early stages of their evolution. In all cases, bidirectional jets are aligned along the $y-$axis and injected into the domain through cylindrical injection zones.}
\label{Fig:Setup} 
\end{figure*}

As anticipated, the majority of elliptical galaxies observed in projection are intrinsically three-dimensional ellipsoidal structures \citep{Binney1985, Sansom1987}. Accordingly, we proceeded to formulate a 3D density distribution using the King's $\beta-$profile \citep{Cavaliere1976}, expressed (in Cartesian system) as,
\begin{align}
    \rho (x',\, y',\, z) &= \rho_{0} \left(1+\frac{(x')^{2}}{a^{2}}+\frac{(y')^{2}}{b^{2}}+\frac{(z)^{2}}{c^{2}}\right)^{-3\beta/2}\\
    x' &\equiv x\,{\rm{cos}}\, \Theta - y\,{\rm{sin}}\, \Theta\\ 
    y' &\equiv x\,{\rm{sin}}\, \Theta + y\,{\rm{cos}}\, \Theta
\end{align}
where, $a$, $b$, and $c$ represent the effective core radii of the ellipsoidal medium, with respective values of $\frac{1}{4}$, $\frac{2}{3}$, and $\frac{1}{3}$ of $L_0$, and $\beta = 0.5$ \citep{Rossi2017}. This configuration produces a triaxial environment of hot gas ($\sim 1.9$ kev), with estimated column densities, $\int n\, ({\rm partiles /cc}) \cdot dl \, ({\rm integration\, length}) \equiv$ $[2.8,\;1.5,\;1.1]\times10^{22}\ \mathrm{cm^{-2}}$
along the major, intermediate, and minor axes, respectively.
The parameter $\Theta$ (the rotation angle measured from the $y$-axis) controls the orientation of the host galaxy in the $x$–$y$ plane, with the $z$-axis serving as the axis of rotation, as illustrated in Fig.~\ref{Fig:Setup}. In line with the focus on investigating how the ambient medium influences jet evolution with varying propagation directions, three specific values of $\Theta$ were selected (keeping other jet-environmental parameters constant):
\begin{enumerate}[label=\alph*)]
    \item "\texttt{maj5}": $\Theta = 5^{\circ}$ to align the jet nearly along the major axis.
    \item "\texttt{intm40}": $\Theta = 40^{\circ}$ for an intermediate trajectory of the jet.
    \item "\texttt{min85}": $\Theta = 85^{\circ}$ for aligning jet-flow along the minor axis.
\end{enumerate}
We note that the galaxy reaches a stable equilibrium initially, where pressure is proportional to density ($P \propto \rho (x',\, y',\, z)$), and hydrostatic equilibrium is maintained by the balance between the pressure gradient ($\nabla P$) and gravitational force ($\rho \textbf{g}$).

We simulate bidirectional jets along the $y$-axis, with the jet flow direction fixed along this axis for all cases. The jet injection region is discretized with 10 cells per jet diameter. The injection is divided into three distinct zones: one injecting the jet along the $+y$ axis, another along the $-y$ axis, and a central `neutral zone' (with no flow) separating the two jets to minimize mutual interference (see Fig.~\ref{Fig:Setup}). The injection zones are modeled as cylindrical structures with radius $r_j$, which is constant for a single simulation but may vary across different cases (see Table~\ref{Tab:Parametric_space}). Each cylinder length is set as $l_j = 120\, \mathrm{pc}$. The jets are injected with an underdense profile, where $\rho_j = 10^{-6} \rho_0$ for cases in Table~\ref{Tab:Parametric_space}. 

The jets are magnetized, with the magnetic field configured as follows for one side of the jet:  
\begin{equation}
    B_{x} = B_{j}\, r\, \sin(\theta), \quad 
    B_{z} = -B_{j}\, r\, \cos(\theta),
\end{equation}
where $(r,\, \theta)$ are the polar coordinates in the $(x,\, z)$ plane, perpendicular to the jet flow, asigned in the injection cylinders (Fig.~\ref{Fig:Setup}). For the opposite side of the jet, the sign of the field is reversed, following the prescription in \citet{Rossi2017}. The strength of the initial magnetic field, $B_j$, is parameterized by $\sigma$, the ratio of magnetic energy to matter energy. This is defined as:
\begin{equation}
    \sigma = \frac{B_j^2}{\Gamma^2 \rho_j h_j},
\end{equation}
where $\rho_j h_j$ is the matter-energy flux (enthalpy), and $\Gamma$ is the bulk Lorentz factor of the jet (subscript $j$ indicates jet properties). The expression for $\rho_j h_j$ follows the Taub-Matthews gas law:
\begin{equation}
    \rho_j h_j = \frac{5}{2}P_j + \sqrt{\frac{9}{4}P_j^2 + \rho_j^2c^4}
\end{equation}
where, $P_j$ is the pressure in the jet while injection. 
Similar to the ambient medium's configurations, we introduced three additional variations for the jet injection cases, as outlined below,
\begin{enumerate}[label=\alph*)]
    \item "\texttt{lowB}": $\sigma = 0.01$, $r_j = 200$ pc, $\Gamma = 3$.
    \item "\texttt{lowB\_varRV}": $\sigma = 0.01$, $r_j = 108$ pc, $\Gamma = 5$.
    \item "\texttt{highB}": $\sigma = 0.1$, $r_j = 200$ pc, $\Gamma = 3$.
\end{enumerate}
to investigate the influence of certain jet parameters (e.g., here, the magnetic field strength, radius and flow speed of the injected jet) on jet evolution. We note that the "\texttt{lowB\_varRV}" case is specifically constructed to examine how enhanced jet collimation (achieved by reducing the injection radius) and increased mechanical thrust (via a higher jet velocity) affect the propagation of a jet directed along the host's minor axis. This configuration provides favorable growth conditions that may enable more rapid jet advancement. The $\sigma$ values adopted in this study are consistent with those used in other works, such as \citet{Mukherjee2020}, on similar scales. To distinguish the jet material from the surrounding ambient medium, we employed a passive scalar tracer ($\mathcal{T}$) with the jet-flow following the standard methodology outlined in \citet{Mignone2013}. The value of $\mathcal{T} = 0$ indicates that the simulation grid is filled with ambient matter, while $\mathcal{T} = 1$ corresponds to the presence of jet matter.

It is important to note that, for all the cases considered, the jet power has been kept constant. The jet power is calculated using the following equation:  
\begin{equation}\label{Eq:Power} 
   Q_j = \pi r_j^2 {\rm v}_j \bigg[ \Gamma (\Gamma - 1)\rho_j c^2 + \frac{\rotatebox[origin=tr]{90}{$\prec$}}{\rotatebox[origin=tr]{90}{$\prec$} - 1}\Gamma^2 P_j + \frac{B_j^2}{4\pi}\bigg]. 
\end{equation}  
Here, ${\rm v}_j$ represents the jet propagation speed during its injection into the domain, which is derived from the $\Gamma$ values. The adiabatic index $\rotatebox[origin=tr]{90}{$\prec$}$ varies between $5/3$ and $4/3$, depending on the temperature of the medium. A combination of these parameters results in seven distinct cases, as summarized in Table~\ref{Tab:Parametric_space}. For all the cases, the jet power has been kept at the same level of $\sim 3 \times 10^{44}$ erg/s (indicating FR I/II transitioning characteristics).

Given that we restrict our discussion of flow dynamics within the galactic scales and considering the mechanical power of the flow noted above, it is important to contextualize our model within the radio-loud/radio-quiet (RL/RQ) dichotomy. RQ sources are broadly associated with disk-dominated (spiral) galaxies and typically attributed to exhibit weak, broadened, intermittent/aborted outflows, while radio-loud sources—predominantly hosted by (core dominated) ellipticals—are characterized by sustained, collimated relativistic jets that contribute significantly to the bolometric luminosity \citep{Capetti2006,Kharb2012}. Within this framework, our models are best interpreted as representative of the radio-loud population, given our injected jets are continuous, relativistic (Table~\ref{Tab:Parametric_space}), well collimated (cylindrical injection zone), and evolve within an environment designed to mimic an elliptical galaxy with tri-axial core. Although the relation between radio luminosity and jet kinetic power is highly uncertain—spanning up to two orders of magnitude \citep[][]{Wilson1995,Birzan2004}—our adopted jet power lies on the lower end of the radio-loud regime. In this regime, there is an absence of a well-defined lower boundary in radio power separating the two populations of sources \citep{Xiao2022}, combined with the presence of variability in both classes \citep{Barvainis2005}, further complicates a strict classification based solely on radio luminosity. In this context, the physical properties of our injected flows remain largely inconsistent with radio-quiet outflows. Our choice of a moderate jet power for this study is further motivated by the need to capture significant jet–environment interaction. More powerful jets ($\gtrsim 5\times 10^{45}$ erg/s) are expected to traverse the host galaxy with minimal coupling \citep{Blandford2019}. In contrast, sources such as GRGs and XRGs, while often morphologically FR II, tend to occupy the zone of the (classical) FR I/II division \citep{Schoenmakers2000,Gillone2016,Dabhade2020_SaganI}, supporting the relevance of our parameter choice. Finally, we note that the minimum spatial resolution achieved in our simulations ($10-20$ pc) limits our ability to probe the detailed physical origin of jet collimation at the launching scale.

\begin{table*}
\caption{Parameters that are varied in a set of seven simulations, showcasing diverse jet-environment conditions.}
\begin{center}
\begin{tabular}{ |c|c|c|c|c|c|c|c| } 
 \hline
 Simulation& Jet radius & Jet velocity & Jet B-field & Jet flow along  & Domain (kpc) & Grid & Remarks\\
 label& ($r_{\rm j}$ in pc) & ($\Gamma$) & ($\sigma$) & environment's & $[x],\,[y],\,[z]$ &  &\\
 \hline
  &  &  & & major axis  & $[-12,\, 12]$ & 576 & low B-field\\ 
 \texttt{lowB\_maj5}&200&3&0.01&($\Theta = 5^{\circ}$)&$[-16,\, 16]$&768&cases $\downarrow$\\
 &&&&&$[-12,\, 12]$&576&\\
 \hline
  &  &  &  & intermediate &  &  & variation in flow\\ 
 \texttt{lowB\_intm40}&200&3&0.01&axis&as above&as above&axis\\
 &&&&($\Theta = 40^{\circ}$)&&&\\
   \hline
    &  &  & &  &  &  & variation in flow\\ 
   \texttt{lowB\_min85}&200&3&0.01&minor axis&as above&as above&axis\\
   &&&&($\Theta = 85^{\circ}$)&&&\\
   \hline
    &  &  & &  & $[-6,\, 6]$ & 556 & variation in $r_j$, $v_j$, \\
   \texttt{lowB\_varRV\_min85}&108&5&0.01&minor axis&$[-16,\, 16]$&1482& but raising same jet\\
   &&&&($\Theta = 85^{\circ}$)&$[-6,\, 6]$&556& power as others\\
   \hline
    &  &  & &  &  &  & \\ 
   \texttt{highB\_maj5}&200&3&0.1&major axis&same as&same as&High B-field\\
   &&&&($\Theta = 5^{\circ}$)&top row&top row&cases $\downarrow$ \\
  \hline
    &  &  & & intermediate & &  & \\
   \texttt{highB\_intm40}&200&3&0.1&axis&same as&same as&variation in flow\\
   &&&&($\Theta = 40^{\circ}$)&top row&top row&axis\\
   \hline
    &  & &  &  &  &  & \\ 
   \texttt{highB\_min85}&200&3&0.1&minor axis&same as&same as&variation in flow\\
   &&&&($\Theta = 85^{\circ}$)&top row&top row&axis\\
   \hline
\end{tabular}

\label{Tab:Parametric_space}
\end{center}
\small
\textbf{Notes.} The first column lists the simulation labels, identifying the unique attributes of each case. The 2nd, 3rd, and 4th columns detail the jet injection radius, velocity, and magnetic field strength, respectively. These parameters are carefully selected to maintain a constant jet power across all cases, allowing the investigation of whether intrinsic parameters, rather than jet power, drive radio galaxy evolution. The 5th column describes the jet's propagation relative to the axis of the ambient medium. The 6th and 7th columns specify the computational domain and grid distribution, ensuring that five grid cells are allocated per jet injection radius. Finally, remarks for each case are included in the last column for clarity. The jet–to–ambient density contrast is fixed at $10^{-6}$ in all of these cases.
\end{table*}

In summary, the suite of seven simulations reported in Table~\ref{Tab:Parametric_space} can be categorized as follows. Aligned with the primary objective of investigating the morphological and spatial evolution of jets as a function of their propagation orientation relative to the host galaxy’s gas distribution, three baseline simulations were performed. In these cases, jets were launched along the major axis (`\texttt{\_maj5}), minor axis (`\texttt{\_min85}), and an intermediate angle between the two (`\texttt{\_intm40}). The jet power was held constant across these runs to isolate the effects of propagation direction. To further examine whether more intrinsic jet parameters—specifically variations in jet radius and flow velocity (while maintaining equal mechanical power)—affect the jet evolution, an additional simulation was conducted (`\texttt{lowB\_varRV\_min85}'). This subset was specifically chosen to explore an initial conditions that may contribute to the formation of the GRG subclass. Building upon the initial three runs, in which magnetic fields were initialized at moderate strengths (`\texttt{lowB}') characterized by the magnetization parameter ($\sigma$), we introduced a higher B-field strength in subsequent simulations (three cases with higher injection values of the B-fields; `\texttt{highB}'). This modification aimed to explore the influence of magnetic instabilities and turbulence on jet collimation and disruption, as well as the resulting impact on synthetic radio morphology \citep[e.g., hotspot vs.  warmspot;][]{Fanaroff1974,Leahy1997,Horton2023}. Together, these seven simulations offer a targeted framework for understanding the interplay between jet parameters, propagation direction, and environmental influence, as detailed in Table~\ref{Tab:Parametric_space}.

In addition to the seven primary simulations listed in Table~\ref{Tab:Parametric_space}, we evolved two of these cases—\texttt{lowB\_varRV\_min85} and \texttt{highB\_min85}—to later times (Appendix~\ref{Sec:Potential dichotomy between classical RGs and giant RGs}) to verify that our main conclusions remain robust to larger spatial extents. While Table~\ref{Tab:Parametric_space} focuses on the key jet–galaxy parameters relevant to the observed trends, Table~\ref{Tab:Counter_examples} includes two additional jet-intrinsic models that probe alternative regions of parameter space, aimed at understanding potential counterexamples. Across all simulations, we varied four jet parameters—its injection radius ($r_j$), injection velocity ($\Gamma$), density contrast ($\rho_j/\rho_0$), and magnetic-field strength ($\sigma$)—while the ambient medium was modified only by changing the jet’s orientation relative to the major axis ($\Theta$).

\subsection{Morphological distinctions}\label{Sec:Morphological distinctions}
This subsection qualitatively interprets the simulation outcomes in terms of their physical evolution (dynamical behavior: \S~\ref{Sec:Dynamical behavior}). It then examines how these outcomes manifest in their observable appearance (radiative imprints: \S~\ref{Sec:Radiative imprints}).
\subsubsection{Dynamical behavior}\label{Sec:Dynamical behavior}
\begin{figure*}
\centering
\includegraphics[width=2\columnwidth]{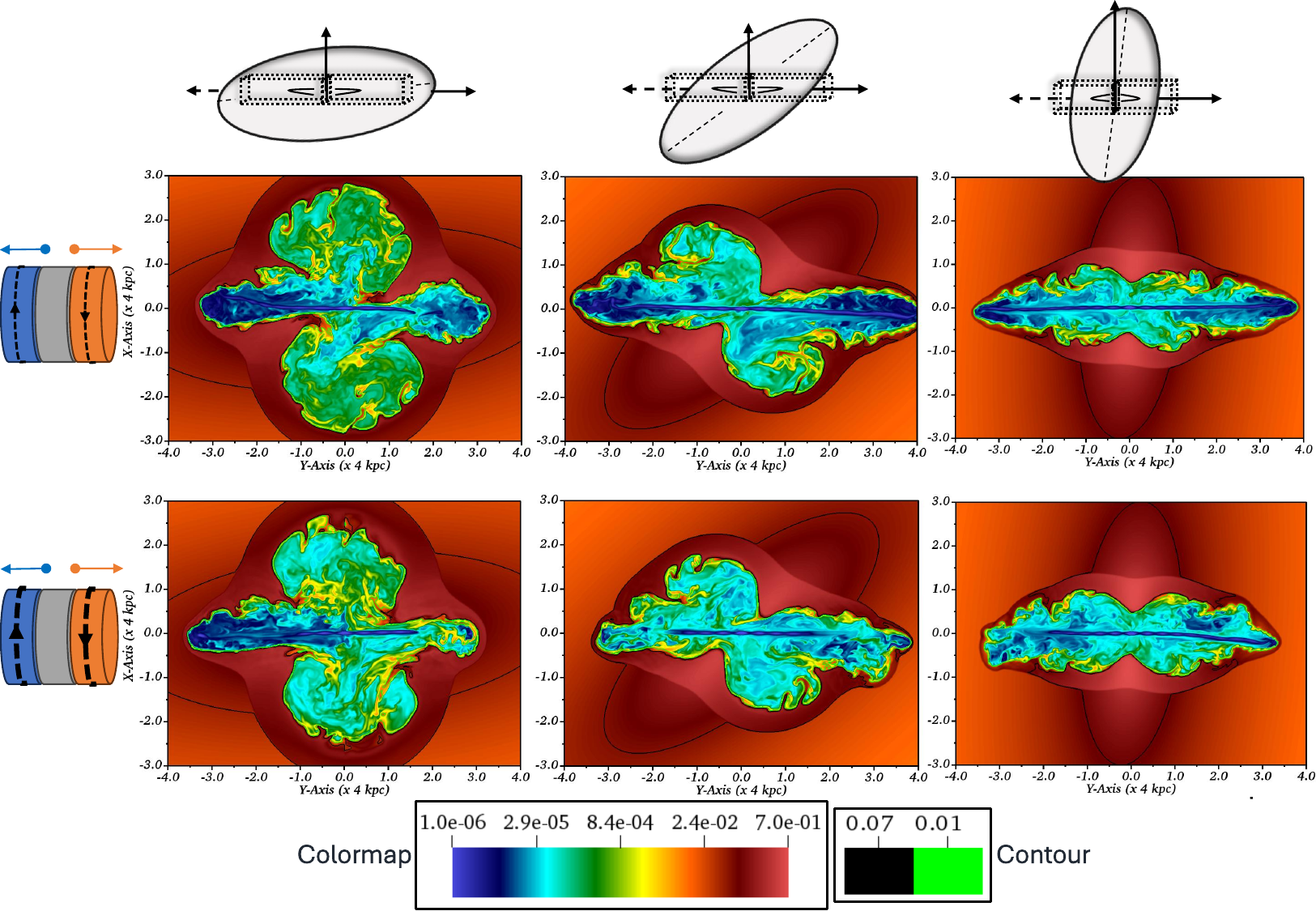}
\caption{Morphological variations (2D slices) resulting from varying jet propagation angles relative to the ambient medium’s axes (\textit{1st col.:} \texttt{\_maj5}, \textit{2nd col.:} \texttt{\_intm40}, \textit{3rd col.:} \texttt{\_min85}) and different jet magnetic field strengths (\textit{top row:} `\texttt{lowB}', \textit{bottom row:} `\texttt{highB}'). The generated structures include an X-shaped radio galaxy (dynamical age: 7.2 Myr), a double-boomerang morphology (dynamical age: 5.2 Myr), and a classical double-lobed radio galaxy (dynamical age: 3.9 Myr), from left to right respectively. The simulations show that while magnetic fields minimally influence large-scale dynamics in these early stages, they play a crucial role in internal processes such as instabilities leading to jet bending and reduction in lateral expansion of the lobes. We note that the jet propagation axis is fixed along $y-$axis for all these cases. The colorscales are normalized with respect to $\rho_0$ (1 amu/cc).}
\label{Fig:Morphology_Bs} 
\end{figure*}

Fig.~\ref{Fig:Morphology_Bs} concisely presents a combination of schematics and simulation slices ($x-y$, $z=0$) of evolved jets for different cases, captured at stages where the jet extents approach the simulation boundary. Given the fixed jet propagation along the $y-$axis in all cases, two key conclusions emerge. First, jets aligned with the major axis—where the surrounding medium presents greater obstruction—tend to advance more slowly than those along the minor axis. In comparison, jets along the minor axis encounter less hindrance and reach comparable extents in nearly half the time (see, Fig.~\ref{Fig:Morphology_Bs}). Second, these variations in environmental obstructions and the orientation of pressure gradient vectors relative to jet propagation significantly influence jet morphology. If similar environmental conditions persist beyond galactic extents, it is likely that the jets will continue to evolve in a comparable manner on larger scales \citep[e.g,][]{Hodges-kluck2011,Giri2023,Giri2025_GRGsim}. Thus, the jet’s early evolution within the host galaxy may act as a tracer for its behavior at larger spatial extents.
A similar line of reasoning has been presented in Section~\ref{Sec:Optical - Radio axes correlation of GRGs},~\ref{Sec:Optical - Radio axes correlation of XRGs}, where the discussion focused specifically on GRGs and XRGs.

When the jet propagates along the direction of greater hindrance—the major axis of the galactic gas distribution—an X-shaped morphology emerges, consistent with the backflow model for XRG formation \citep{Capetti2002}. Alternatively, when the jet propagates along the minor axis of the host galaxy, it evolves as a linear structure resembling classical double-lobed radio galaxies. This latter outcome, characterized by the absence of wing-like features, further explains observational findings of why XRGs rarely exhibit jet propagation aligned with the minor axis (see Section~\ref{Sec:Optical - Radio axes correlation of XRGs}). Morphological differences due to variations in magnetic field strength appear minimal dynamically for XRGs; however, stronger magnetic tension clearly restricts the lateral expansion of both the primary (active) and secondary (wing) lobes. The influence of magnetic fields is more pronounced in the case of double-lobed structures; stronger fields introduce instabilities, such as kink modes, which can lead to noticeable jet bending \citep[e.g.,][]{Mignone2013,Mukherjee2020}. This characteristic of jet bending hints that such double-lobed sources are more likely to evolve into classical radio galaxies rather than GRGs, as the latter requiring sustained collimation and minimal disruption \citep{Oei2024_7Mpc}.  

\begin{figure}
\centering
\includegraphics[width=\columnwidth]{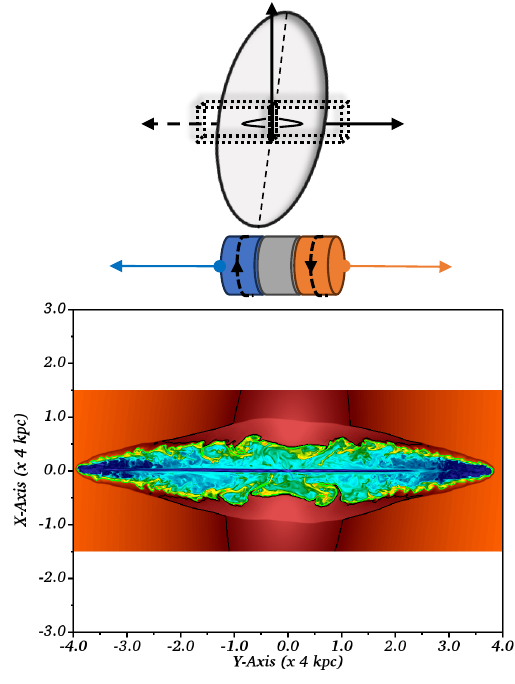}
\caption{A higher jet speed and reduced injection radius, while keeping the jet power constant, result in a rapidly advancing jet with a narrow nose-cone morphology, reaching a total extent of 32 kpc in just 2.6 Myr. The jet propagates along the minor axis of the environment. Such a configuration (`\texttt{lowB\_varRV\_min85}') is likely a prime candidate for forming a giant radio galaxy in later stages of their temporal evolution. The colorscale is same as in Fig.~\ref{Fig:Morphology_Bs}.}
\label{Fig:Morphology_RV} 
\end{figure}

To investigate the role of sustained collimation and minimal disruption in the spatial evolution of the injected jets, we performed the simulation `\texttt{lowB\_varRV\_min85}'. In this run, the jet power remains same like the other cases, but the jet radius is reduced to minimize ambient disruption, and the jet speed is increased to match the power of the other cases, resulting in a corresponding increase in jet thrust (Table~\ref{Tab:Parametric_space}). The result, shown in Fig.~\ref{Fig:Morphology_RV}, reveals that the jet evolves significantly faster, about 1.5 times quicker compared to the suspected classical double-lobed cases (`\texttt{lowB\_min85}' and `\texttt{highB\_min85}'). After a time of 2.6 Myr since injection, the narrow nose-cone shape of the jet head (Fig.~\ref{Fig:Morphology_RV}) indicates rapid flow and unhindered growth to larger scales \citep{Todo1992,Stone2000}. This simulation, which also propagates along the minor axis ($\sim 3\times$ faster than major-axis cases), further suggests why GRGs are likely to evolve along the minor axis of the host galaxy {\color{ForestGreen}(refer to Section~\ref{Sec:Optical - Radio axes correlation of GRGs}).}

One of the most intriguing case arises when the jet propagates along an intermediate direction between the major and minor axes of the host galaxy. This configuration generates a double-boomerang structure (Fig.~\ref{Fig:Morphology_Bs}; \textit{2nd col.}), as the backflowing material from the jet head follows the pressure gradient vector, which has its highest magnitude along the minor axis. Since the minor axis, for this case, closely aligns with the jet's propagation direction (Fig.~\ref{Fig:Setup}), it initiates strong bending. Our findings support the jet bending demonstrated in the simulations by \citet{Rossi2017}, where the jet flow is directed $30^{\circ}$ from the galactic major axis. The observational predictions of \citet{Kraft2005,Hodges-Kluck2010,Gopal-krishna2022} and \citet{Patra2025}, which support a backflow-driven origin for double-boomerang radio galaxies, appear to be well motivated in this context.
Assessing such configuration's appearance in sky-projected emission maps is crucial, as intensity maps capture the integrated three-dimensional structure, offering a more realistic view than the two-dimensional slices used in Fig.~\ref{Fig:Morphology_Bs}. We note that due to the intermediate level of environmental hindrance in this case, the jet reaches a comparable extent to other runs in an intermediate evolutionary time ($\sim$ 5.2 Myr).

An important insight from Fig.~\ref{Fig:Morphology_Bs} emerges when examining the ambient density distribution: the jet imparts feedback by depositing mechanical and thermal energy into the surrounding medium. The bow shock at the jet head and the extended shock envelope around the cocoon clearly trace this interaction between the thermal ambient gas and the non-thermal jet plasma. A canonical example of jet–interstellar medium interaction is observed in the Teacup Galaxy \citep{Audibert2023}, indicating the presence of such co-evolutionary mechanisms and underscoring their potential importance in the broader context of galaxy evolution. Notably, the XRG case—owing to its more isotropic distribution of non-thermal material—produces a correspondingly more isotropic feedback pattern compared to classical doubles (e.g., the kinetic energy transferred to the ambient medium is nearly twice as high between the two cases). This has implications for the long-standing (isotropic) heating problem in galaxy groups and clusters, where AGN jets, despite being intrinsically bipolar, appear to heat their environments more uniformly. Bent or deflected sources may therefore provide enhanced and more widespread feedback, a possibility of growing relevance in current studies of large-scale cosmic structures' thermodynamics \citep{Cielo2018,Giri2023}. A future study examining the detailed feedback effects of these distinct jet topologies on the large-scale ambient medium is planned, given their particular importance on galaxy-group scales \citep{Cavaliere2008}.

\subsubsection{Radiative imprints}\label{Sec:Radiative imprints}
To convert the 3D dynamical data cubes into emissivity cubes, we followed the methodology detailed in \citet{Meenakshi2023}, which assumes that 10\% of the internal energy ($E_n$) is converted into non-thermal particle energy. This assumption enables the calculation of the particle spectral normalization, $N_0$, using a power-law index $p = 2.2$ and electron (charge $e$ and mass $m_e$) Lorentz factors ranging from $10^2$ to $10^6$, as follows,
\begin{equation}
    N_0 = \frac{0.1 E_n}{m_e c^2 \int_{10^2}^{10^6} \gamma^{-p+1}d\gamma}
\end{equation}
The synchrotron emissivity can thereafter be expressed as,
\begin{equation}
\begin{split}
    \mathcal{J'}_{\rm syn} = N_0 \frac{3^{p/2}e^2\nu'^{-(p -1)/2}\left| \vec{B'} \times \vec{\hat{n'}} \right|^{(p+1)/2}}{2c(p + 1)} \\
    \mathcal{G} \left( \frac{p}{4} + \frac{19}{12} \right) \mathcal{G} \left( \frac{p}{4} - \frac{1}{12} \right) \left( \frac{e}{2\pi m_e c}\right)^{(p+1)/2} 
\end{split}
\end{equation}
The primed quantities (emissivity, emitting synchrotron frequency, magnetic field, and line-of-sight vector) are defined in the rest frame attached to the radio galaxy, and are converted to the observer’s frame using the transformation equations provided in \citet{Vaidya2018}. $\mathcal{G}$ represent the gamma-function.
The intensity maps were then generated by integrating the emissivity data cubes along the line of sight (z-axis), thereby projecting the three-dimensional structures onto the sky plane ($X-Y$) as follows,
\begin{equation}
\begin{split}
    I_{\nu} (\nu, X, Y) &= \int_{-\infty}^{\infty} \mathcal{J}_{\rm syn}  \, \, dZ 
\end{split}
\end{equation}
The observation frequency ($\nu$) is set at 1 GHz, aligning with the operating bands of several contemporary radio telescopes. The maps were smoothed with a circular Gaussian beam of 0.45 arcsec to provide a more realistic representation of the emission features at GHz frequency. The methodology described above has been employed by \citet{Meenakshi2023} to jetted sources spanning a broad range of radio powers during their early stages of evolution, and has subsequently been extended to the study of giant radio galaxy jets by \citet{Giri2025_GRGEm} to examine their large-scale implications, highlighting the broad applicability of the assumptions across different domains.

\begin{figure*}
\centering
\includegraphics[width=2\columnwidth]{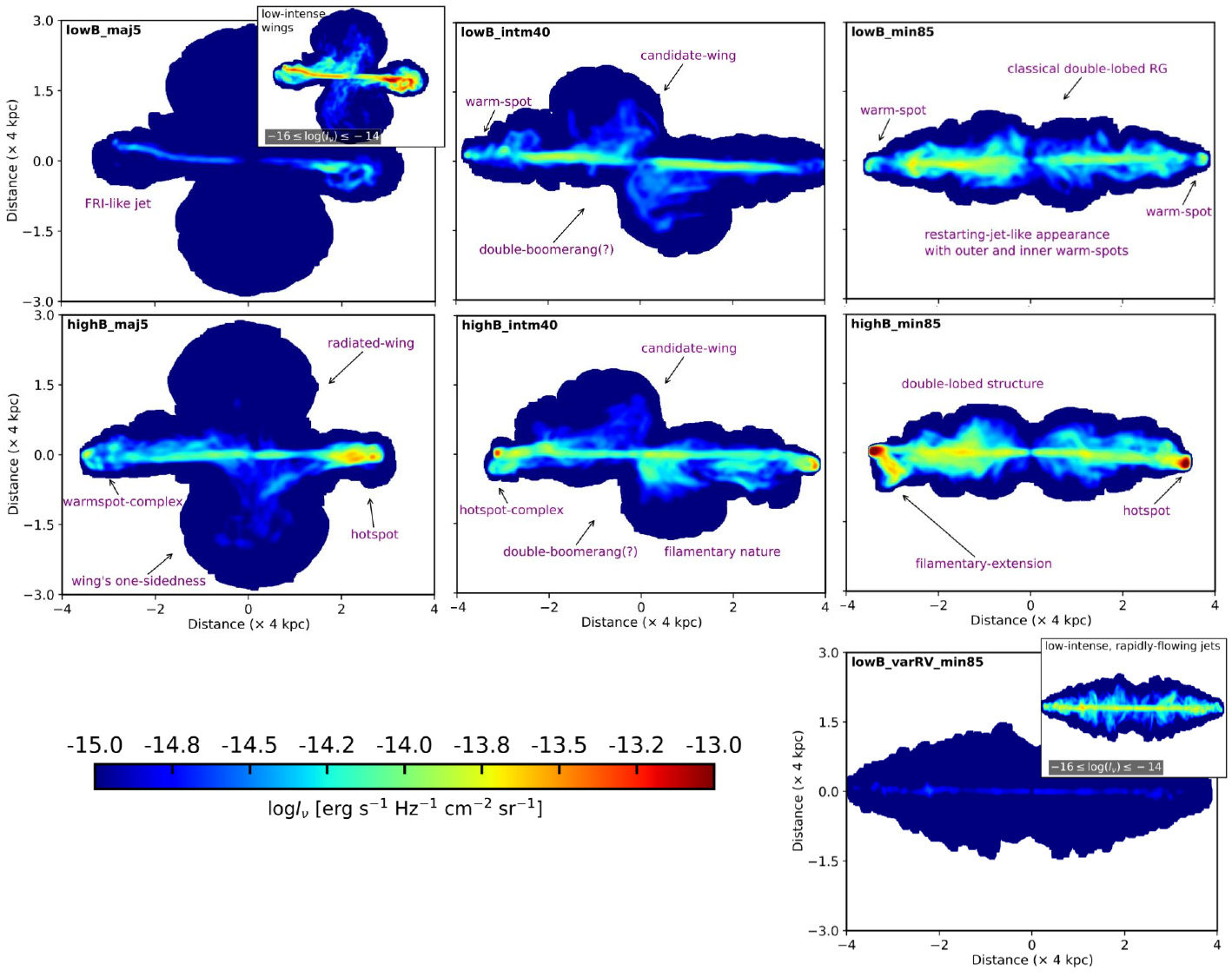}
\caption{1 GHz sky-projected intensity maps (smoothed with 0.45-arcsec circular beams) from seven simulations, viewed along the $z-$axis. Distinct emission features in each case are marked with arrows, several of which remain key questions for contemporary radio telescopes. Notably, the lack of hotspots in low-magnetization jets—contrasting with their presence in high-magnetization cases—offers crucial insights into the conditions required for hotspot formation in radio galaxies. This contrast is especially significant given that all simulations share identical jet power.  While the magnetic field may play a limited role dynamically, it proves critical in shaping the radiative appearance of these sources.}
\label{Fig:Emission_col} 
\end{figure*}

In Fig.~\ref{Fig:Emission_col}, we present the intensity maps for the seven simulated jet-evolution scenarios at their respective evolved times, as shown in Figs.~\ref{Fig:Morphology_Bs} and \ref{Fig:Morphology_RV}.

Fig.~\ref{Fig:Emission_col} highlights two key points. First, while sky-projected emission maps capture the underlying dynamical morphology, the visibility of finer structures is strongly influenced by observational sensitivity. Under a default colorscale we choose ($-15 \leq \log(I_{\nu}) \leq -13$), the XRG cases resemble FR-I-like morphologies in low-magnetization scenarios and FR-II-like jets with asymmetric wings in high-magnetization cases. However, applying a deeper intensity cutoff ($-16 \leq \log(I_{\nu}) \leq -14$), as shown in the insets, reveals the full X-shaped morphology otherwise hidden below the detection thresholds of ($-15 \leq \log(I_{\nu}) \leq -13$). A similar effect is seen in the bottom row, where the rapidly flowing low-magnetization jet shows extended features only in the deeper map. These results underscore the importance of observing sensitivity of radio telescopes—as emphasized in recent studies by \citet{Yang2019, Sebokolodi2020, Mahatma2023, Koribalski2025}—for accurately interpreting such complex radio galaxy structures \citep[also, see,][]{Parma1985}.

The second insight is particularly critical and offers a valuable implication for radio galaxy studies: the strength of the magnetic field directly influences the observability and radiative characteristics of jet-driven structures. A comparison of the XRG simulations with low and high magnetization reveals that the average tracer-weighted \citep{Mignone2013} magnetic field strength increases from 9.6 $\mu$G to 19.1 $\mu$G in the evolved stages, significantly enhancing the visibility of extended structures. Even within the same low-magnetization group, the classical double-lobed sources appear more prominent than the XRGs, corresponding to a higher average magnetic field (18.6 $\mu$G vs. 9.6 $\mu$G). This disparity (within the same low-magnetization group) arises despite identical initial injection conditions, indicating that the larger volume occupied by XRGs leads to a lower average B-field due to magnetic dilution over expanded structures (see also, Fig.~\ref{Fig:B-Dyn_3D}).
This trend is further supported by the double-boomerang cases, where the average magnetic field evolves to the intermediate values of 15.6 $\mu$G and 20.8 $\mu$G for low- and high-magnetization runs, respectively. 

A key observational consequence of this magnetic field dependence is the formation of prominent hotspots at jet termini in high-B-field scenarios, contrasting with warm, diffuse features \citep[warm-spots;][]{Leahy1997} in lower-B-field counterparts. Given that all simulations start with the same jet power and evolve under similar environmental conditions for each morphology category, our results suggest that the classical FR I/II dichotomy—sometimes attributed to more intricate factors (e.g., host galaxy mass \citep{Mingo2019}, jet composition \citep{Croston2018})—may in fact also be intrinsically linked to the magnetic field strength within these systems.

\begin{figure}
\centering
\includegraphics[width=\columnwidth]{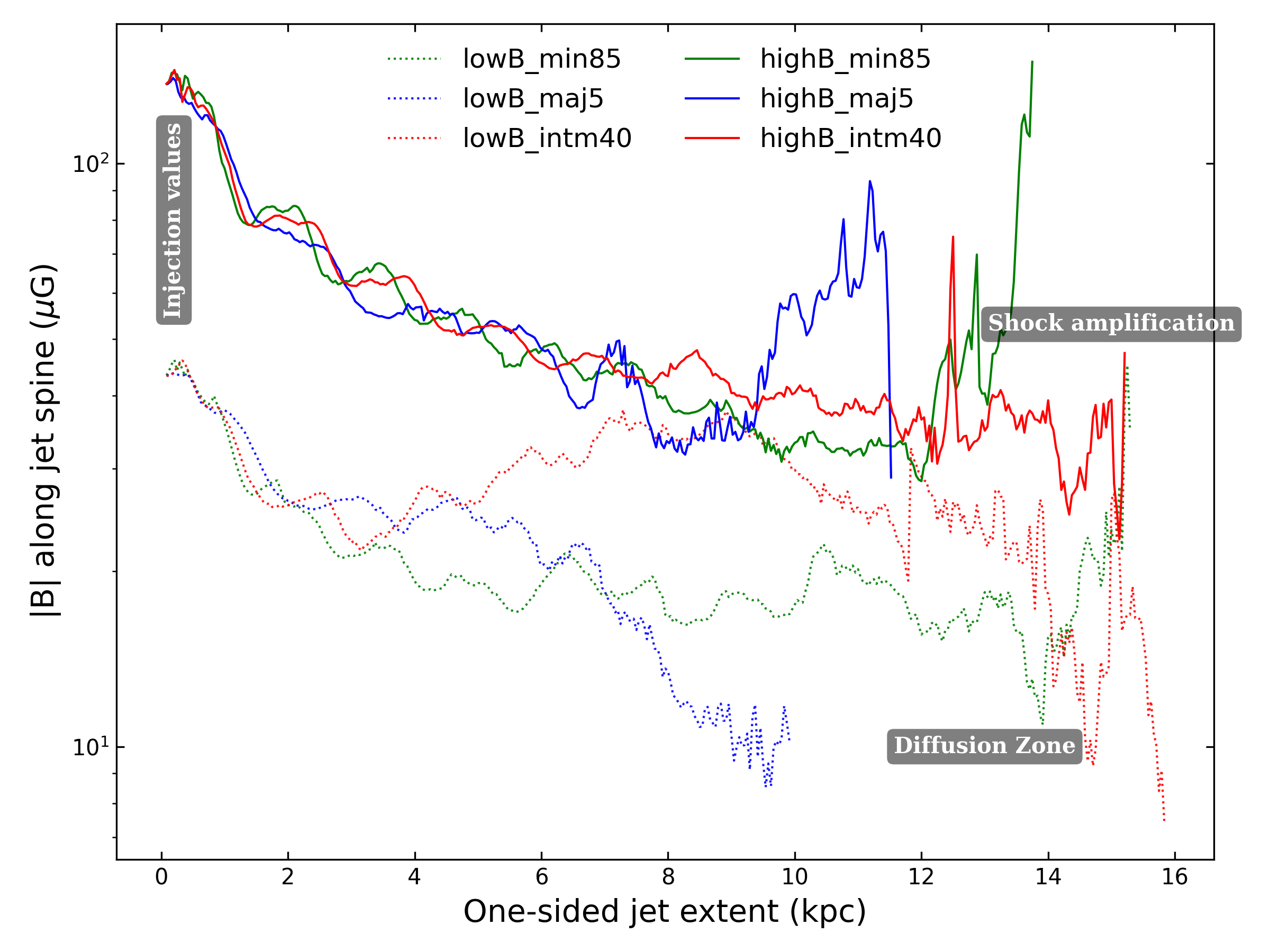}
\caption{Variation of the magnetic field strength along the jet axis, measured from the jet base, for two categories of simulations with high and low magnetization. The efficient transport of magnetic field in the high-B cases and its subsequent amplification near the jet head are clearly visible. In contrast, the low-B cases exhibit a more diffusive distribution, as the jet becomes more susceptible to small-scale MHD instabilities. The snapshot corresponds to the same evolutionary time shown in Fig.~\ref{Fig:Morphology_Bs}.}
\label{Fig:Hotspot_B} 
\end{figure}

In this context, the results inferred here are not unexpected, but rather represent a natural consequence of the findings discussed, for example, in \citet{Mignone2010,Mukherjee2020} and \citet{Rossi2024}. These studies have shown that, at higher levels of magnetization, short-wavelength Kelvin–Helmholtz modes are significantly stabilized, magnetic diffusion becomes less efficient, and the magnetic field remains more strongly confined within the jet. Such confinement and suppression of the small-scale MHD instabilities within the jet spine \citep{Biskamp1998,Biskamp2000}, allowing the jet thrust to be maintained efficiently up to the jet head. As a consequence, the magnetic flux is largely preserved along the propagation length and is transported to the termination region, where compression leads to an amplification of the magnetic field. This amplification is expected to be weaker in low-magnetization cases, where the magnetic field undergoes stronger diffusion during propagation \citep{Wang2023}. To simply elaborate this, we plot the magnetic-field distribution along the jet spine (from jet base to head) for the low- and high-magnetization cases of Table.~\ref{Tab:Parametric_space} and illustrated in Fig.~\ref{Fig:Hotspot_B}. The distributions clearly show a pronounced spike in the magnetic field strength at the jet termination regions in the high-magnetization cases, whereas the low-magnetization cases exhibit a much more diffusive field distribution along the jet (see also, Appendix~\ref{Sec:Role of magnetization in modulating radio galaxy properties}).

From a dynamical standpoint, \citet{Rossi2024} recently pointed out that higher magnetization hinders the transition toward a lobed morphology in radio galaxies. Building upon this argument, we additionally suggest that the enhanced magnetic field strength at the terminal shock (driven by higher injection power and the resilience of the jet spine against deformation) facilitate the formation of prominent hotspots. 
Our results can therefore be interpreted as a radiative manifestation of previously identified dynamical effects associated with enhanced magnetization. We note that the injected magnetic field in our simulations is purely toroidal. The inclusion of a poloidal component at the injection region may further influence these effects, potentially strengthening the contrast between hotspot- and warmspot-like structures. Finally, however, it should also be emphasized that our analysis focuses on the early evolutionary stages of radio galaxies. Whether these conclusions remain valid on the much larger spatial and temporal scales relevant for fully developed radio jets \citep{Gurkan2022} remains an open question and should be explored in future work.

\begin{figure}
\centering
\includegraphics[width=\columnwidth]{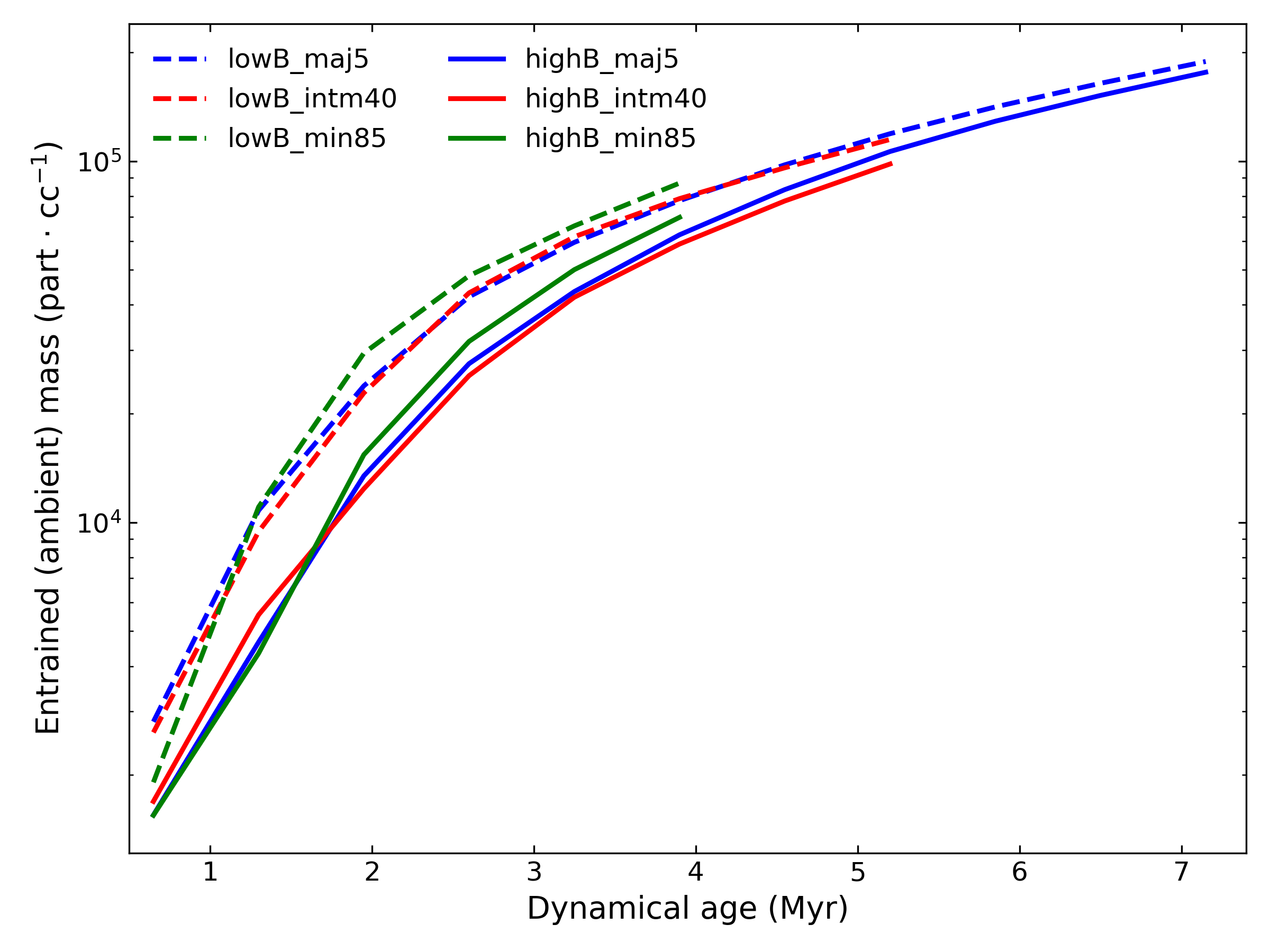}
\caption{Entrained thermal environmental mass within the jet cocoon, expressed in particles $\cdot$ cm$^{-3}$ (in log-scale), shown as a function of the evolutionary time of the jetted structures for different simulation cases.}
\label{Fig:Entrained_Mass} 
\end{figure}

To verify that we are indeed observing the magnetic stabilization effect of the cocoon discussed earlier, we further examined the entrainment of ambient material into the cocoon for the low- and high-magnetization cases. The results, shown in Fig.~\ref{Fig:Entrained_Mass}, indicate a reduced level of thermal matter entrainment within the cocoon for the higher B-field jets. Furthermore, in Section~\ref{Sec:Characterizing the intrinsic turbulence levels}, we discuss in more detail the stabilizing influence of stronger magnetic fields on cocoon turbulence. Taken together, these results further support the interpretation presented above.

The integrated flux values for the XRGs, double-boomerang, and double-lobed sources in the low-magnetization simulations are 0.24 Jy, 0.80 Jy, and 1.02 Jy, respectively (adopting a redshift of 0.05). In the corresponding high-magnetization cases, the fluxes increase to 0.87 Jy, 0.99 Jy, and 1.71 Jy, reflecting the enhanced radiative efficiency driven by stronger magnetic fields. The rapidly propagating jet case shown in the bottom panel of Fig.~\ref{Fig:Emission_col} exhibits a comparatively low integrated flux of 0.07 Jy. Several morphologically significant features emerge across the radio maps and are annotated with arrows in Fig.~\ref{Fig:Emission_col}. These include warm-spot complexes in XRGs indicative of unstable jet heads \citep{Horton2023,Horton2025}, filamentary extensions from hotspots in double-lobed sources \citep{Fanaroff2021}, and structures resembling restarting jets, where outer warmspots are followed by compact enhanced inner emission regions \citep[see,][for a similar perspective]{Young2024}. Additionally, back-flowing stream splitting resembling filamentary texture are evident—features that are relevant to the interpretation of high-resolution observations from contemporary radio telescopes \citep{Condon2021,Velovic2023,Wezgowieck2024}.

A notable issue in the simulated radio maps of the double-boomerang morphology is the relatively shorter extension of the secondary lobes (compared to the active lobe), falling short of what is typically observed in prominent winged sources of such topology. The simulated structures in Fig.~\ref{Fig:Emission_col} rather represent candidate wings with a wing-to-active-lobe length ratio of $< 0.8$ \citep{Cheung2007}. This raises questions about the adequacy of the current model (back-flow scenario) in reproducing the extended wing structures characteristic of such (prominent) double-boomerang systems. While the study by \citet{Hodges-kluck2011} demonstrated that a highly ellipsoidal ambient medium can produce a prominent double-boomerang morphology, the existence of such extreme ellipticity, particularly on larger scales, remains to be verified. Our obtained result therefore underscores the importance of exploring alternative formation scenarios for this subclass of X-shaped radio galaxies, such as jet precession models as utilised in \citet{Gower1982} and \citet{Gong2011}.

\subsubsection{The curious case of double-boomerang RGs}
The origin of double-boomerang morphology remains particularly challenging to disentangle, as its curved radio lobe structures can be interpreted through multiple frameworks, most notably the back-flow model \citep{Leahy1984,Rossi2017,Cotton2020} and the jet precession model \citep{Gower1982,Parma1985,Gong2011}. This ambiguity is further compounded by individual case studies that successfully reproduce the observed morphology using either approach. For instance, sources such as 3C 315 \citep{Gower1982, Hodges-kluck2011} and NGC 326 \citep{Gower1982, Hodges-kluck2012} have been modeled under both interpretations, while other such sources like 3C 223.1 exhibit spectral characteristics that support a diverse theoretical frameworks \citep{Dennett-thorpe2002, Lal2005, Gopal-krishna2022}. Given that the jet precession can account for many of the observed morphological and radiative complexities of such sources \citep{Horton2020,Bruni2021,Nandi2021,Nolting2023,Sebastian2024}, we sought to investigate whether a similar scenario could explain the features observed in our simulated double-boomerang case.
\begin{figure}
\centering
\includegraphics[width=\columnwidth]{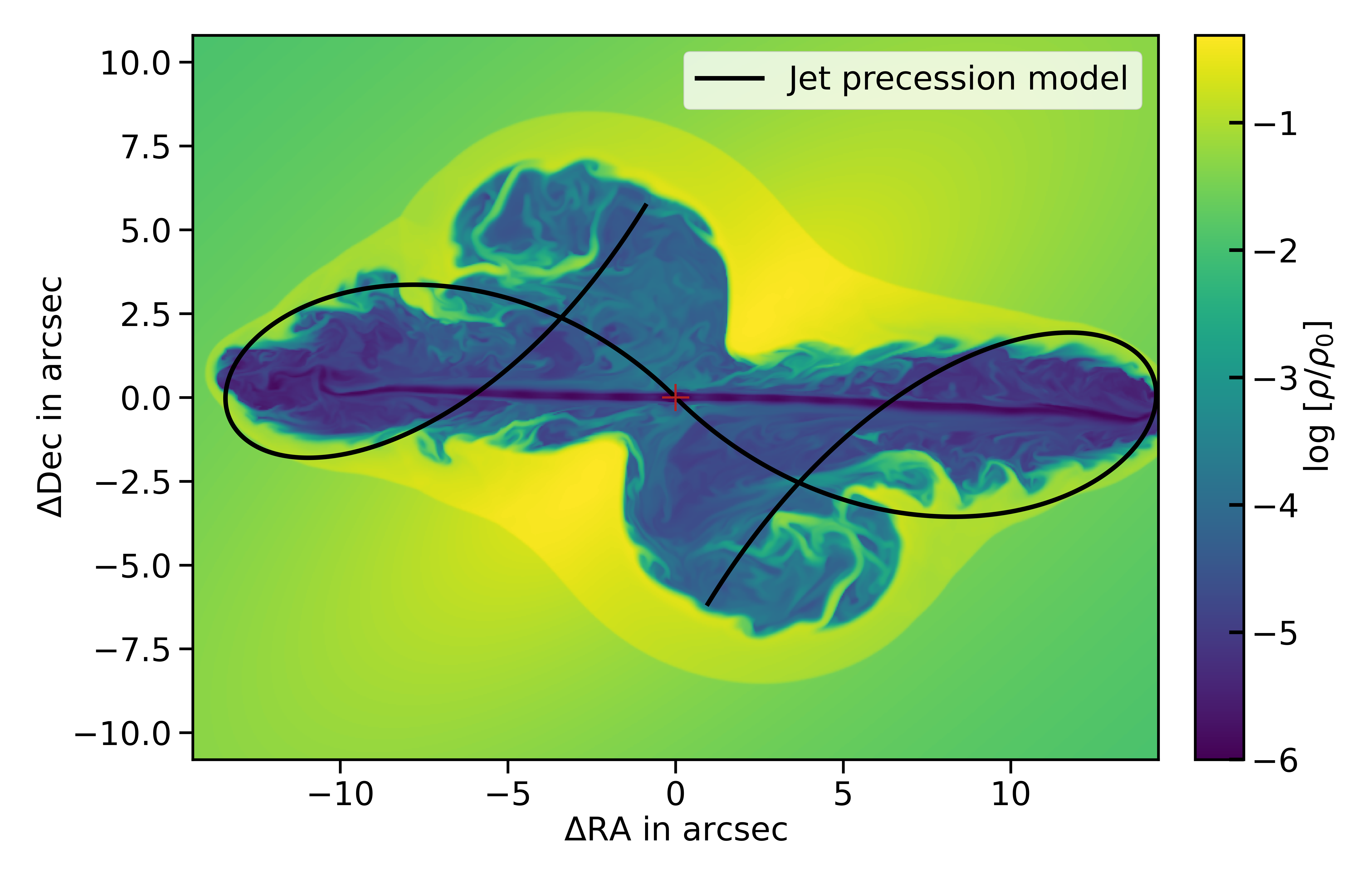}
\caption{Composite image showcasing simulated double-boomerang structure from the backflow model (density slice in colormap), overlaid with an analytical jet trajectory (black solid line) generated utilising a jet precession model. Despite arising from distinct mechanisms, both structures represent resembling morphologies, highlighting the challenge in identifying the origin of XRGs based solely on their topological features. Each structure originates from the central coordinate (red marker) in the RA–Dec frame (assuming redshift 0.05). Notably, the two models share comparable lobe propagation speeds and evolutionary timescales.}
\label{Fig:Backflow-precession} 
\end{figure}

To analytically trace the jet trajectory resulting from precessional motion—potentially driven by complex AGN processes \citep[see][for discussions on the triggers of jet precession]{Dennett-thorpe2002,Giri2024}—we adopt the geometric framework of \citet{Hjellming1981}. Originally formulated to describe the precessing jets of the microquasar SS 433, this model has since been effectively applied to extragalactic systems exhibiting precessing jets \citep{Kharb2006,Steenbrugge2008,Rubinur2017,Kharb2019}.

The \citet{Hjellming1981} model incorporates several free parameters that govern the jet trajectory, including the jet or lobe’s bulk expansion speed ($v_l$), the precession angle ($\psi$), the precession period ($P$), the inclination angle ($\mathcal{I}$)—i.e., the angle between the lobe and the observer’s line of sight—the total dynamical age of the structure ($t_{\rm dyn}$), and a rotation angle ($\chi$) that adjusts the projected orientation of the structure on the plane of the sky (RA–Dec). Given the broad parameter space involved, it is essential to constrain some of the variables using observational or simulation-based priors. We initiated this process by anchoring a few parameters using outputs from the simulation of the double-boomerang morphology formed through the back-flow scenario. Specifically, we fixed the dynamical age around $t_{\rm dyn} \sim 5.2$ Myr (Fig.~\ref{Fig:Morphology_Bs}) and varied lobe expansion speed around $v_l \sim 0.029c$ (average lobe expansion speed of the simulated structure). Additionally, motivated by previous studies \citep[e.g.,][]{Gong2011,Misra2025}, we considered precession periods on the values of several million years. We then iteratively adjusted these initial values, along with the remaining free parameters, to obtain a visually consistent match between the analytically generated jet morphology (precession model) and the simulated structure (back-flow model).
The final set of parameters that yielded the closest morphological resemblance with the back-flow generated structure (refer to Fig.~\ref{Fig:Backflow-precession}) are: $v_l = 0.037c$, $\psi = 19^{\circ}$, $P = 10$ Myr, $\mathcal{I} = -23^{\circ}$ ($\equiv 337^{\circ}$), $t_{\rm dyn} = 5.3$ Myr, and $\chi = 135^{\circ}$ (adopted a redshift of 0.05).

This resemblance between the analytically predicted and simulated morphologies underscores why XRGs (in general) remain inherently challenging to interpret in terms of their physical origin \citep[see,][]{Joshi2019,Giri2023}. Our results suggest that jet precession models, when calibrated appropriately, can mimic backflow-induced structures with considerable fidelity. This result is particularly relevant in light of the recent study by \citet{Sebastian2024}, which focused on X-shaped radio sources on much larger scales, thereby reinforcing the broader applicability of our findings. This highlights the need for caution when attributing morphology to a particular mechanism. The analytical parameters derived here may serve as valuable inputs for future numerical simulations incorporating precessing jets, enabling tests of the model's robustness in fully 3D MHD environments. Indeed, recent work by \citet{Giri2022_Sshape} demonstrated the viability of such analytical formulations (albeit under simplified assumptions) in numerically modeling other winged radio sources.

\subsection{Morphological quantification}\label{Sec:Morphological quantification}
Beyond the simulation snapshots discussed above, it is equally important to investigate the quantitative evolution of these systems, which we perform in this subsection through measurements of the active lobe's spatial growth and the turbulence level within the outflows.

\subsubsection{Characterizing the length--lobe speed--age evolution}
\begin{figure*}
\centering
\includegraphics[width=\columnwidth]{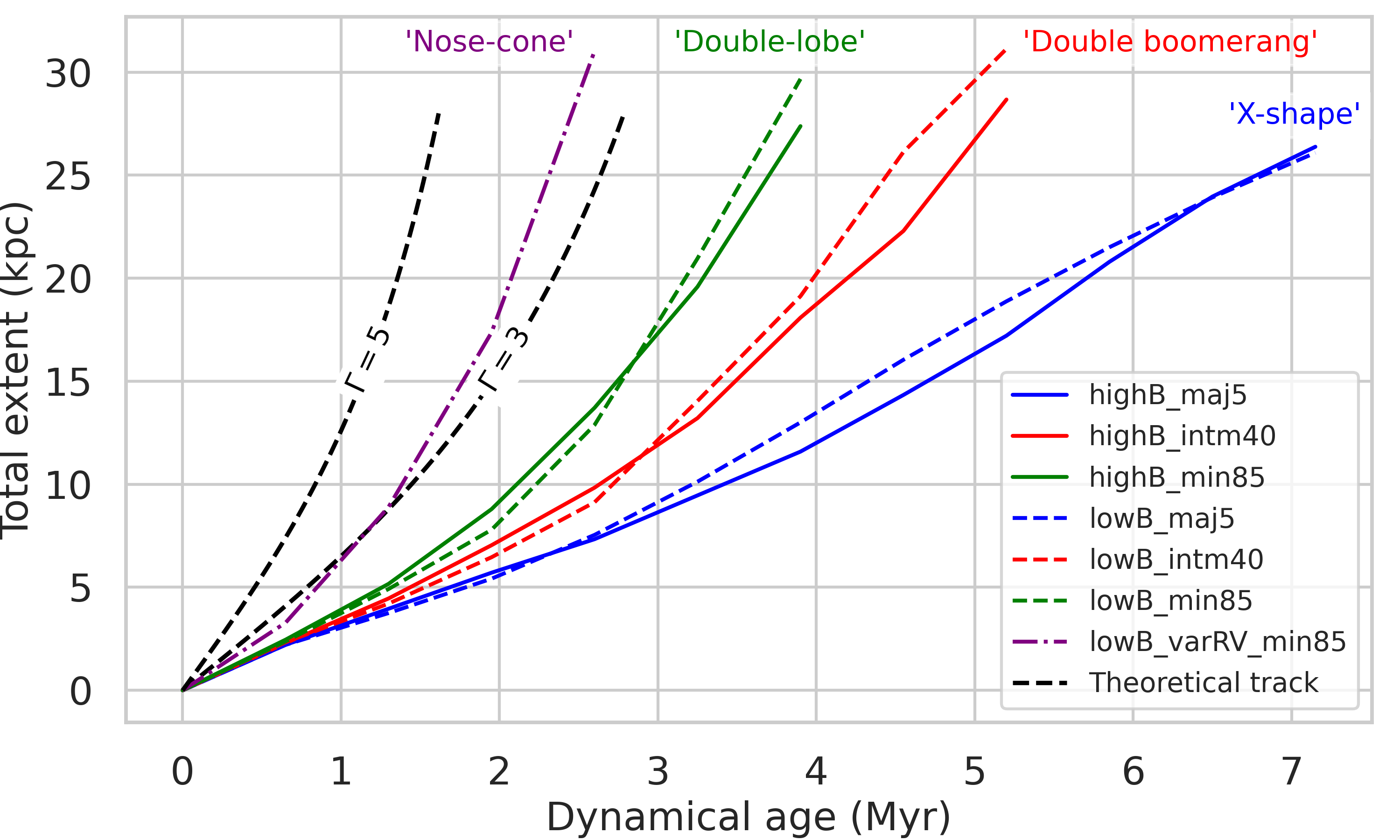}
\includegraphics[width=\columnwidth]{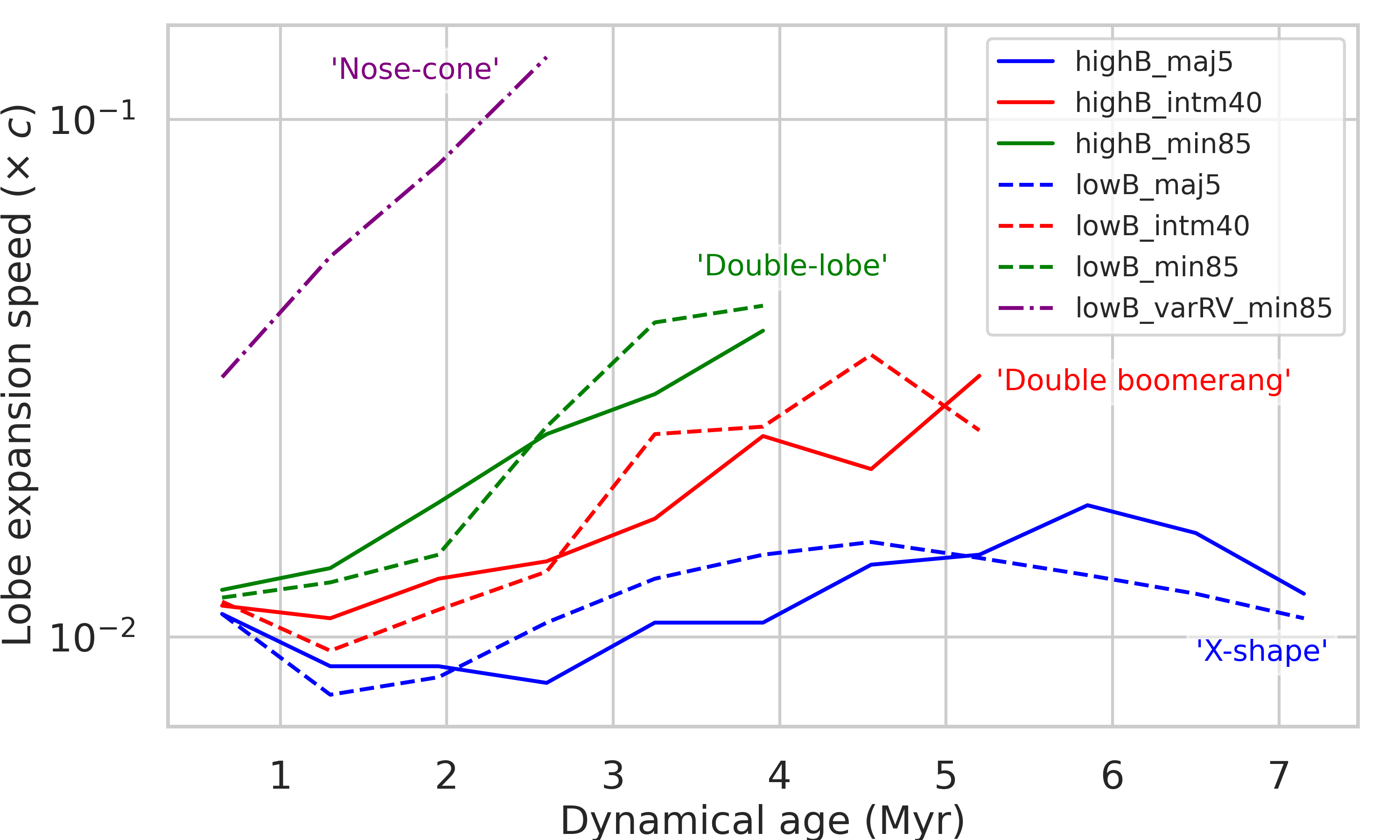}
\caption{(\textit{Left}:) Length–Age evolution diagram for the various morphological classes observed in our simulations. Theoretical expectations (for different jet Lorentz factors) are overlaid, illustrating the deviations between the fully 3D simulated cases and the simplified 1D analytical models. (\textit{Right}:) The associated evolution of lobe expansion speed as a function of source age for different subclasses of radio galaxies. The observed variations emphasize the role of jet thrust, (de)collimation, and environmental pressure gradients. Despite a constant jet power across simulations, the distinct behavior of nose-cone structures indicates that jet radius and speed may be more fundamental parameters (than jet power). }
\label{Fig:L-T} 
\end{figure*}

In Fig.~\ref{Fig:L-T}, we present the evolution of the total extent of the active lobes over time, alongside the lobe expansion speed, as determined within the computational domain using tracer diagnostics. The extents of the bidirectional lobe heads are measured as the maximum distance traveled by the jet head from the injection position. We measure it by locating the tracer values $\mathcal{T} \geq 10^{-7}$ \citep{Mukherjee2020}. 

Fig.~\ref{Fig:L-T} clearly demonstrates that jet propagation along different directions within the galactic environment leads to varying advance speeds (subject to different levels of jet frustration), thereby influencing the overall extent of radio galaxies from their earliest evolutionary stages. The distinct evolutionary tracks of classical XRGs, double-boomerang structures, and double-lobed morphologies in the length--lobe speed--age diagram highlight that magnetic fields play a dynamically minor role during the early expansion phases. Nevertheless, their influence on structural intricacies, such as the induction of magnetic instabilities, is hinted (as also discussed earlier) and likely impacts the larger-scale evolution (see, Appendix~\ref{Sec:Potential dichotomy between classical RGs and giant RGs}). This interpretation is further supported by the observation that, despite all simulations maintaining comparable jet powers, the `Nose-cone' structure exhibits significantly faster propagation (clearer in the lobe speed--age diagram), as attributed to its higher degree of jet collimation and minimal hindrance within the environment.

The simulated lobe extent evolution is compared to theoretical tracks for relativistic jets with varying Lorentz factors ($\Gamma \equiv 3,\, 5$), following the analytical formulation of \citet{Marti1997}. All simulated structures fall below the theoretical predictions in their spatial evolution (note: the `Nose-cone' shape has injection $\Gamma$ value of 5), indicating that the 1D theoretical framework overestimates jet growth. This highlights the need for theoretical modifications \citep{Bromberg2011}, as the 3D evolution shows jet decollimation and deformation, which impact its propagation speed. This behavior has been indicated and demonstrated in \citet{Rossi2017} for a similar XRG simulation.

\subsubsection{Characterizing the intrinsic turbulence levels}\label{Sec:Characterizing the intrinsic turbulence levels}
Measuring the turbulence level in each of our simulated scenarios is essential for understanding how jet–environment interactions influence internal cocoon dynamics across different classes of radio sources. Turbulent motions are believed to generate localized shock sites, which together shape the internal energy, the efficiency of particle energization and the resulting radiative signatures \citep{Giri2022_XRG,Kundu2022,Dubey2024}.

We build upon the method adopted by \citet{Dubey2023} to quantify turbulence levels, wherein local turbulence is estimated by computing the deviation of each cell’s value (e.g., $X_i$) from the local average ($\overline{X}_i$) (at a single snapshot in time), calculated within a cubic neighborhood centered on the cell of interest. These deviations are then summed over the entire cocoon region, identified using a jet tracer threshold of $\mathcal{T} \ge 10^{-2}$ (the threshold ensures we focus on the main jet-lobe structure, and excludes overlapping ambient cells, while performing spatial averages of $X_i$; \citet{Dubey2023}). The cumulative turbulence measure is subsequently normalized by the total value of the parameter $X$ in that region, yielding a normalized turbulence level ($\delta X_N$) that allows meaningful comparison across different radio galaxy cases. The derivation is expressed as follows,
\begin{equation}\label{Eq:Turb}
    \delta X_N = \frac{\sum_i (X_i - \overline{X_i})}{\sum_i X_i}
\end{equation}
where, the chosen parameter for estimating turbulence is the magnetic field ($\delta B_N$), due to its direct relevance to observational diagnostics of radio sources \citep{sullivan_2009,matthews_2020,Baidoo2023}.

\begin{figure}
\centering
\includegraphics[width=\columnwidth]{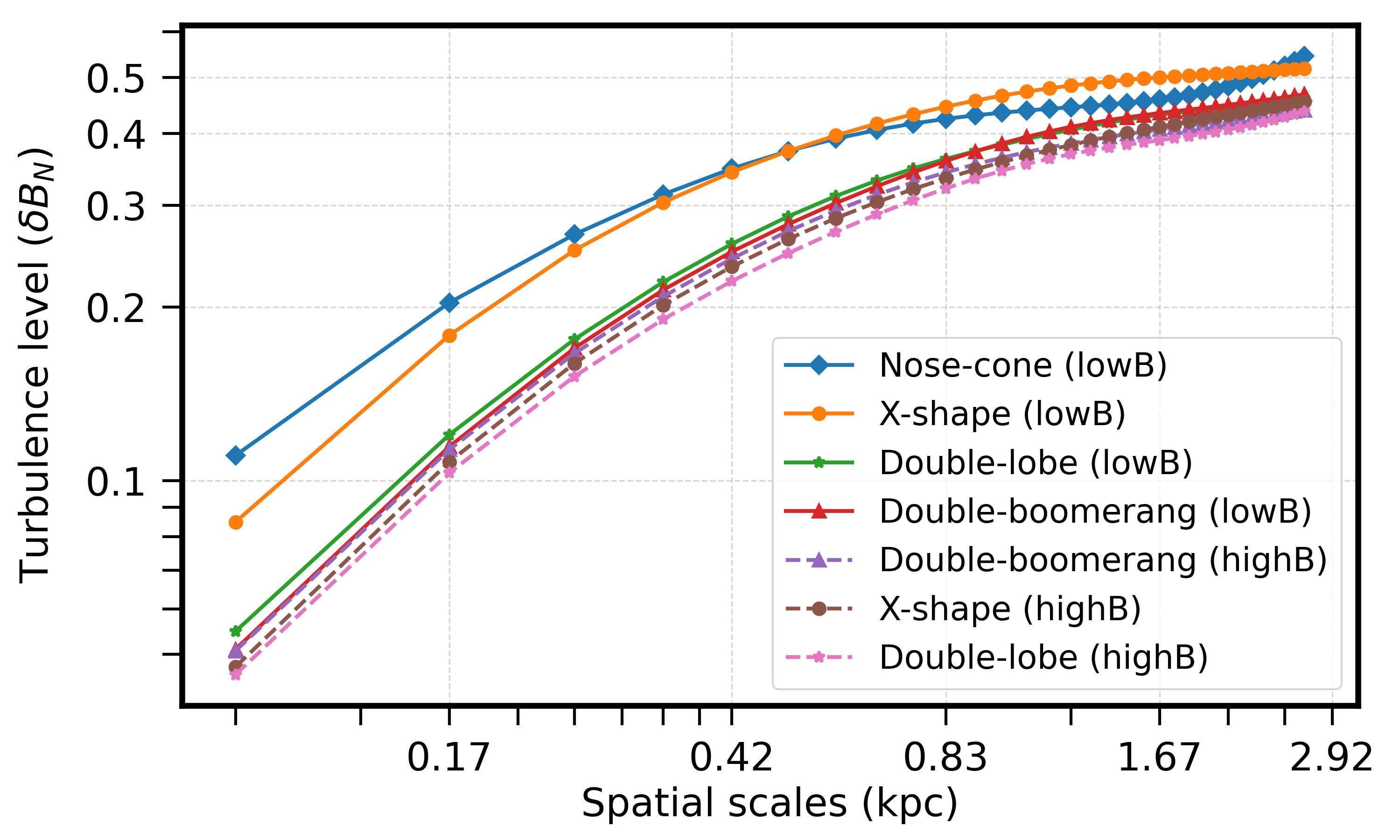}
\caption{The level of magnetic turbulence, evaluated across various length scales (defined by the length of the cubic neighborhood used to estimate local magnetic fluctuations), is shown for all seven simulations. The figure notably illustrates that beyond a certain characteristic size ($\sim 1$ kpc), the turbulence amplitude reaches a plateau, indicating the physical scales below which magnetic inhomogeneities manifest and evolve (for the present simulations). The stabilizing effect of stronger magnetic fields on turbulence is particularly evident here. The calculations were performed only once each simulations reached their late evolutionary stage, i.e., when the jets approached the domain boundary (Fig.~\ref{Fig:Morphology_Bs}, \ref{Fig:Morphology_RV}).}
\label{Fig:Turbulence_level} 
\end{figure}

We present the variation of $\delta B_N$ for our simulations in Fig.~\ref{Fig:Turbulence_level}, plotted as a function of the characteristic length used to compute the local magnetic field average. This spatial scale corresponds to the length of the cubic neighborhood over which local field averages are determined, thereby characterizing how the level of turbulence evolves across different extents within the radio cocoon. The results clearly show that the magnetic irregularities increases from smaller to larger neighborhoods, up to a threshold—approximately 1 kpc—beyond which the turbulence amplitude stabilizes, consistent with the observational findings of \citet{Baidoo2023}, who reported magnetic field fluctuations below 1.5 kpc in Hydra A. The observed plateau likely marks a transition to a regime where magnetic variations are dominated by global structures rather than local processes. Further studies using simulations of more extended radio galaxies (beyond galactic scales) are needed to determine whether this characteristic scale remains unchanged or shifts to larger lengths.

The figure also reveals that simulations with weaker field strengths (\textit{solid lines} in Fig.~\ref{Fig:Turbulence_level}) exhibit more pronounced irregularity compared to those with stronger magnetization (\textit{dashed lines} in Fig.~\ref{Fig:Turbulence_level}), where the system appears more coherent \citep[e.g.,][]{Sironi2011,Comisso2019}. This trend highlights the stabilizing influence of enhanced magnetic stress on cocoon dynamics. The magnitude of these fluctuations, however, remains non-negligible, as reflected in their relative magnitudes. Interestingly, Fig.~\ref{Fig:Turbulence_level} does not indicate a consistent pattern in turbulence strength across different jet models. For example, both highly obstructed jets (such as in the X-shaped morphology) and less impeded ones (e.g., in the nose-cone scenario) display comparable variation amplitudes, suggesting that jet frustration alone does not govern the degree of magnetic irregularity within the lobes.

\begin{figure}
\centering
\includegraphics[width=\columnwidth]{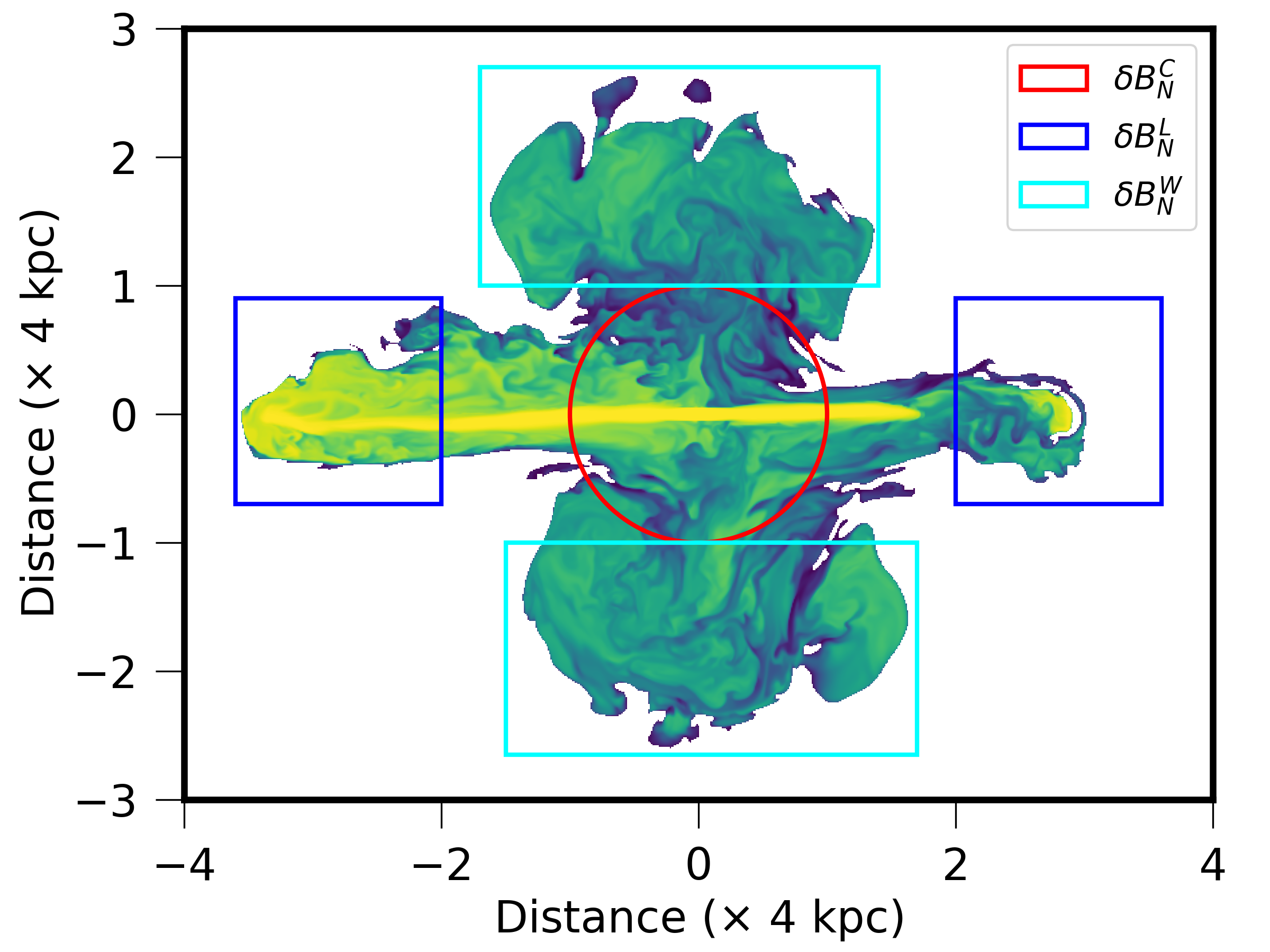}
\caption{2D colormap of the jet tracer distribution ($z = 0$) is shown for the X-shaped radio structure, illustrating the segmentation of the source into three distinct regions: the active lobe ($L$), the wing ($W$), and the central region ($C$) where (back)flow redirection occurs. This regional dissection is performed to quantify the turbulence magnitude within each zone, providing insight into how jet-induced irregularities propagate and evolve throughout the overall flow zones.}
\label{Fig:X-turbulence} 
\end{figure}

An important extension of our analysis is assessing the turbulence magnitude specifically within the X-shaped radio structure. Rather than averaging over the entire jet-cocoon structure, we divide the source into distinct regions—namely the active lobes, wings, and the central area where (back)flow material undergoes diversion—and evaluate the magnetic irregularity within each domain separately. This segmentation approach is illustrated in Fig.~\ref{Fig:X-turbulence}, where we report the resulting turbulence measures as $\delta B_N^C$ for the central region, $\delta B_N^L$ for the active lobes, and $\delta B_N^W$ for the wings, following Eq.~\ref{Eq:Turb}. 

The motivation for this assessment arises from several recent studies of X-shaped radio sources that report comparable or even flatter spectral indices in the wings relative to the active lobes. This anomalous behavior has been attributed to turbulence (consequently by random shocks) re-energizing the radiating particles \citep[e.g.,][]{Gopal-krishna2022}. Previously, such spectral properties were cited to challenge the general validity of the backflow model in explaining XRG morphologies \citep{Lal2019}. However, more recent observational and simulation-based investigations suggest that localized particle acceleration processes may be responsible for this anomaly \citep[e.g.,][]{Giri2022_XRG,Patra2023,Patra2025}.

We find clear evidence of turbulence within the wing region, with the corresponding magnitudes summarized in Table~\ref{Tab:Turbulence_level}.

\begin{table}[ht]
\caption{Evaluation of turbulent flow across distinct areas within XRGs.}
\centering
\begin{tabular}{|l|c|c|c|c|}
\hline
Simulation & Spatial & $\delta B_N^L$ & $\delta B_N^C$ & $\delta B_N^W$ \\
label & scales (kpc) & (Lobe) & (Center) & (Wing) \\
\hline
\multirow{3}{*}{\texttt{lowB\_maj5}} 
& 0.17 & 0.28 & 0.21 & 0.14 \\
& 0.67 & 0.48 & 0.42 & 0.37 \\
& 2.67 & 0.61 & 0.54 & 0.49 \\
\hline
\multirow{3}{*}{\texttt{highB\_maj5}}
& 0.17 & 0.23 & 0.15 & 0.08 \\
& 0.67 & 0.46 & 0.36 & 0.25 \\
& 2.67 & 0.65 & 0.50 & 0.41 \\
\hline
\end{tabular}
\label{Tab:Turbulence_level}
\small
\tablefoot{At 7.2 Myr into the evolution of the XRG, when the wing, active lobe, and central region are distinctly identifiable, we segment these areas as illustrated in Fig.~\ref{Fig:X-turbulence}, and quantify the corresponding turbulence strength ($\delta B_N$ values) across a range of spatial scales within each region. A consistent trend $\delta B_N^L > \delta B_N^C > \delta B_N^W$ is observed, independent of B-field strength or spatial scale of measurement.}
\end{table}

The measured values indicate that the active lobe is the principal site of magnetic fluctuations, driven by the combined influence of the jet and backflow dynamics \citep{Mukherjee2020,Mukherjee2021}. These disturbances are advected toward the central region, where they are subsequently redirected into the wing cavities. This redirection of flow appears to be the dominant mechanism responsible for the development of turbulence within the wings, in agreement with the findings of \citet{Giri2022_XRG}, who identified signatures of mild shocks in these regions \citep[see also,][]{Patra2025}. Our analysis shows that increasing the jet magnetization leads to a reduction in the turbulence amplitude within the wings; however, the turbulent energy remains non-negligible, varying up to 50\% (or more on smallest scales) of that in the active lobe. The decline in turbulence in wing region (compared to the primary lobe) also suggests that the magnetic field progressively aligns with the flow along the wings, providing a plausible explanation for the higher fractional polarization and more ordered polarization field line distribution observed in wing regions compared to the active lobes in XRGs \citep[cf.][]{Giri2024, Cotton2020}.

\subsection{Numerical insights into the counterexamples}\label{Sec:Numerical insights into the Counterexamples}

Although strong trends exist between optical (galactic)–radio (jet) orientations in GRGs, XRGs, classical doubles, and double-boomerang sources, well-documented counterexamples remain. Some XRGs show active jets along the minor axis with wings aligned to the major axis—opposite to the canonical geometry—and their origin remains debated \citep{Joshi2019}. Likewise, many major-axis jets do not produce X-shaped structures, as evidenced by classical doubles (Section~\ref{Sec:Radio - Optical Axes Correlation of Extended Radio Galaxies}). Thus, despite the strong correlations demonstrated in Section~\ref{Sec:Morphological distinctions}, exploring parameter regimes that address these exceptions is necessary.

\begin{figure}
\centering
\includegraphics[width=\columnwidth]{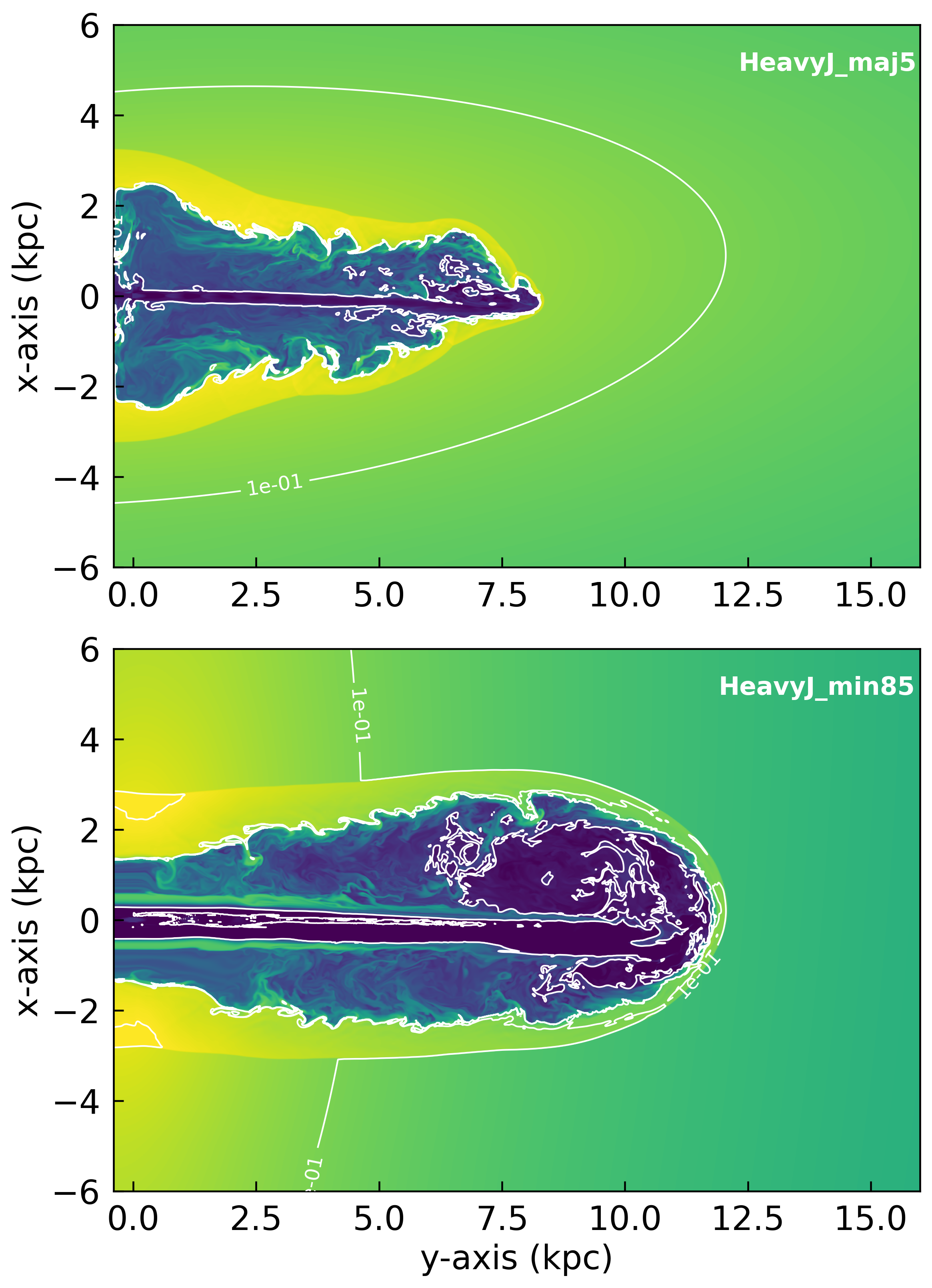}
\caption{Two simulations (labels shown in the inset; also in Table~\ref{Tab:Counter_examples}) conducted to investigate counterexample morphologies. ``\texttt{HeavyJ\_maj5}'' demonstrates a case where a jet propagating along the host’s major axis fails to produce prominent wings or secondary lobes (1.6 Myr). ``\texttt{HeavyJ\_min85} illustrates a jet launched along the minor axis (of the host) that becomes rapidly hindered, leading to decollimation and the formation of a lobe-like structure (1.7 Myr). The ambient medium’s shape is outlined with white density contour (with inline values in amu/cc), while internal contours—derived from the jet tracer—highlight the jet spine and backflowing structures for clarity.}
\label{Fig:Counter-examples} 
\end{figure}

The first category of XRG counterexamples has been investigated by \citet{Giri2023}, who argue that minor-axis active lobes with major-axis wings require jet reorientation. Here, we examine the later category of counterexamples (as noted above): a jet flow along the major-axis (of host) that does not develop a prominent X-shape. The simulation ``\texttt{HeavyJ\_maj5}'' in Table~\ref{Tab:Counter_examples}, uses a slightly higher jet to ambient density contrast with appropriately reduced radius and injection speed to maintain the same jet power as in our baseline models (Table~\ref{Tab:Parametric_space}). As shown in Fig.~\ref{Fig:Counter-examples}, this case exhibits strongly suppressed lateral growth of secondary structures despite the jet flowing along the major axis.

On the other hand, analyses of classical doubles with sizes $\lesssim 200$ kpc (e.g. the 3C sample) show a random distribution of jet orientations relative to the host galaxy’s principal axes \citep{Saripalli2009}. This indicates that, even for jets launched along the host's minor axis, long-term collimation is not guaranteed; decollimation can arise through diffusion and flow broadening over time. An example of such behaviour is seen in the ``\texttt{HeavyJ\_min85}'' case (Table~\ref{Tab:Counter_examples}), which produces a decollimated, lobe-like structure in its spatio-temporal evolution, as demonstrated in Fig.~\ref{Fig:Counter-examples}.

\begin{table*}
\caption{Exploring the parameter space for explaining a subset of counterexample sources that deviate the optical-radio axes traits.}
\begin{center}
\begin{tabular}{ |l|c|c|c|c|c|c|c| } 
 \hline
 Simulation& Jet radius & Jet velocity & Jet density & Jet flow along  & Domain (kpc) & Grid & Remarks\\
 label& ($r_{\rm j}$ in pc) & ($\Gamma$) & contrast ($\rho_j/\rho_0$) & environment's & $[x],\,[y],\,[z]$ &  &\\
 \hline
  &  &  & & major axis  & $[-6,\, 6]$ & 1154 & Same jet power as\\ 
 \texttt{HeavyJ\_maj5}& 52 & 2 & $5 \times 10^{-5}$ &($\Theta = 5^{\circ}$)&$[-0.4,\, 16]$&1578& `\texttt{lowB\_maj5}'\\
 &&&&&$[-6,\, 6]$& 1154 & \\
 \hline
  &  &  &  &  &  &  & Same jet power as \\ 
   \texttt{HeavyJ\_min85}& 52 & 2 & $5\times 10^{-5}$ &minor axis&as above&as above&`\texttt{lowB\_min85}'\\
   &&&&($\Theta = 85^{\circ}$)&&&\\
   \hline
\end{tabular}

\label{Tab:Counter_examples}
\end{center}
\small
\textbf{Notes.} Two additional simulations beyond those listed in Table~\ref{Tab:Parametric_space} were performed to investigate the possible origins of the counterexample sources. In these cases, the jet power was kept fixed (consistent with Table~\ref{Tab:Parametric_space}), while intrinsic jet parameters were varied: a slightly heavier jet (``\texttt{HeavyJ}'') was introduced, requiring corresponding reductions in jet radius and propagation speed to maintain the same power (Eq.~\ref{Eq:Power}). Both simulations were run with a magnetization parameter of $\sigma = 0.01$ and the jet radius has been resolved using 5 cells.
  
\end{table*}

Motivated by these simulation cases that reproduce counterexample morphologies and traits, we examined their occurrence in observations by performing a $\Delta$PA analysis—the position-angle difference between the radio jet axis (PA$_{\rm radio}$) and the optical major axis of the host galaxy (PA$_{\rm optical}$). Similar analyses exist for classical doubles and XRGs (as discussed in Section~\ref{Sec:Radio - Optical Axes Correlation of Extended Radio Galaxies},~\ref{Sec:Optical - Radio axes correlation of XRGs}), which we do not repeat here. The lack of equivalent statistical studies for GRGs motivated us to perform a $\Delta$PA analysis for this category of RGs.

Starting from the \citet{Dabhade2020_LotssI} GRG sample, we derived radio position angles from LOFAR 144 MHz images and optical position angles from Legacy Survey DR10 $g,\,r,\,z$-band data. Reliable PA$_{\rm optical}$ measurements were obtained for 50 sources, selected to be non-crowded, sufficiently elliptical $(1 - b/a > 0.1)$, and with semi-major axes larger than the average seeing ($\sim 1.5''$). Only sources with robust PA$_{\rm radio}$ measurements were retained. Optical PAs were determined by fitting ellipses to the galaxy isophotes, excluding the outermost isophotes (to avoid contamination) and those smaller than the seeing; the final PA$_{\rm optical}$ was taken as the mean of the remaining isophotes. The $\Delta$PA was computed as the acute angle between the two, restricted to the range $0$–$90^\circ$. Radio PAs were obtained from flux-weighted second moments of the emission and cross-checked using the line connecting the host galaxy to the lobe brightness peaks. For mildly bent or asymmetric sources, the average lobe orientation was used, with an estimated uncertainty of $5$–$10\%$ (we rejected complex radio morphologies).

\begin{figure}
\centering
\includegraphics[width=\columnwidth]{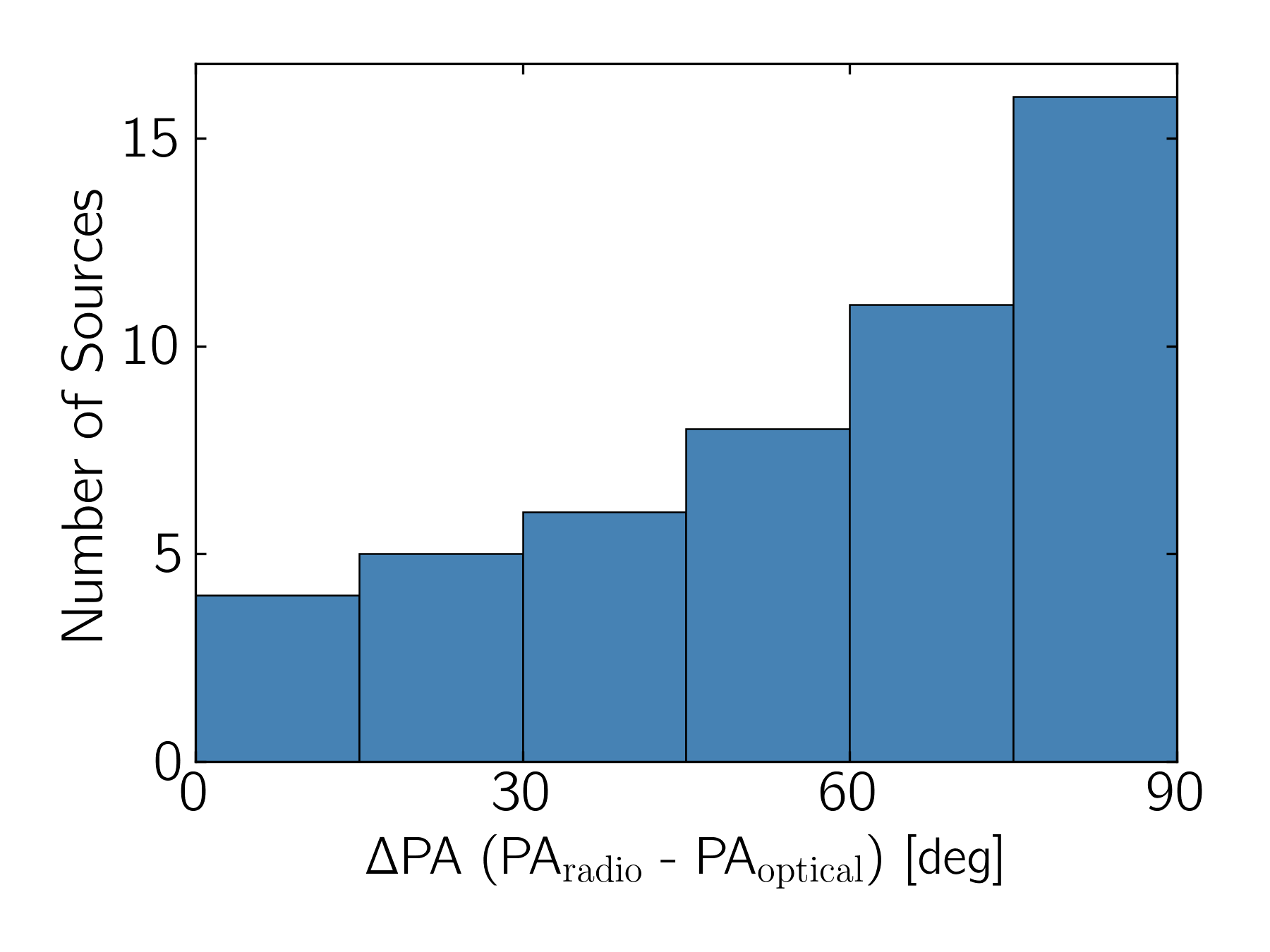}
\caption{Distribution of $\Delta\mathrm{PA}$, the position--angle difference between the radio axis (PA$_{\rm radio}$: jet direction) and the optical major axis (PA$_{\rm optical}$) of the host galaxy for GRGs. The $\Delta\mathrm{PA}$ distribution indicates that a small fraction of GRG sources deviate from the expected alignment trends with the host's principal axis (Section~\ref{Sec:Optical - Radio axes correlation of GRGs}), highlighting the presence of counterexamples (e.g., sources with $\Delta$PA $\lesssim 45^{\circ}$).}
\label{Fig:Delta_PA} 
\end{figure}

Fig.~\ref{Fig:Delta_PA} presents the $\Delta$PA distribution for the GRG sample. This histogram demonstrates that, despite the trend discussed in Section~\ref{Sec:Optical - Radio axes correlation of GRGs}—where GRG jets preferentially align with the host’s minor axis (which we also observed here)—clear counterexamples also exist (e.g., sources with $\Delta$PA $\lesssim 45^{\circ}$). Such diversity indicates that GRG jet orientations are influenced not only (primarily) by jet–ambient coupling but also by additional factors such as capacity of suppressing MHD instabilities \citep{Oei2024_7Mpc}, prolonged evolutionary histories \citep{Gurkan2022} or episodic (restarting) jet activity \citep{Dabhade2025}. These possibilities warrant further investigation through numerical modeling on much larger spatial and temporal scales. We also highlight that the jet power adopted in our simulations 
($3 \times 10^{44}\,\mathrm{erg\,s^{-1}}$) is lower than the values 
typically inferred for the majority of giant radio galaxies, 
most of which belong to the powerful FR\,II class ($\sim$90\%). 
Such systems likely host jets with substantially higher mechanical 
thrust, potentially strong enough to overcome ambient resistance 
even when propagating along the host galaxy's major axis, which may affect the jet propagation to several tens of kpc only.

\section{Summary}
\label{Sec:Summary}
Extended extragalactic radio jets exhibit a wide range of growth
histories, reflecting the complex and coupled interaction be-
tween the jet and its ambient environment. This study is tailored
to focus on understanding jet evolution under such intertwined
configurations, specifically concentrating on the initial evolutionary phases when the jet remains confined within the host
galaxy’s extent. The primary objective is to assess the impact of
a triaxial host geometry on jet propagation—introducing varying
degrees of jet frustration—and systematically varying internal jet parameters while maintaining constant jet power. To this end,
we conduct a detailed analysis of the resulting morphological
structures, radiative properties, and turbulence characteristics of
the cocoon, based on a suite of high-resolution numerical
simulations.

The observational traits synthesized and re-analyzed here have shown that different subclasses of radio galaxies tend to align with specific axes of their triaxial host galaxies: giant radio galaxies preferentially propagate along the minor axis, X-shaped radio galaxies along the major axis, and the double boomerang subclass—characterized by curved lobes—along the intermediate axis. Our suite of numerical simulations reproduces these trends, reinforcing the significant role of the host galaxy’s pressure gradient in shaping the spatial transformation and morphological evolution of radio jets, through its modulation of resistance or guidance to jet propagation.

This study also reveals a potential divergence in the growth dynamics between classical double-lobed radio galaxies—typically extending over hundreds of kiloparsecs—and the less common giant radio galaxies, which evolve on megaparsec scales. Given that both of these sources exhibit preferential propagation along the minor axis of the host galaxy, corresponding to paths of reduced hindrance, our simulations indicate that a modest increase (by a factor of $\sim$ 2) in jet collimation and axial thrust (keeping jet power fixed) significantly enhances the propagation efficiency, resulting in jets that evolve nearly twice as fast in spatiotemporal dimensions. While this difference may appear subtle at smaller scales, it becomes markedly influential over large-scale evolution, supporting the idea that GRGs can be explained within the standard paradigm of jet dynamics, without requiring fundamentally different physical processes.

Similarly, our results show that the `double boomerang' subclass of radio galaxies (characterized by curved lobes) can, in principle, arise independently from jet precession and backflow-driven scenarios, making it challenging to conclusively determine their origin based solely on morphology. However, our assessment also reveals that the backflow model tends to struggle in producing wing structures that are highly curved, well collimated and more extended than the active lobes. This limitation suggests that observed sources with such features can more consistently be explained by a jet undergoing precessional motion, rather than by classical backflow dynamics. We note that these two scenarios—jet precession and backflow-driven dynamics—may also operate simultaneously, further increasing the complexity of the interactions between jet and the surrounding environment.

These different categories of radio morphology, owing to their strong coupling with the surrounding environment, also play a key role in regulating feedback. Although this aspect has not been explicitly modeled in the present work, initial indications point toward measurable differences in how these source types impact their environments, motivating future studies.

The level of (magnetic) turbulence within the cocoon structures formed in our simulations does not exhibit a consistent trend with the type of jet propagation model employed. Instead, it correlates more strongly with the degree of jet magnetization: higher magnetic strengths tend to suppress turbulence. Our inquiry reveals that magnetic field fluctuations vary over spatial scales of $\lesssim 1$ kpc, beyond which the turbulence level reaches a plateau, indicating the influence of global dynamics rather than local effects (requiring its validation through larger-scaled simulations). In X-shaped radio galaxies, active lobes are more turbulent than the wings, where the wings gain turbulence primarily from deflected backflow at the center. Despite being lower, wing turbulence remains significant, explaining their dynamically active nature and providing a likely explanation for the observed higher fractional polarization and more ordered field distributions in the wings.

This study also examines sources that deviate from the typical optical–radio axis correlations, identifying regions of parameter space where such exceptions (counter-exampled sources) can arise without invoking additional physical mechanisms. These include cases where a jet aligned with the host’s major axis does not develop prominent secondary lobes or wings, as well as jets launched along the minor axis that fail to remain collimated and instead form decollimated, lobe-like structures.

Finally, our synthetic 1 GHz emission maps reveal several intricate features that raise compelling questions for current and upcoming radio observations. For instance: Do the weak or absent radiative signatures in wings and rapidly advancing lobes (as in potential GRG cases) suggest that XRGs and GRGs are more common than previously thought? It may simply be that the detection sensitivity threshold of these contemporary telescopes needs to be improved \citep{Yang2019,Koribalski2025}. Are the filamentary structures in radio lobes a consequence of flow diversion driven by environmental asymmetries or are they inherently magnetized structures? Perhaps they are both—but this remains to be confirmed through dedicated investigation \citep{Brienza2021,Rudnick2022}. Lastly, is hotspot formation (or disruption) more fundamentally linked to jet magnetization besides dependence on other plausible conditions \citep{Croston2018,Rossi2024}?
 
\subsection*{Present limitation and future extension}

Our present work focuses on the early, galaxy-scale evolution of radio jets, a phase that plays a crucial role in setting the initial conditions for their subsequent large-scale development. While radio galaxies ultimately evolve to scales well beyond their host galaxies, extending this framework to larger spatial scales therefore represents an important next step. 

A natural extension thus is to apply herein developed methodologies to larger-scale simulations that capture both galactic and extragalactic environments, for example, by adopting double–$\beta$ density profiles to model the multi-scale ambient medium \citep{ONeill2010}. 

It is also important to recognize that these systems are inherently multiscale, and current numerical constraints prevent us from fully resolving the dynamical intricacies across these scales. In addition, the emission prescription employed here is minimalistic and does not account for the local shocks that arise in the diffuse lobes and  wings as a result of turbulent re-acceleration and mixing. As a future extension, we plan to employ the recently developed GPU-accelerated MHD version of the PLUTO code \citep{Rossazza2025}. This will enable simulations at unprecedented spatial scales and resolutions through GPU acceleration. In addition, the particle module \citep{Suriano2026}—based on advected Lagrangian tracers that represent non-thermal electron populations—will allow the development of more sophisticated and physically consistent emission models, forming a natural next step of this work. 

Observationally, we further recognize an important limitation associated with the assumed jet propagation direction, which can be significantly affected by jet reorientation phenomena. While restarted jets are often found to broadly align with their earlier propagation axes \citep{Mahatma2019,Dabhade2025}, even across (suspected) three activity cycles \citep{Cotton2020,Chavan2023,Rarivoarinoro2026}, there exist notable cases where the jet axis undergoes substantial reorientation between epochs \citep[e.g.,][]{Liu2019,Nandi2021,Ubertosi2021,Rao2023}. Such systems require careful treatment, ideally through the identification of relic, diffusively cooling plasma from previous jet activity. This approach has now become increasingly feasible with the availability of deep, low-frequency observations, particularly with the public release of the third data release from LOFAR \citep{Shimwell2026}.

\begin{acknowledgements}
GG thanks Sushmita Agarwal for organizing an open discussion session at Indian Institute of Technology Indore, India, which sparked the idea of pursuing such study from a numerical perspective. GG and KT acknowledges financial support from the South African Department of Science and Innovation's National Research Foundation under the ISARP RADIOMAP Joint Research Scheme (DSI-NRF Grant Number 150551). GG acknowledges the travel support from Gianluigi Bodo and Paola Rossi (for a visit to Osservatoria Astrofisico di Torino - INAF) through Ob.Fu. 1.05.01.89.01 and `Ricerca di base gruppo extragalattico'. SK gratefully acknowledge support by the Science and Technology Facilities Council (STFC) through grant ST/X001067/1.
The authors acknowledge the Centre for High Performance Computing (CHPC), South Africa, for providing computational resources to this research project (\url{https://www.chpc.ac.za/}). We acknowledge the use of the ilifu cloud computing facility – \url{https://www.ilifu.ac.za/}, a partnership between the University of Cape Town, the University of the Western Cape, Stellenbosch University, Sol Plaatje University, the Cape Peninsula University of Technology and the South African Radio Astronomy Observatory. The ilifu facility is supported by contributions from the Inter University Institute for Data Intensive Astronomy (IDIA – a partnership between the University of Cape Town, the University of Pretoria and the University of the Western Cape), the Computational Biology division at UCT and the Data Intensive Research Initiative of South Africa (DIRISA).
\end{acknowledgements}

%
%


\bibliographystyle{aa} 
\bibliography{sample1} 

\begin{appendix}
\section{Potential dichotomy between classical RGs and giant RGs}\label{Sec:Potential dichotomy between classical RGs and giant RGs}
This work alongside offers preliminary insights into the broader question of why giant radio sources constitute only a small subset of classical double radio sources. While both extended doubles ($\gtrsim 200$ kpc) and GRGs ($\gtrsim 700$ kpc) tend to align with the minor axis of the host galaxy—suggesting an initial shared environmental influence—their divergence in spatio-temporal evolution is likely shaped by long-term differences in jet spine stability and dynamics \citep{Oei2024_7Mpc}.

To further examine the influence of individual jet parameters on evolutionary outcomes, we extended the simulations of `\texttt{highB\_min85}' and `\texttt{lowB\_varRV\_min85}' (Table~\ref{Tab:Parametric_space})—representing potential classical doubles and GRGs, respectively (Fig.~\ref{Fig:Morphology_Bs}, \ref{Fig:Morphology_RV}). Due to increased computational demands, however, we tracked the one-sided jet evolution up to 32 kpc, a scale typically associated with emission-line nebulae in jetted elliptical galaxies \citep{Baum1989}. A summary of the simulation details is provided in Table~\ref{Tab:CRG-GRG}. 

\begin{table*}
\caption{Parameters that differ between the two simulations selected for extended evolution.}
\begin{center}
\begin{tabular}{ |c|c|c|c|c|c|c|c| } 
 \hline
 Simulation& Jet radius & Jet velocity & Jet B-field & Jet flow along  & Domain (kpc) & Grid & Remarks\\
 label& ($r_{\rm j}$ in pc) & ($\Gamma$) & ($\sigma$) & environment's & $[x],\,[y],\,[z]$ &  &\\
 \hline
  &  &  & &  & $[-6,\, 6]$ & 556 & \\ 
  \texttt{lowB\_varRV\_min85}&108&5&0.01& minor axis&$[-1.2,\, 32]$&1538& giant-RG\\
   &&&&($\Theta = 85^{\circ}$) &$[-6,\, 6]$&556& (candidate)\\
   \hline
    &  &  &&&$[-8,\, 8]$&384& \\  
   \texttt{highB\_min85}&200&3&0.1&minor axis &$[-0.4,\, 32]$&778&classical-RG\\
   &&&&($\Theta = 85^{\circ}$)&$[-8,\, 8]$&384&(candidate)\\
   \hline
\end{tabular}

\label{Tab:CRG-GRG}
\end{center}
\small
\textbf{Notes.} Both simulations propagate through similar environmental condition and share identical jet power, but differ in their internal jet parameters. The parameters align with those listed in Table~\ref{Tab:Parametric_space}, except for the domain size, which has been adjusted to track the extended (spatio-temporal) evolution of the one-sided jet arm. A resolution of 5 grid cells per jet radius was maintained for each run, consistent with the earlier simulations.
\end{table*}

\begin{figure*}
\centering
\includegraphics[width=2\columnwidth]{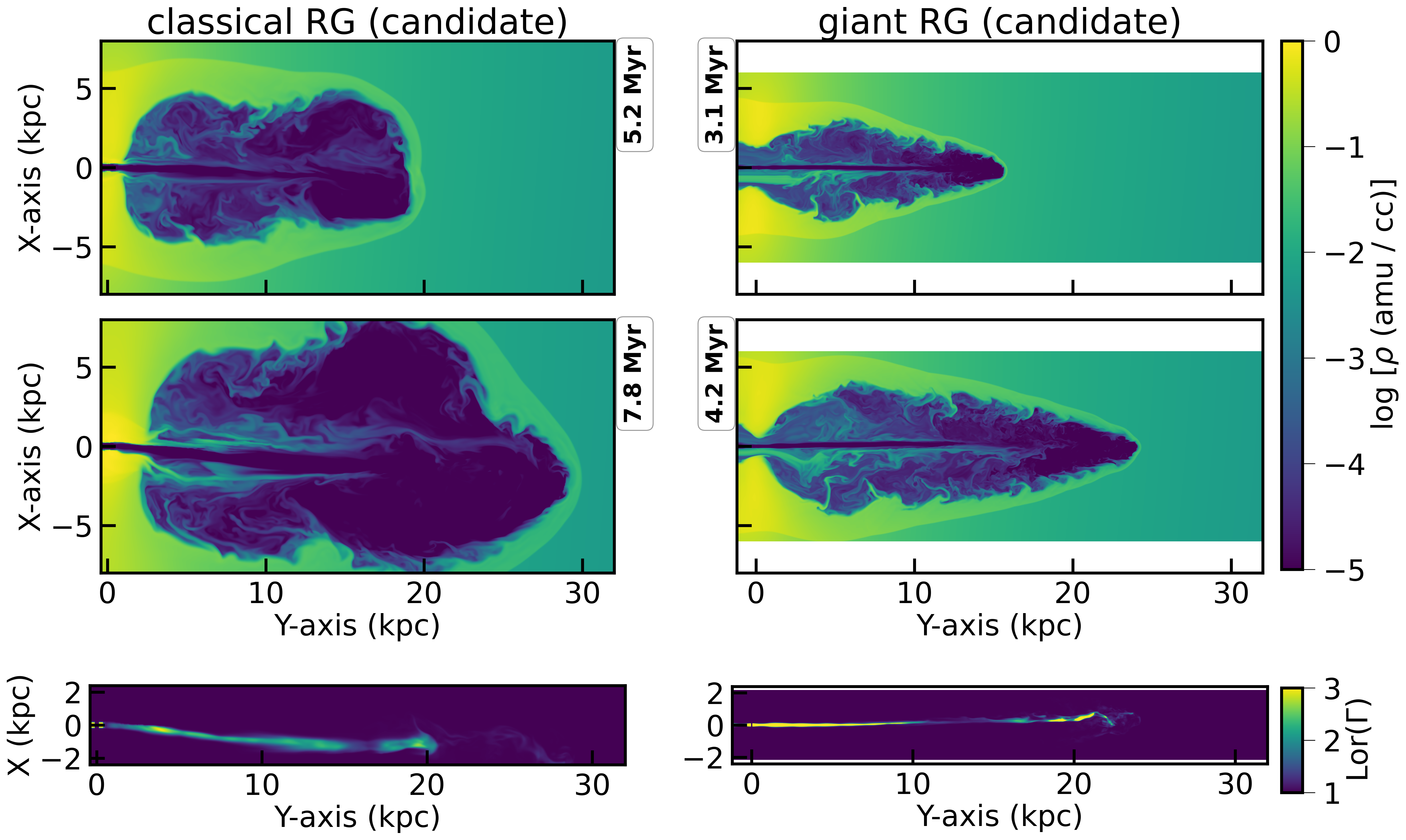}
\caption{Comparison between two simulations with identical jet power but differing intrinsic parameters—jet radius, velocity, and magnetic field—highlighting their impact on the spatio-temporal evolution of the jets (\textit{left:} `\texttt{highB\_min85}'; \textit{right:} `\texttt{lowB\_varRV\_min85}'). The top two rows show density slices (at $z=0$), illustrating the larger-scale jet morphology over an extended evolution period. The bottom row presents corresponding Lorentz factor slices, emphasizing jet-beam's (de)collimation and thrust in shaping different classes of radio galaxies.}
\label{Fig:CRG-GRG} 
\end{figure*}

Fig.~\ref{Fig:CRG-GRG} shows the longer-term evolution of the injected jets (compared to the relevant cases shown in Fig.~\ref{Fig:Morphology_Bs}, \ref{Fig:Morphology_RV}), where one case dissipates rapidly into a broad cocoon, while the other remains collimated with sustained thrust up to its head location, forming an arrowhead-shaped cocoon (also refer to as nose-cone structure). When comparing temporal scales over which jets evolve to similar spatial extents, the collimated jet exhibits more rapid growth, indicating the interpretation that the extreme linear sizes of GRGs can arise naturally within conventional jet dynamical models. From Fig.~\ref{Fig:CRG-GRG}, the typical lobe expansion speed for the GRG candidate is estimated to be $\sim 0.025c$, consistent with previous findings \citep[e.g.,][]{Machalski2011}, whereas the classical double candidate exhibits a slower expansion of $\sim0.013c$, nearly half that of the GRG case (injection jet power is same for both the cases). 

The formation of a broad cocoon in the potential classical double-lobed case (`\texttt{highB\_min85}') arises from the jet's tendency to develop lateral wiggling, driven by increased magnetic strength of the jet beam inducing kink-mode instabilities \citep{Mignone2013} and amplified further by its relatively lower flow speed \citep[e.g.,][]{Mukherjee2020}. These effects not only bend the jet spine but also decelerate the overall propagation by dispersing the jet flow into a relatively wider opening angle. In contrast, the `\texttt{lowB\_varRV\_min85}' simulation—despite having the same jet power—features a narrower jet radius and higher flow speed (Table~\ref{Tab:CRG-GRG}). These properties suppress environmental disruption, enhance collimation, and allow the jet to sustain its velocity up to the head. These distinctions are captured in the Lorentz factor distribution maps shown in the bottom row of Fig.~\ref{Fig:CRG-GRG}, highlighting the persistence of a high-speed, well-collimated jet spine in the GRG-like scenario \citep[aligning with expectations of, e.g.,][]{Oei2024_7Mpc}.

\section{Role of magnetization in modulating radio galaxy properties}\label{Sec:Role of magnetization in modulating radio galaxy properties}

The polarized synchrotron emissivity can be expressed as \citep[following,][]{Meenakshi2023,Giri2025_GRGEm},
\begin{equation}
\begin{split}
    \mathcal{J'}_{\rm pol} = N_0 \frac{3^{p/2}e^2\nu'^{-(p -1)/2}\left| \vec{B'} \times \vec{\hat{n'}} \right|^{(p+1)/2}}{8c} \\
    \mathcal{G} \left( \frac{p}{4} + \frac{7}{12} \right) \mathcal{G} \left( \frac{p}{4} - \frac{1}{12} \right) \left( \frac{e}{2\pi m_e c}\right)^{(p+1)/2} 
\end{split}
\end{equation}
The parameters governing the above formulation are consistent with the assumptions outlined in Section~\ref{Sec:Radiative imprints}. Analogous to the construction of the intensity map via line-of-sight integration along the $z-$axis, we compute the Stokes parameters $\mathcal{Q}_{\nu}$ and $\mathcal{U}_{\nu}$ through a similar line-of-sight integration.

\begin{equation}
\begin{split}
    \mathcal{Q}_{\nu} (\nu, X, Y) &= \int_{-\infty}^{\infty} \mathcal{J}_{\rm pol}\,\, cos2\chi  \, \, dZ \\
    \mathcal{U}_{\nu} (\nu, X, Y) &= \int_{-\infty}^{\infty} \mathcal{J}_{\rm pol}\,\, sin2\chi  \, \, dZ
\end{split}
\end{equation}
where, 
\begin{equation}
\begin{split}
    cos2\chi  = \frac{q_{\small X}^2 - q_{\small Y}^2}{q_{\small X}^2 + q_{\small Y}^2}; \,\,\, sin2\chi  = -\frac{2\, q_{\small X} \, q_{\small Y}}{q_{\small X}^2 + q_{\small Y}^2}
\end{split}
\end{equation}
and,
\begin{equation}
\begin{split}
    q_{\small X} = (1 - \beta_z)B_x - \beta_xB_z;\,\,\, q_{\small Y} = (1 - \beta_z)B_y - \beta_yB_z
\end{split}
\end{equation}

\begin{figure*}
\centering
\includegraphics[width=0.67\columnwidth, trim=150 70 150 70, clip]{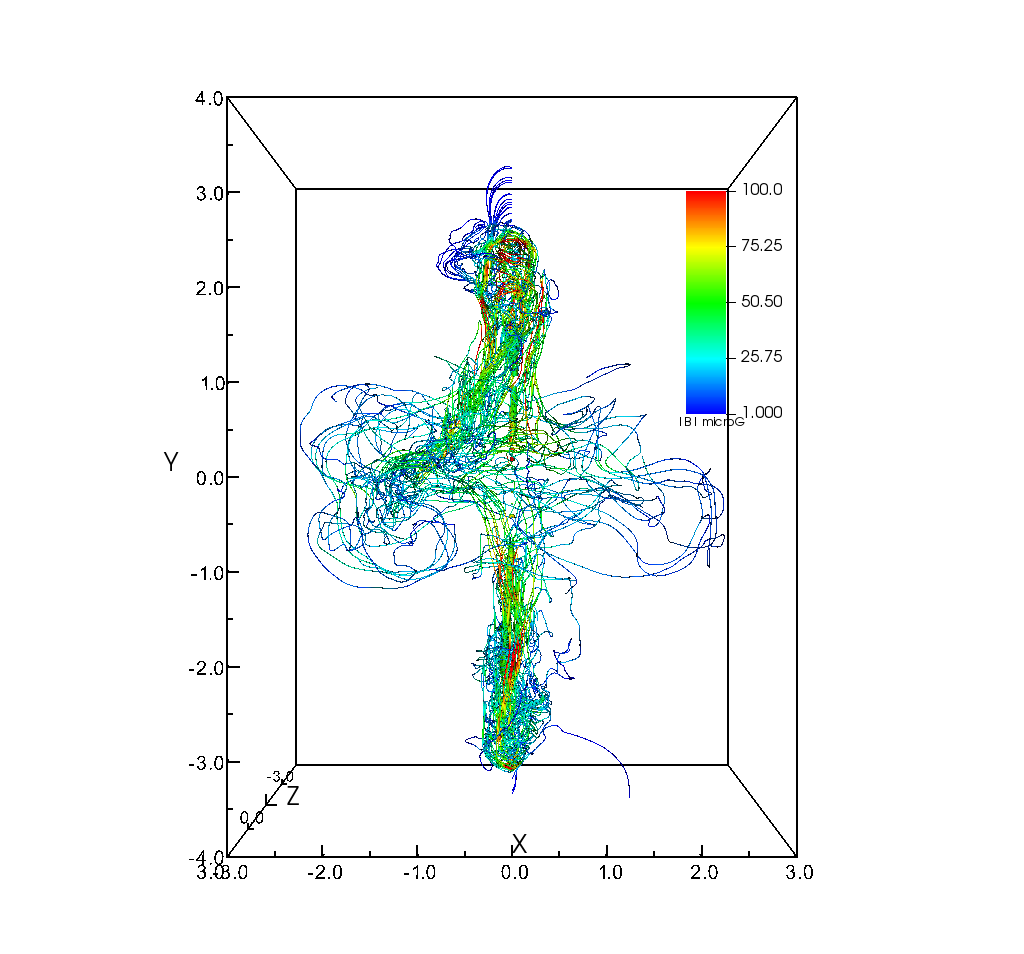}
\includegraphics[width=0.67\columnwidth, trim=150 70 150 70, clip]{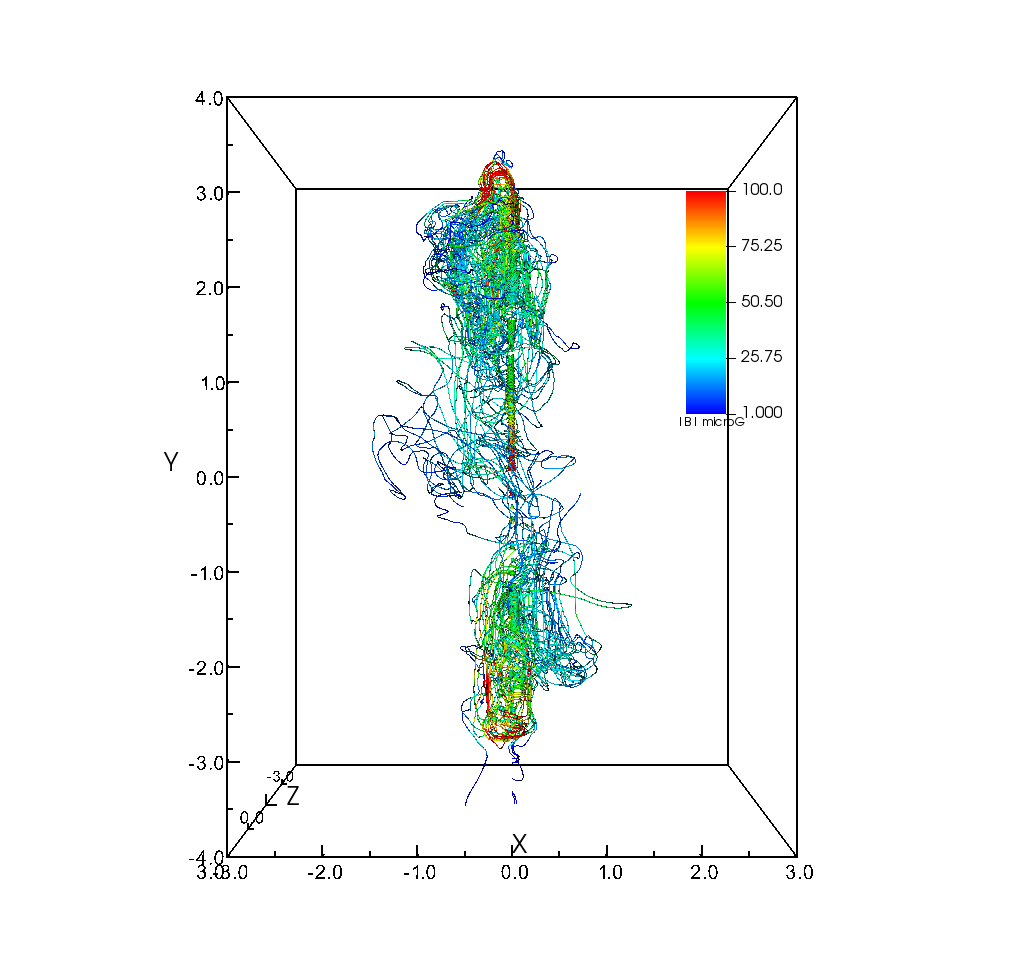}
\includegraphics[width=0.67\columnwidth, trim=150 70 150 70, clip]{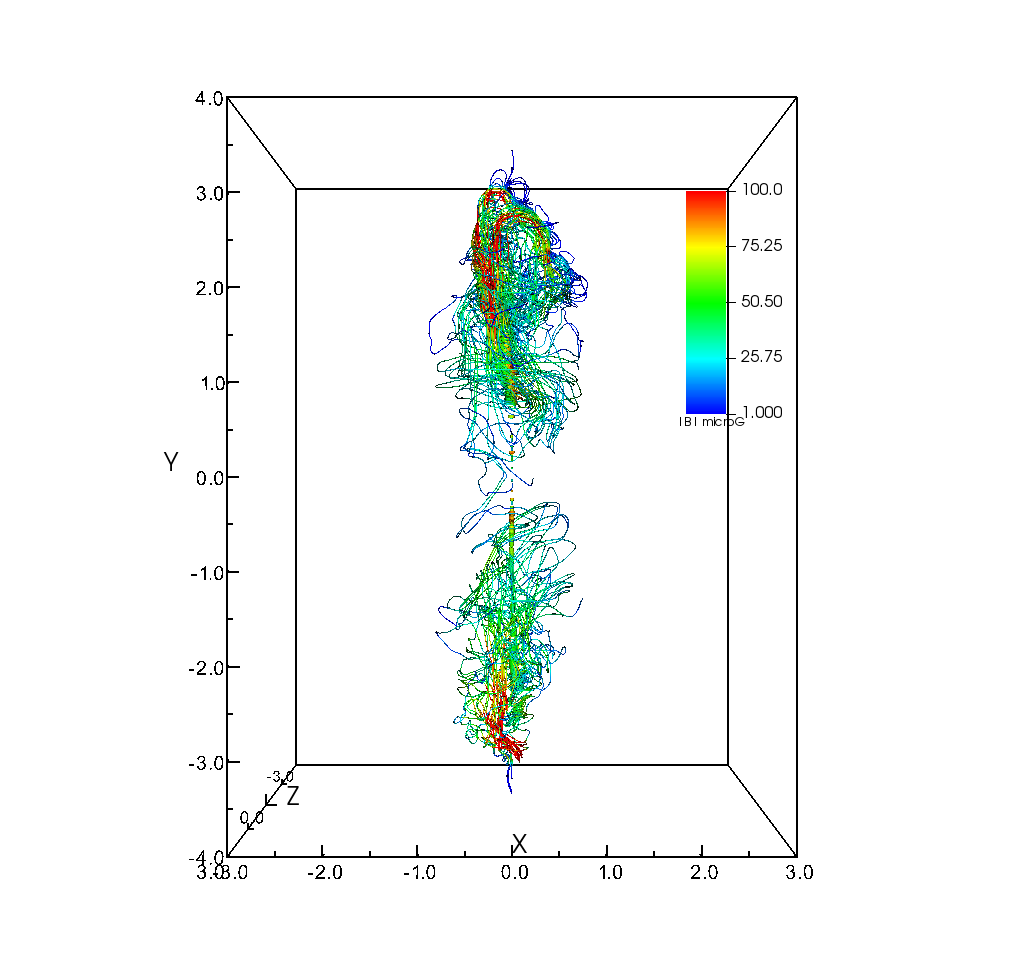}\\
\includegraphics[width=0.67\columnwidth, trim=150 70 150 70, clip]{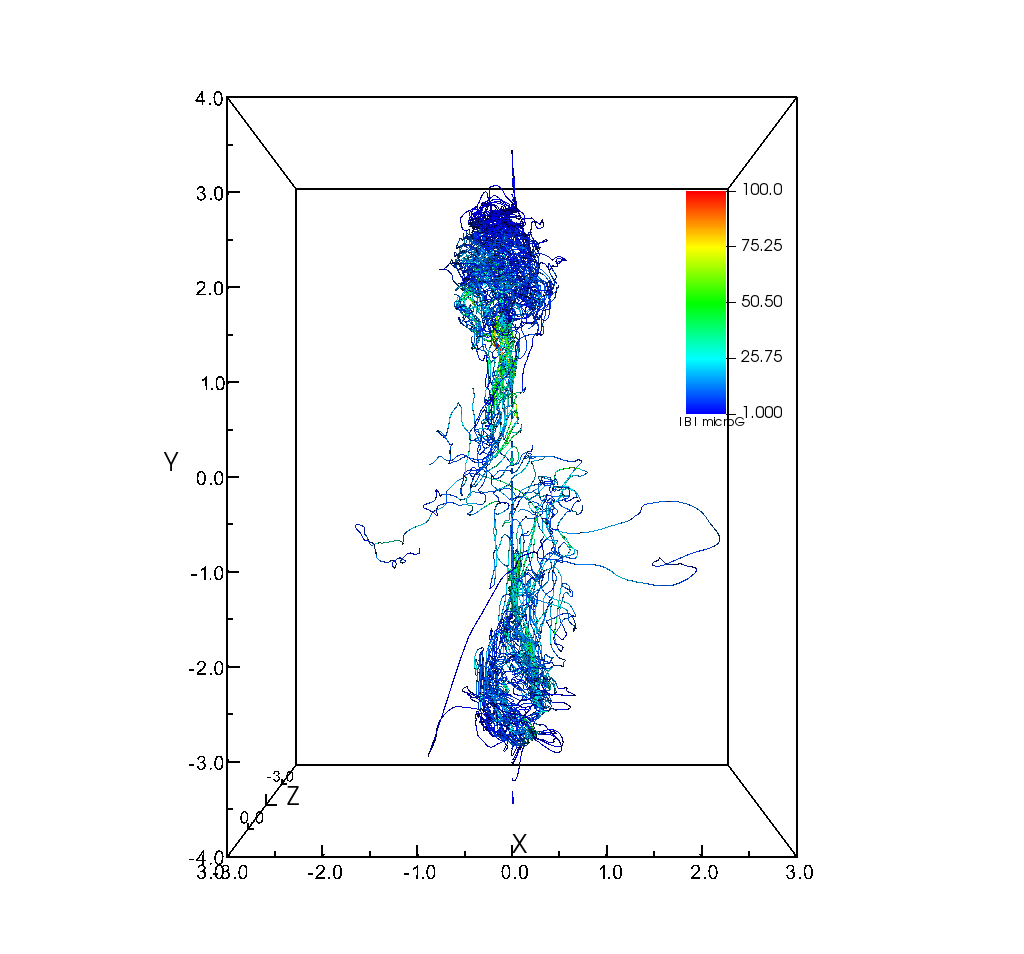}
\includegraphics[width=0.67\columnwidth, trim=150 70 150 70, clip]{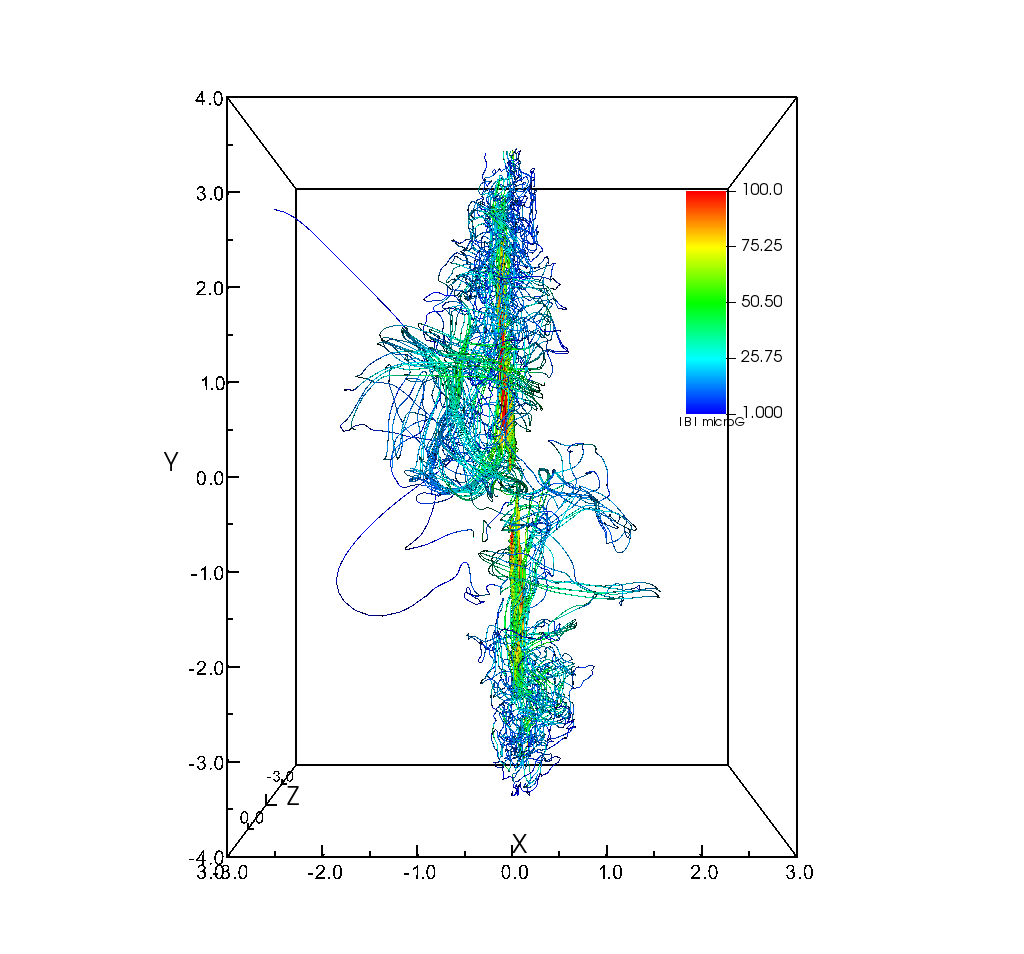}
\includegraphics[width=0.67\columnwidth, trim=150 70 150 70, clip]{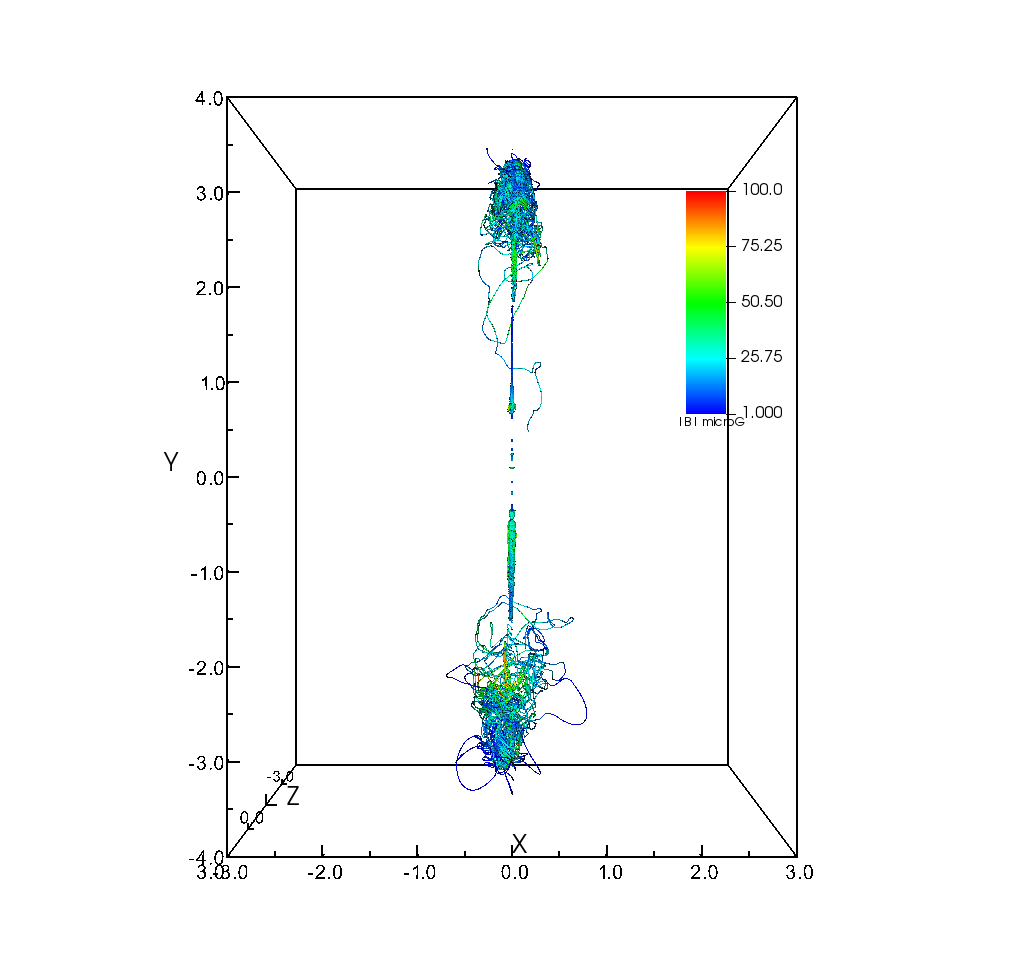}
\caption{3D volume renderings of the magnetic field line distribution for the high-B (\textit{top panels}) and low-B (\textit{bottom panels}) cases are shown, with jet propagation angles $\Theta = 5^\circ,\, 40^\circ,\, 85^\circ$ from \textit{left} to \textit{right} (Table~\ref{Tab:Parametric_space}). A common color scale ($|B|$ in $\mu$G) is adopted to enable direct comparison of field strength and spatial distribution. Although the magnetic field remains dynamically subdominant within the explored parameter regime, it plays a critical role in shaping the emission properties. In particular, the high-B cases exhibit enhanced magnetic field concentration at the terminal shocks and a stronger presence within secondary structures, leading to observable differences in radio galaxy morphology. Regardless of magnetization level, the tangled field topology highlights the complexity of the cocoon environment, supporting the development of turbulence and providing favorable conditions for particle acceleration.
}
\label{Fig:B-Dyn_3D} 
\end{figure*}

\begin{figure*}
\centering
\includegraphics[width=0.67\columnwidth, trim=0 10 120 0, clip]{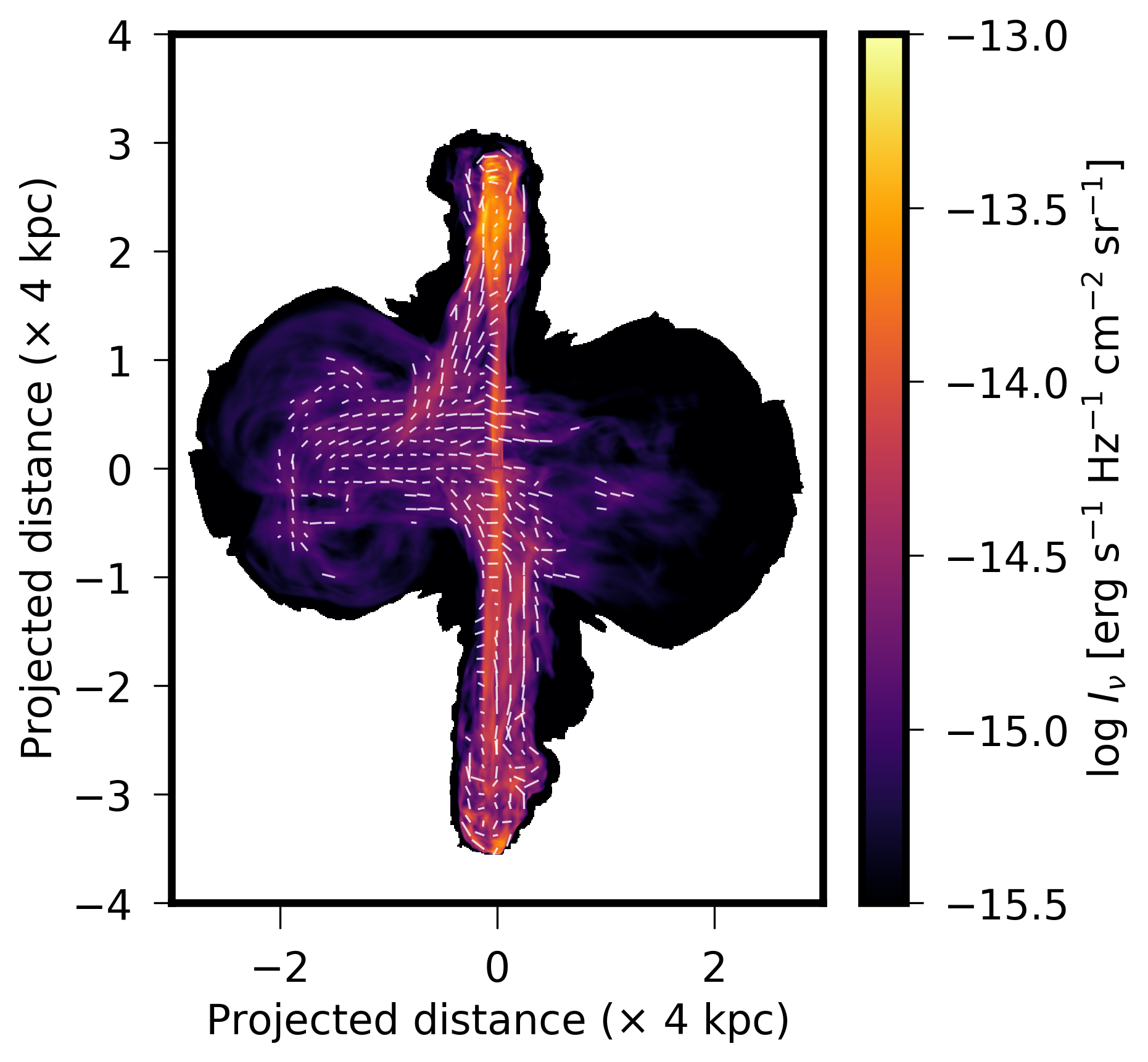}
\includegraphics[width=0.67\columnwidth, trim=0 10 120 0, clip]{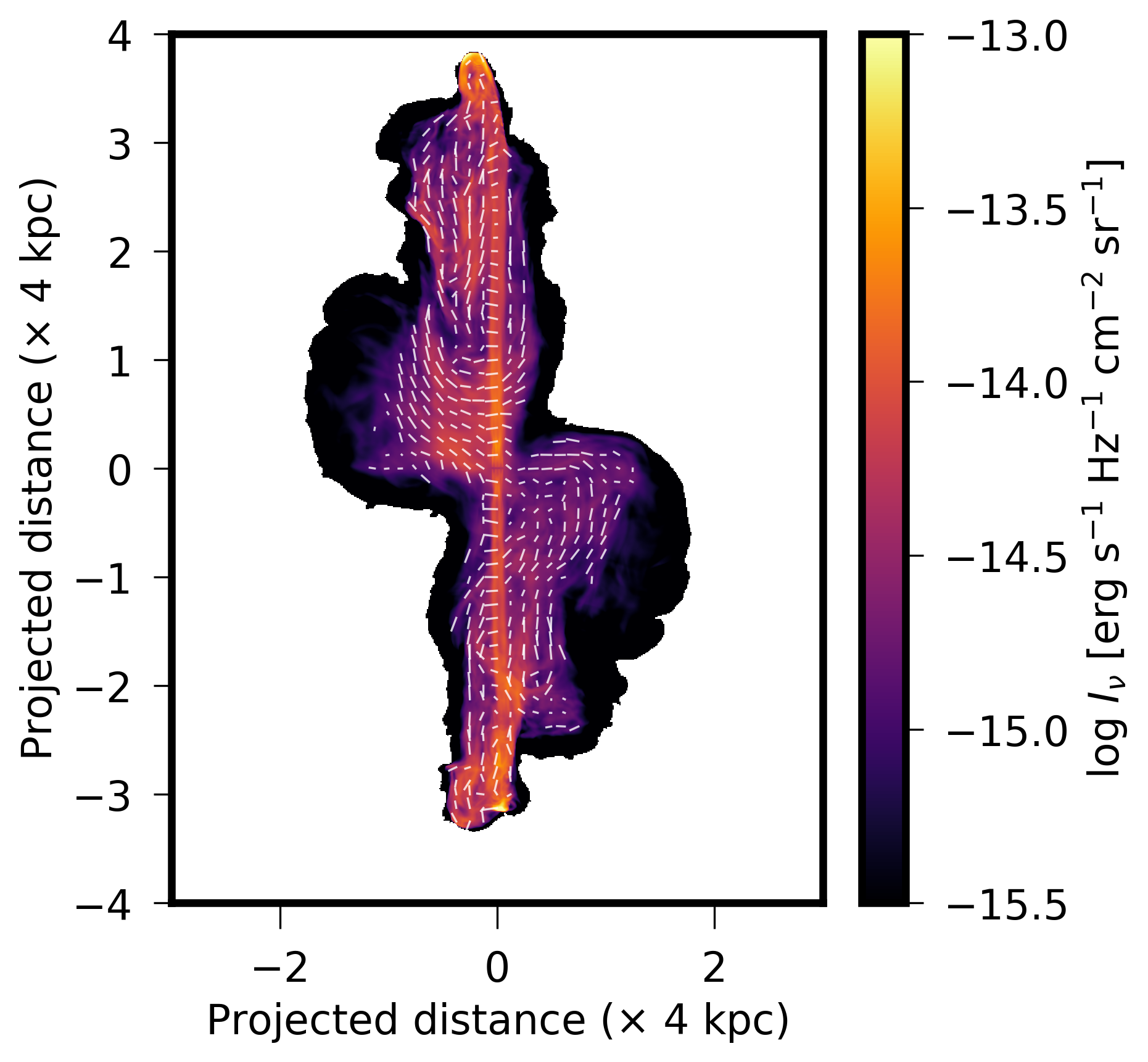}
\includegraphics[width=0.67\columnwidth, trim=0 10 120 0, clip]{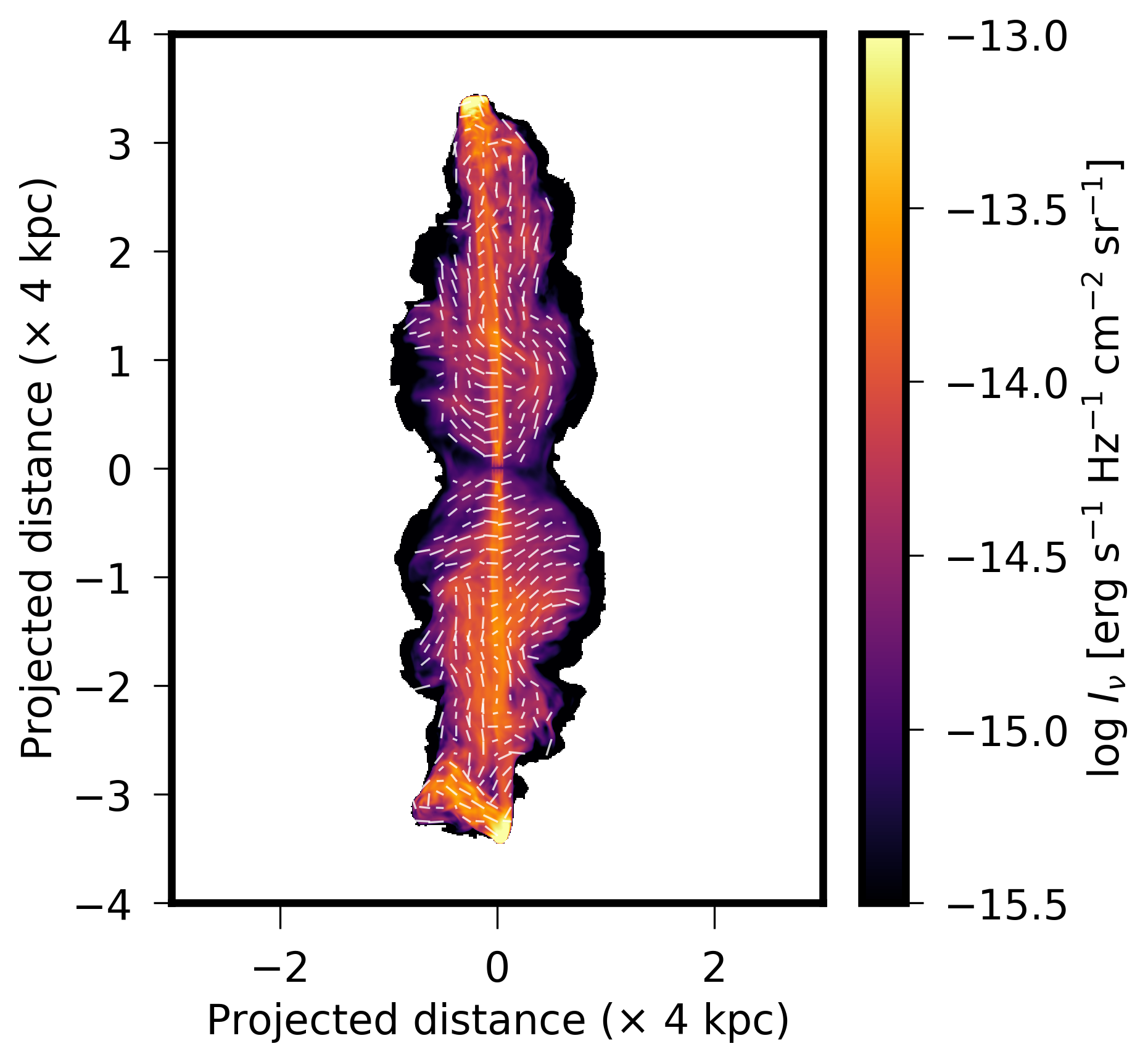}\\
\includegraphics[width=0.67\columnwidth, trim=0 10 120 0, clip]{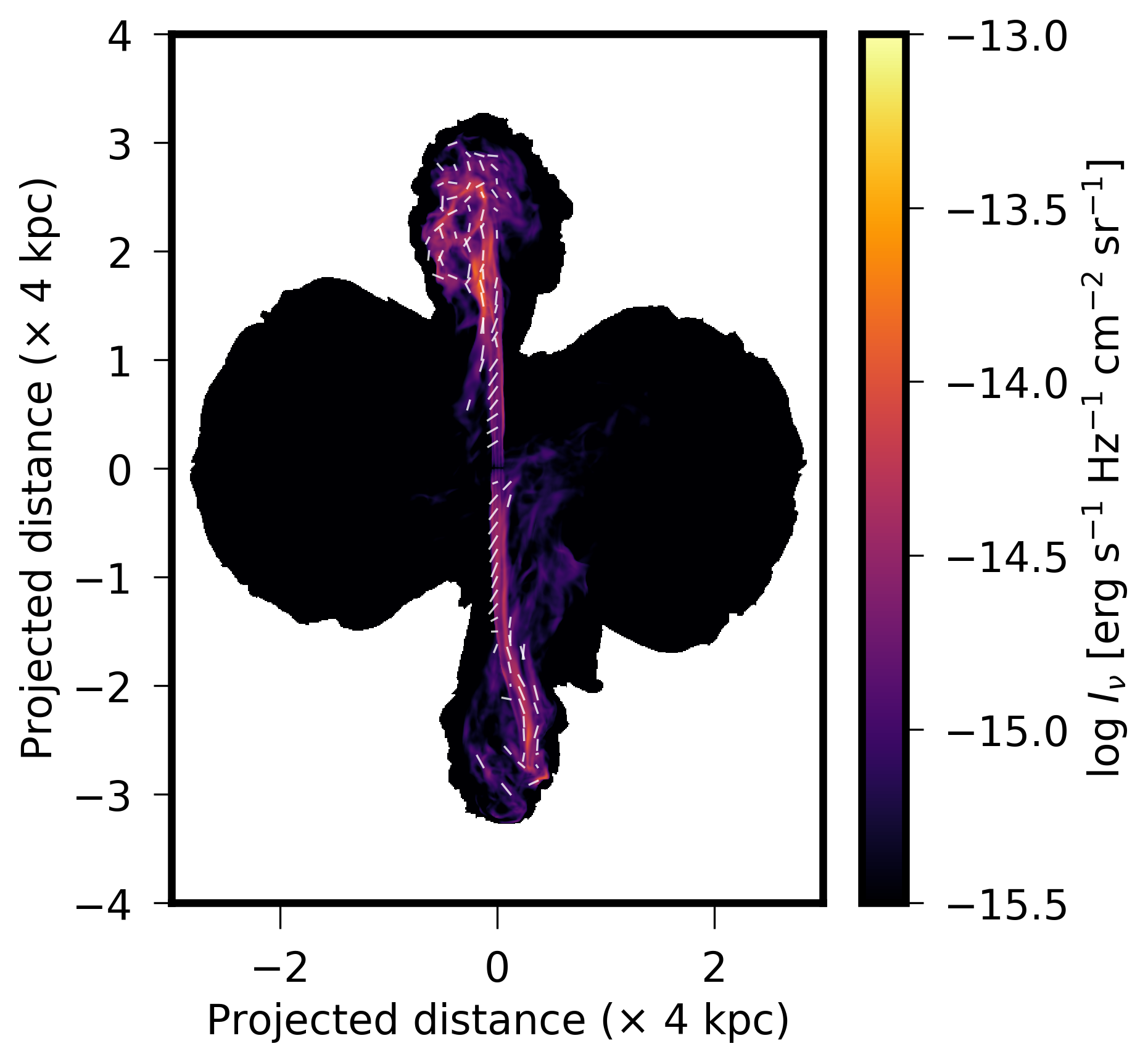}
\includegraphics[width=0.67\columnwidth, trim=0 10 120 0, clip]{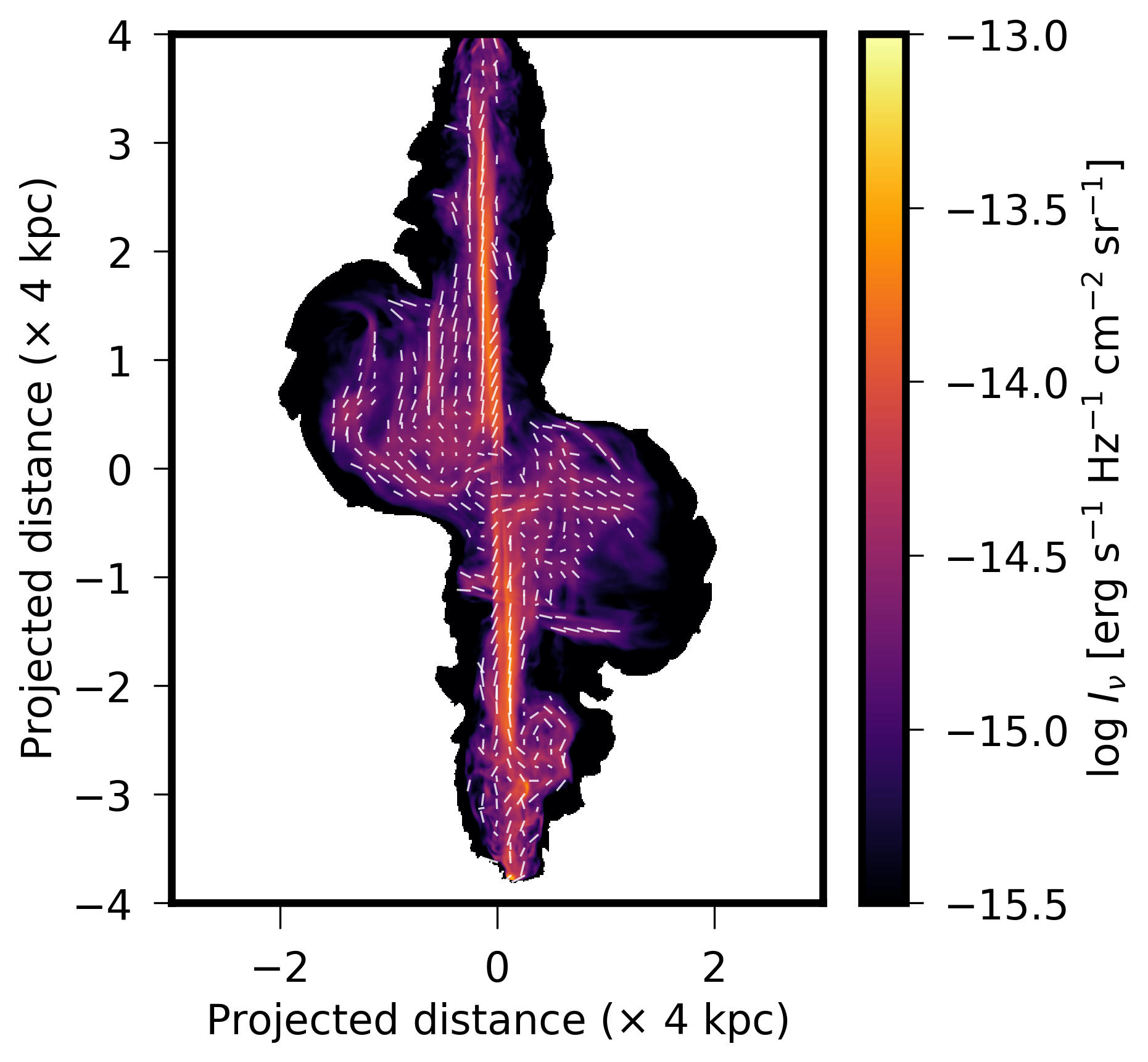}
\includegraphics[width=0.67\columnwidth, trim=0 10 120 0, clip]{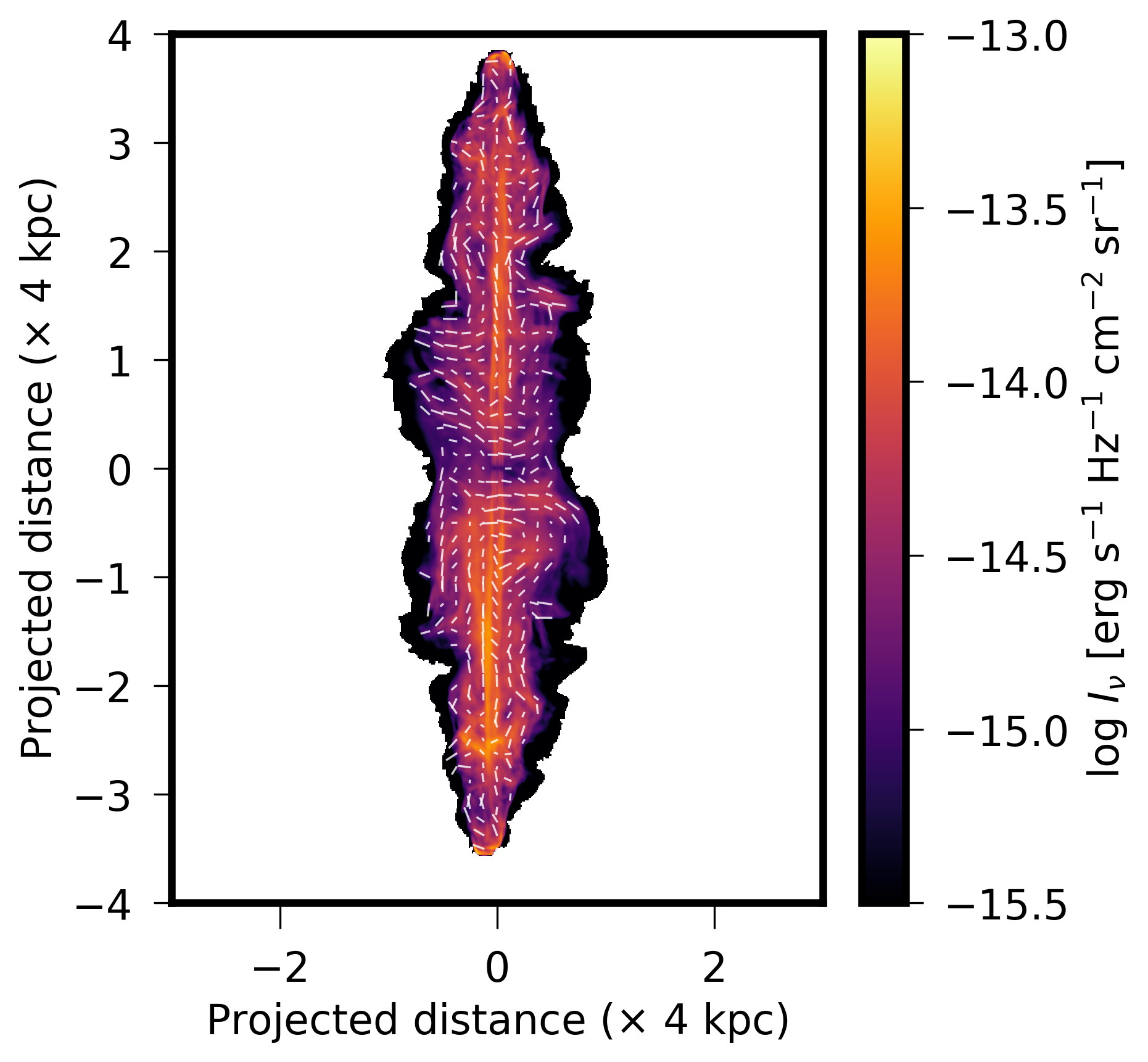}
\caption{Projected magnetic field orientations, derived from polarized emission (line lengths are proportional to fractional polarization values), are overlaid on the intensity colormap (using the same color-scale range as Fig.~\ref{Fig:Emission_col}). The resulting distribution closely mirrors the broad-scale magnetic field structure seen in the 3D volume-rendered maps (Fig.~\ref{Fig:B-Dyn_3D}). The evolution from an initially toroidal field configuration (at injection) to a helicoidal and subsequently poloidal structure is particularly evident along the jet spine. This field is further compressed at the hotspot regions, a feature that is especially pronounced in the high-magnetization cases (\textit{top panel}). Within the cocoon and secondary structures, the magnetic field appears to broadly align with the underlying flow patterns \citep[e.g.,][]{Black1992,Baghel2023}.}
\label{Fig:B-Em_Pol} 
\end{figure*}

To provide a broad understanding of the complexity associated with magnetic fields in radio galaxies, we first present a 3D volume rendering of the magnetic field distribution (Fig.~\ref{Fig:B-Dyn_3D}), followed by its projected representation derived from polarized emission (Fig.~\ref{Fig:B-Em_Pol}). A direct comparison of these figures highlights that the intrinsic three-dimensional magnetic field structure is significantly more complex than the projected polarization pattern. The latter reflects a line-of-sight integrated view, where the contributions from highly tangled magnetic field components partially cancel out. This intrinsic complexity is expected to manifest in the fractional polarization (FP) distribution; however, a detailed analysis of the fractional polarization is deferred to future work at more evolved scales \citep{Gurkan2022,Bhukta2022}, where such effects can be examined more systematically. At present, we plotted the B-field lines in Fig.~\ref{Fig:B-Em_Pol}, whose length is proportional to the estimated FP values.

From the dynamical perspective (Fig.~\ref{Fig:B-Dyn_3D}), higher magnetization leads to more efficient transport of magnetic flux along the jet spine toward the jet head, resulting in enhanced magnetic field amplification at the termination region. This behavior is consistently reflected in the polarization maps, where the magnetic field vectors appear more compressed (line alignment perpendicular to jet flow) in the high-magnetization cases. In contrast, lower magnetization cases exhibit a mixed behavior: sometimes with similar pattern of compression at jet termini (e.g., the classical double case), and sometimes more diffusive, flow-aligned structures (e.g., in the X-shaped morphologies). 

We note here that the differences in magnetic field strengths within the secondary structures (when compared between low and high magnetization cases) further contribute to features such as one-sided lobes or fading cocoons, as also illustrated in Fig.~\ref{Fig:Emission_col}. Regardless of the absolute field strength, the inherently chaotic nature of the flow and magnetic field suggests that these regions may serve as potential sites for weak particle re-acceleration, supporting conclusions of \citet{Mukherjee2021,Giri2022_XRG,Kundu2022}.

We further highlight three key aspects. First, the emergence of filamentary, loop-like, and transversely threaded structures (Fig.~\ref{Fig:B-Em_Pol}) in the emission maps points to magnetically dominated features \citep[][]{Meenakshi2024,Giri2025_GRGEm} that are of particular relevance for contemporary observational facilities in radio-bands. Second, the initially injected toroidal magnetic field evolves into helicoidal and, further downstream the jet, poloidal configurations, consistent with compression driven by shocks and shear \citep{Laing1981}. Third, magnetohydrodynamic instabilities that induce jet bending or wobbling can give rise to complex hotspot chain, indicating that such features should not be unambiguously interpreted as signatures of jet precession \citep[e.g.,][]{Horton2025}. We are currently formulating a dedicated follow-up simulation campaign aimed at testing these effects under a broader range of physical conditions.

\end{appendix}

\end{document}